\documentclass[aps,prx,a4paper,superscriptaddress,twocolumn,amsmath,amssymb,longbibliography]{revtex4-1}

\usepackage{bm}
\usepackage{graphicx,epsfig}
\usepackage[usenames]{color}
\usepackage[english]{babel}

\usepackage{array}
\usepackage{dsfont}
\usepackage{multirow}
\usepackage{csquotes}
\usepackage{enumitem}

\usepackage{orcidlink}

\MakeOuterQuote{"}
\allowdisplaybreaks

\begin{document}

\newcommand{\B}{\mathcal{B}}
\newcommand{\E}{\mathcal{E}}
\newcommand{\G}{\mathcal{G}}
\newcommand{\Lag}{\mathcal{L}}
\newcommand{\M}{\mathcal{M}}
\newcommand{\N}{\mathcal{N}}
\newcommand{\U}{\mathcal{U}}
\newcommand{\R}{\mathcal{R}}
\newcommand{\F}{\mathcal{F}}
\newcommand{\V}{\mathcal{V}}
\newcommand{\C}{\mathcal{C}}
\newcommand{\I}{\mathcal{I}}
\newcommand{\s}{\sigma}
\newcommand{\up}{\uparrow}
\newcommand{\dw}{\downarrow}
\newcommand{\h}{\hat{H}}
\newcommand{\himp}{\hat{h}}
\newcommand{\g}{\mathcal{G}^{-1}_0}
\newcommand{\D}{\mathcal{D}}
\newcommand{\A}{\mathcal{A}}
\newcommand{\projs}{\hat{\mathcal{S}}_d}
\newcommand{\proj}{\hat{\mathcal{P}}_d}
\newcommand{\K}{\textbf{k}}
\newcommand{\Q}{\textbf{q}}
\newcommand{\T}{\tau_{\ast}}
\newcommand{\io}{i\omega_n}
\newcommand{\eps}{\varepsilon}
\newcommand{\+}{\dag}
\newcommand{\su}{\uparrow}
\newcommand{\giu}{\downarrow}
\newcommand{\0}[1]{\textbf{#1}}

\newcommand{\og}[1]{\textcolor{blue}{[\textbf{OG}: #1]}}
\newcommand{\sg}[1]{\textcolor{orange}{[\textbf{SG}: #1]}}
\newcommand{\dagga}{{\phantom{\dagger}}}
\newcommand{\ca}{c^{\phantom{\dagger}}}
\newcommand{\cc}{c^\dagger}

\newcommand{\pa}{{p}^{\phantom{\dagger}}}
\newcommand{\pc}{{p}^\dagger}

\newcommand{\fa}{f^{\phantom{\dagger}}}
\newcommand{\fc}{f^\dagger}

\newcommand{\da}{{d}^{\phantom{\dagger}}}
\newcommand{\dc}{{d}^\dagger}

\newcommand{\bF}{\mathbf{F}}
\newcommand{\bD}{\mathbf{D}}

\newcommand{\bR}{\mathbf{R}}
\newcommand{\bQ}{\mathbf{Q}}
\newcommand{\bq}{\mathbf{q}}
\newcommand{\bqp}{\mathbf{q'}}
\newcommand{\bk}{\mathbf{k}}
\newcommand{\bh}{\mathbf{h}}
\newcommand{\bkp}{\mathbf{k'}}
\newcommand{\bp}{\mathbf{p}}
\newcommand{\bL}{\mathbf{L}}
\newcommand{\bRp}{\mathbf{R'}}
\newcommand{\bx}{\mathbf{x}}
\newcommand{\bX}{\mathbf{X}}
\newcommand{\by}{\mathbf{y}}
\newcommand{\bz}{\mathbf{z}}
\newcommand{\br}{\mathbf{r}}
\newcommand{\Ima}{{\Im m}}
\newcommand{\Rea}{{\Re e}}
\newcommand{\Pj}[2]{|#1\rangle\langle #2|}
\newcommand{\ket}[1]{\vert#1\rangle}
\newcommand{\bra}[1]{\langle#1\vert}
\newcommand{\setof}[1]{\left\{#1\right\}}
\newcommand{\fract}[2]{\frac{\displaystyle #1}{\displaystyle #2}}
\newcommand{\Av}[2]{\langle #1|\,#2\,|#1\rangle}
\newcommand{\av}[1]{\langle #1 \rangle}
\newcommand{\Mel}[3]{\langle #1|#2\,|#3\rangle}
\newcommand{\Avs}[1]{\langle \,#1\,\rangle_0}
\newcommand{\eqn}[1]{(\ref{#1})}
\newcommand{\Tr}{\mathrm{Tr}}

\newcommand{\bba}{b^{\phantom{\dagger}}}
\newcommand{\bbc}{b^\dagger}

\title{Unifying Variational and Dynamical Quantum Embedding: From Ghost Gutzwiller Approximation to Dynamical Mean-Field Theory}

\author{Samuele Giuli}
\affiliation{Center for Computational Quantum Physics, Flatiron Institute, New York, New York 10010, USA}

\author{Tsung-Han Lee}
\affiliation{Department of Physics, National Chung Cheng University, Chiayi 62102, Taiwan}

\author{Yong-Xin Yao}
\affiliation{Ames National Laboratory, Ames, Iowa 50011, USA}
\affiliation{Department of Physics and Astronomy, Iowa State University, Ames, Iowa 50011, USA}

\author{Gabriel Kotliar}
\affiliation{Physics and Astronomy Department, Rutgers University, Piscataway, New Jersey 08854, USA}
\affiliation{Condensed Matter Physics and Materials Science Department, Brookhaven National Laboratory, Upton, New York 11973, USA}

\author{Andrei E. Ruckenstein}
\affiliation{Physics Department, Boston University, Boston, Massachusetts 02215, USA}

\author{Olivier Gingras}
\affiliation{Center for Computational Quantum Physics, Flatiron Institute, New York, New York 10010, USA}
\affiliation{Université Paris-Saclay, CNRS, CEA, Institut de physique théorique, 91191, Gif-sur-Yvette, France}

\author{Nicola Lanat\`a} 
\altaffiliation{Corresponding author: nxlsps@rit.edu}
\affiliation{School of Physics and Astronomy, Rochester Institute of Technology, Rochester, New York 14623, USA}
\affiliation{Center for Computational Quantum Physics, Flatiron Institute, New York, New York 10010, USA}

\date{\today}

\begin{abstract}

Dynamical and variational frameworks have long been viewed as distinct paradigms.
In particular, in quantum embedding (QE) frameworks, dynamical mean-field theory (DMFT) captures nonperturbative dynamical correlations through a frequency-dependent self-energy, while the Gutzwiller approximation (GA) is formulated in terms of a variationally optimized ground-state wavefunction.
Here we bridge these perspectives, proving that the ghost-Gutzwiller approximation (ghost-GA), which also admits a density-matrix-matching QE formulation known as ghost density matrix embedding theory (ghost-DMET), becomes strictly equivalent to DMFT in the limit of infinitely many auxiliary bath modes.
This formal unification has immediate consequences. In particular, it yields a rigorous finite-temperature extension of ghost-GA and shows that the physical Green's function can be determined from static expectation values of the embedding Hamiltonians, providing a route to computational studies of competing phases in strongly correlated matter with DMFT-level accuracy, while bypassing the need to calculate dynamical spectra with conventional impurity solvers.
More broadly, it shows that the variational ghost-GA, the density-matrix-matching ghost-DMET formulation, and the dynamical DMFT description are not separate constructions, but complementary formulations of the same QE structure, thereby providing a concrete formal basis for future controlled extensions beyond DMFT.

\end{abstract}

\maketitle

\section{Introduction}

Strongly correlated electrons~\cite{Kotliar-Science} generate a wide range of collective phenomena, from metal-insulator transitions~\cite{Mott-book} to unconventional superconductivity~\cite{highTc-1,highTc-2,highTc-3,highTc-4,highTc-5,highTc-6,highTc-7,highTc-8} and magnetism, that remain difficult to predict from first principles~\cite{corr-structure-6,NSTC2011,Prasankumar2026Path}.
QE methods address this challenge by treating a selected set of local degrees of freedom with an interacting solver while describing the rest of the system through an effective environment determined self-consistently.
Among them, dynamical mean-field theory (DMFT)~\cite{dmft_book,Anisimov_DMFT,Held-review-DMFT,DMFT,xidai_impl_LDA+DMFT,LDA+U+DMFT} provides a nonperturbative description of local dynamical correlations and has become a standard framework for correlated materials, especially when combined with density-functional theory (DFT) in the DFT+DMFT approach~\cite{Metzner-Vollhardt-PRL,DMFT,dmft_book,Anisimov_DMFT,Held-review-DMFT,xidai_impl_LDA+DMFT,LDA+U+DMFT}.
However, accounting for these dynamical effects entails a significant computational cost, which often hinders the efficient simulation of complex materials.

In contrast, static QE schemes formulate the problem in terms of static expectation-value self-consistency conditions, offering higher computational efficiency at the cost of sacrificing the resolution of frequency-dependent spectral features.
Standard examples include the Gutzwiller approximation (GA)~\cite{Gutzwiller3,Our-PRX,lanata-barone-fabrizio,mybil,GA-infinite-dim,Fang,Ho,Gmethod,Bunemann,Attaccalite,Gebhard-FL}, the rotationally invariant slave-boson (RISB) theory~\cite{Kotliar-Ruckenstein,rotationally-invariant_SB,Lanata2016}, and density matrix embedding theory (DMET)~\cite{DMET,DMET-spectral,DMET-qchem,Bulik-DMET,dmet-new-2,dmet-new-3,dmet-new-4,dmet-new-5,dmet-new-6,dmet-thesis}.
%
%
A systematic variational pathway to enhance the accuracy from a static QE perspective is the ghost-GA~\cite{Ghost-GA,ALM_g-GA,gRISB,gDMET,Mejuto2023}, which augments the GA variational space by incorporating auxiliary ``ghost'' fermionic degrees of freedom---a recurrent theme in many-body theory, including extensions to DMET~\cite{Booth-ghost,Booth-ghost2}, matrix product states (MPS)~\cite{itensor,block2,DMRG-REVIEW,DMRG-original-White-PRL,DMRG-original-White-PRB,DMRG_PhysRevB.104.115119} and projected entangled pair states~\cite{Vaestrate-VRG}, the ancilla qubit technique~\cite{ghost-Subir}, and recent extensions of neural network states~\cite{ghost-NeuralNetworks}.
Specifically, the ghost-GA ansatz is constructed by applying a variationally-optimized local non-perturbative map to a non-interacting Slater determinant living in the enlarged Hilbert space generated by auxiliary ghost fermionic modes.

A remarkable property of the ghost-GA is that it describes in this fashion not only the ground state, but also provides an explicit wavefunction representation for the excited states. Specifically, it portrays both the heavy quasiparticles and the Hubbard bands as the image of non-interacting fermionic excitations defined in the extended Hilbert space, which are connected to the physical system via the same non-perturbative map calculated by optimizing the ground-state variational energy~\cite{Ghost-GA,ALM_g-GA,Gebhard-FL}. This construction presents suggestive analogies with the concepts of hidden fermions~\cite{ghost-HF} and hidden Fermi liquids~\cite{Ghost-Anderson}.
Recently, this emergent quasiparticle picture has also enabled a rigorous wavefunction perspective for correlated topological phases~\cite{Pasqua-2025}, showing that the ghost-GA can describe topologically nontrivial Hubbard bands hosting their own protected edge states.
The same wavefunction perspective has also proved fruitful in other contexts, including the exciton Mott transition~\cite{Guerci}, neutral spinon excitations that re-emerge as heavy-fermion bands by proximity~\cite{gGA-PhysRevB.111.125110}, interaction-driven altermagnetism and its tunable transport properties~\cite{Altermagnetism-gGA}, and a rigorous bridge from fully interacting chemistry to an effective one-body quasiparticle language, including a reformulation of Woodward--Hoffmann rules~\cite{Mejuto2026}.

Since both ghost-GA and DMFT are formulated by neglecting nonlocal correlation contributions that vanish in the limit of infinite coordination number, a fundamental link between these theories is expected.
Nevertheless, dynamical and variational embeddings appear to be mutually exclusive: DMFT is intrinsically dynamical, encoding the physics of spectral weight transfer and quasiparticle lifetimes through a continuous frequency-dependent self-energy; whereas variational methods, by definition, do not explicitly contemplate these dynamical fluctuations within their formalism, optimizing instead a single variational ground state.
This raises a fundamental question: Is it possible for such a static wavefunction to strictly encode the complete dynamical information of the Green's function, or is the frequency dependence irretrievably lost in the variational formalism?
In this work, we resolve this apparent dichotomy by proving that ghost-GA becomes strictly equivalent to DMFT in the limit of infinitely many ghost modes.
This equivalence implies that the same local variational map that defines the ghost-GA ground state also maps the elementary single-particle excitations of the auxiliary non-interacting reference system onto a complete representation of the physical one-particle spectrum, so that the full frequency-dependent structure of the DMFT self-energy is encoded within static variational parameters.

This correspondence has profound immediate consequences. In particular, it opens the possibility of using highly efficient ground-state impurity solvers for calculating both ground-state observables and spectral properties.
Besides MPS~\cite{itensor,block2,DMRG-REVIEW,DMRG-original-White-PRL,DMRG-original-White-PRB,DMRG_PhysRevB.104.115119}, already employed within the ghost-GA/g-RISB framework for multiorbital models and realistic materials~\cite{TH2,TH3}, this includes neural quantum states (NQS)~\cite{carleo2017solving,sharir2022neural,chen2024empowering,ghost-NeuralNetworks,GOLDSHLAGER2024113351,levine2019quantum,yu2024solving,Zhou-2025}, methods based on coupled-cluster (CC) theory~\cite{CC1,CC2,CC3,CCSD-T-1,CCSD-T-2,CCSD-T-3,Sun2026-stochastic-CCSD}, variational impurity solvers based on superpositions of Gaussian states~\cite{BravyiGosset2017,Bauer-impurity-solver,Hogan2025EfficientQuantumImplementationDMFT}, quantum-assisted methods~\cite{Sriluckshmy2025,AVQITE,Error-mitigation-GPR,Chen2025,Kirby2026ObservationImprovedAccuracy,Rigo2025OperatorLanczosNQS}, and machine learning frameworks~\cite{surrogate-Marius,Linear-foundation-model-ghostGA-2025,Rende-foundation-NNQS,Zaklama2025AttentionBased-foundation-attention}.
Furthermore, our formalism yields a principled finite-temperature extension of ghost-GA that can be equivalently obtained from a functional reformulation of DMFT, whose stationarity conditions can be fully formulated in terms of static self-consistency conditions.
This result opens a route to accurate finite-temperature studies of competing phases in strongly correlated matter at substantially lower computational cost.


The manuscript is organized as follows.
In Sec.~\ref{subsec:gGA} we introduce the ghost-GA framework and set the notation.
In Sec.~\ref{sec:lagrange_formulation} we present the central theoretical result of this work, namely the functional reformulation of ghost-GA that underpins the connection between the static QE perspective of Ref.~\cite{gDMET} and the dynamical one of DMFT~\cite{LDA+U+DMFT}.
In Sec.~\ref{sec:derivation_new_lagrange} we show that this reformulation is equivalent to the conventional ghost-GA formulation of Refs.~\cite{Ghost-GA,ALM_g-GA}.
In Sec.~\ref{sec:equivalence_gga_dmft_Binfty} we establish the connection between ghost-GA and DMFT by proving that, in the infinite-bath limit, the ghost-GA fixed point converges to the DMFT fixed point.
In Sec.~\ref{sec:finiteT_functional_BK} we develop the finite-temperature extension of the formalism and connect it to the DMFT functional~\cite{LDA+U+DMFT}.
In Sec.~\ref{sec:numerics} we present numerical validations and benchmarks.
Finally, in Sec.~\ref{sec:conclusions} we summarize the main results and discuss their implications.

\section{The ghost-GA formalism}
\label{subsec:gGA}

This section introduces the ghost-GA formalism and the notation employed in its formulation. In Sec.~\ref{sec:def_hubbard} we define the multi-orbital Hubbard Hamiltonian and its block structure. In Sec.~\ref{sec:gga_setup} we outline the ghost-GA variational ansatz and formalism, as developed in Refs.~\cite{Ghost-GA,ALM_g-GA}.

\subsection{Multi-orbital Hubbard Hamiltonian}
\label{sec:def_hubbard}

We consider a general multi-orbital Hubbard Hamiltonian written as
\begin{align}
\h=\hat{H}_0+\hat{H}_{\mathrm{int}}
\,,
\label{eq:H_split}
\end{align}
where the interaction part is assumed local,
\begin{align}
\hat{H}_{\mathrm{int}}=\sum_{i=1}^{\N}\hat{H}^{i}_{\mathrm{int}}
\,,
\label{eq:Hint_def}
\end{align}
and the one-body part is
\begin{align}
\hat{H}_0
=
\sum_{i,j=1}^{\N}
\sum_{\alpha=1}^{\nu_i}\sum_{\beta=1}^{\nu_j}
[h_0]_{i\alpha,j\beta}\,\cc_{i\alpha}\ca_{j\beta}
\,.
\label{eq:H0_def}
\end{align}
Here $i=1,\dots,\N$ labels system fragments, each hosting $\nu_i$ physical fermionic modes with annihilation operators $\ca_{i\alpha}$ ($\alpha=1,\dots,\nu_i$), including both spin and orbital degrees of freedom.

We represent the one-body matrix $h_0$ in block form as
\begin{align}
h_0=
\begin{pmatrix}
\eps_{1} & t_{12} & \dots & t_{1\N}\\
t_{21} & \eps_{2} & \dots & \vdots\\
\vdots & \vdots & \ddots & \vdots\\
t_{\N 1} & \dots & \dots & \eps_{\N}
\end{pmatrix}
\,,
\label{eq:h0_block}
\end{align}
where $\eps_i\in\mathbb{C}^{\nu_i\times\nu_i}$ are Hermitian local one-body blocks and $t_{ij}\in\mathbb{C}^{\nu_i\times\nu_j}$ ($i\neq j$) are the hopping blocks satisfying $t_{ji}=t_{ij}^\dagger$ (equivalently, $h_0=h_0^\dagger$).
We also define the hopping matrix $t$ as the block matrix collecting the off-diagonal blocks of $h_0$:
\begin{align}
t&=
\begin{pmatrix}
\mathbf{0} & t_{12} & \dots & t_{1\N}\\
t_{21} & \mathbf{0} & \dots & \vdots\\
\vdots & \vdots & \ddots & \vdots\\
t_{\N 1} & \dots & \dots & \mathbf{0}
\end{pmatrix}
\,,
\label{eq:block_t}
\end{align}
and the corresponding block-diagonal local one-body matrix
\begin{align}
\eps=
\begin{pmatrix}
\eps_{1} & \mathbf{0} & \dots & \mathbf{0}\\
\mathbf{0} & \eps_{2} & \dots & \vdots\\
\vdots & \vdots & \ddots & \vdots\\
\mathbf{0} & \dots & \dots & \eps_{\N}
\end{pmatrix},
\qquad
h_0=\eps+t.
\label{eq:eps_def}
\end{align}

\begin{figure}[t]
  \centering
  \includegraphics[width=\columnwidth]{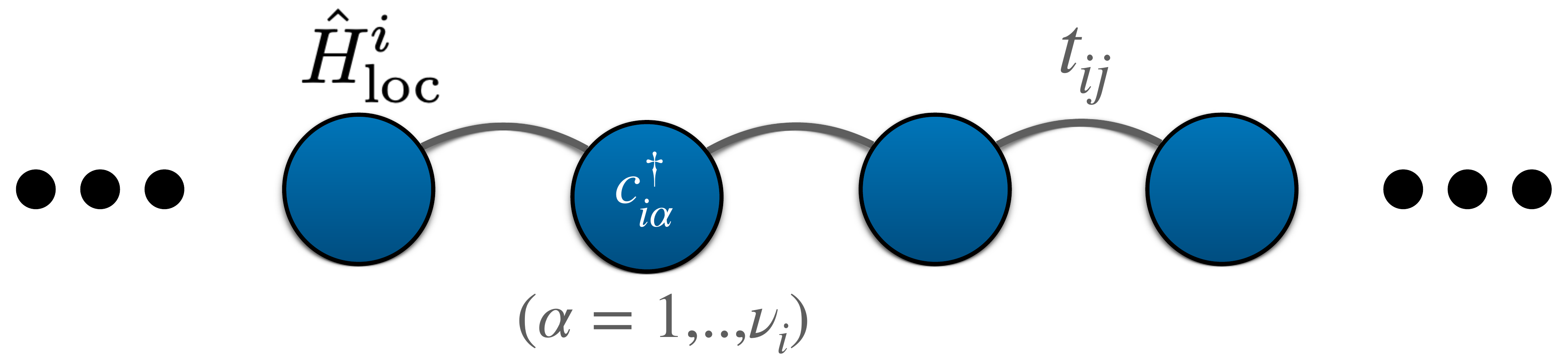} 
  \caption{Schematic representation of the multi-orbital Hubbard lattice Hamiltonian.}
  \label{Figure1}
\end{figure}

For later convenience, we group the local one-body terms and the local interactions into a single local operator
\begin{align}
\hat{H}^{i}_{\mathrm{loc}}\big[\cc_{i\alpha},\ca_{i\alpha}\big]
=
\sum_{\alpha,\beta=1}^{\nu_i}[\eps_{i}]_{\alpha\beta}\,\cc_{i\alpha}\ca_{i\beta}
+\hat{H}^{i}_{\mathrm{int}}
\,,
\label{eq:Hloc_def}
\end{align}
so that Eq.~\eqref{eq:H_split} can be rewritten in the following form used throughout this work:
\begin{align}
\h
&=\sum_{\substack{i,j=1\\ i\neq j}}^{\N} 
\sum_{\alpha=1}^{\nu_i}\sum_{\beta=1}^{\nu_j}
[t_{ij}]_{\alpha\beta}\,\cc_{i\alpha}\ca_{j\beta}
+\sum_{i=1}^{\N}\hat{H}^{i}_{\mathrm{loc}}\big[\cc_{i\alpha},\ca_{i\alpha}\big]
\,,
\label{eq:H_def}
\end{align}
schematically represented in Fig.~\ref{Figure1}.

\subsection{The ghost-GA variational framework}
\label{sec:gga_setup}

The ghost-GA is a theoretical framework based on the variational principle. Specifically, the variational energy is minimized over multiconfigurational states of the form
\begin{align}
\ket{\Psi_G}
&=\hat{\mathcal P}_G\,\ket{\Psi_0},
\label{eq:gGA_ansatz}
\end{align}
where $\ket{\Psi_0}$ is a single-particle wavefunction and $\hat{\mathcal P}_G$ is an operator, both to be variationally determined. 
A key feature of ghost-GA is that, differently from the standard GA, $\ket{\Psi_0}$ is constructed in an auxiliary Fock space of tunable dimension, and $\hat{\mathcal P}_G$ provides a local embedding map into the physical Hilbert space.

Specifically, the physical many-body Hilbert space is the tensor product
$\mathcal{H}=\bigotimes_{i=1}^{\N}\mathcal{H}_i$,
where $\mathcal{H}_i$ is the local Fock space generated by $\{\cc_{i\alpha}\}_{\alpha=1}^{\nu_i}$ acting on a local vacuum $\ket{0}$ annihilated by all $\ca_{i\alpha}$. A convenient orthonormal basis of $\mathcal{H}_i$ is given by the set of Fock states:
\begin{align}
\ket{\Gamma,i}
&=[\cc_{i1}]^{q_1(\Gamma)}\cdots[\cc_{i\nu_i}]^{q_{\nu_i}(\Gamma)}\ket{0}
\,,
\label{eq:Gamma_def}
\end{align}
where $\Gamma=0,\dots,2^{\nu_i}-1$, $q_\alpha(\Gamma)\in\{0,1\}$ denotes the $\alpha$-th occupation number of the state labeled by $\Gamma$ (equivalently, the $\alpha$-th digit of $\Gamma$ in binary form).

For each fragment $i$ we introduce $B\nu_i$ auxiliary (ghost) fermionic modes $\fa_{ia}$ ($a=1,\dots,B\nu_i$), where $B$ is a positive integer. The auxiliary many-body Hilbert space is
$\tilde{\mathcal{H}}_{\mathrm{ghost}}=\bigotimes_{i=1}^{\N}\tilde{\mathcal{H}}_i$,
where $\tilde{\mathcal{H}}_i$ is the local auxiliary Fock space generated by $\{\fc_{ia}\}_{a=1}^{B\nu_i}$ acting on a local auxiliary vacuum $\ket{0}$ annihilated by all $\fa_{ia}$.
A convenient orthonormal basis of $\tilde{\mathcal{H}}_i$ is
\begin{align}
\ket{n,i}
&=[\fc_{i1}]^{q_1(n)}\cdots[\fc_{iB\nu_i}]^{q_{B\nu_i}(n)}\ket{0}
\,,
\label{eq:n_def}
\end{align}
where $n=0,\dots,2^{B\nu_i}-1$,  $q_a(n)\in\{0,1\}$ denotes the $a$-th occupation number of the state labeled by $n$ (i.e., the $a$-th digit of $n$ in binary form).

We assume that $\ket{\Psi_0}\in\tilde{\mathcal{H}}_{\mathrm{ghost}}$, and that $\ket{\Psi_G}\in\mathcal{H}$ is obtained from $\ket{\Psi_0}$ through the embedding map $\hat{\mathcal P}_G$, which is assumed to have a local-product structure:
\begin{align}
\hat{\mathcal P}_G&=\prod_{i=1}^{\N}\hat{\mathcal P}_i
\,,
\end{align}
where each $\hat{\mathcal P}_i$ is a general linear map from the local auxiliary Fock space $\tilde{\mathcal{H}}_i$ to the local physical Fock space $\mathcal{H}_i$. In the bases \eqref{eq:Gamma_def} and \eqref{eq:n_def}, we parametrize it as
\begin{align}
\hat{\mathcal P}_i
&=\sum_{\Gamma=0}^{2^{\nu_i}-1}\sum_{n=0}^{2^{B\nu_i}-1}
[\Omega_i]_{\Gamma n}\,\ket{\Gamma,i}\bra{n,i}
\,,
\label{eq:Pi_def}
\end{align}
where $[\Omega_i]_{\Gamma n}$ are complex variational parameters. 
Note that, despite the traditional terminology, the operators $\hat{\mathcal P}_i$ are not assumed to be projectors in the operatorial sense; they are general local embedding maps implementing the variational ansatz \eqref{eq:gGA_ansatz}.

Given $\ket{\Psi_G}$, the corresponding variational energy is
\begin{align}
\E = \Av{\Psi_G}{\h}
=\Av{\Psi_0}{\hat{\mathcal P}_G^\dagger\,\h\,\hat{\mathcal P}_G}
\,,
\label{eq:var_energy_def}
\end{align}
to be minimized both with respect to $\ket{\Psi_0}$ and the entries of the rectangular matrices $\Omega_i\in\mathbb{C}^{2^{\nu_i}\times 2^{B\nu_i}}$.

\subsubsection{Variational ansatz for normal states}

In this work we restrict to particle-number conserving variational states, i.e., we consider $\ket{\Psi_G}$ that are eigenstates of the physical number operator
\begin{equation}
\hat{N}= \sum_{i=1}^{\N}\sum_{\alpha=1}^{\nu_i}\cc_{i\alpha}\ca_{i\alpha}
\,.
\end{equation}
Within the parametrization \eqref{eq:Pi_def}, this restriction can be enforced by requiring that the coefficients $[\Omega_i]_{\Gamma n}$ satisfy the selection rule
\begin{align}
N(n)-N(\Gamma)=m_i
\quad
\forall\,\Gamma,n\ \ \text{s.t.}\ \ [\Omega_i]_{\Gamma n}\neq 0
\,,
\label{eq:mi_rule}
\end{align}
where
\begin{align}
N(\Gamma) &= \sum_{\alpha=1}^{\nu_i} q_\alpha(\Gamma)
\,,
\\
N(n) &= \sum_{a=1}^{B\nu_i} q_a(n)
\,,
\end{align}
and the $m_i$ are fixed integer values that, in principle, could be chosen arbitrarily, with each choice corresponding to a different variational ansatz.
As clarified in previous work~\cite{Ghost-GA,ALM_g-GA}, setting $B$ odd and $m_i=\frac{B-1}{2}\,\nu_i$ usually produces the optimal solution.

\subsubsection{Gutzwiller constraints and Gutzwiller approximation}
\label{sec:gutz_constraints}

The evaluation of the variational energy in Eq.~\eqref{eq:var_energy_def} is simplified by imposing, for each fragment $i$, the Gutzwiller constraints
\begin{align}
\Av{\Psi_0}{\hat{\mathcal P}_i^\dagger\hat{\mathcal P}_i}&=1
\,,
\label{eq:GC1}
\\
\Av{\Psi_0}{\hat{\mathcal P}_i^\dagger\hat{\mathcal P}_i\,
\fc_{ia}\fa_{ib}}
&=\Av{\Psi_0}{\fc_{ia}\fa_{ib}}
\,,
\label{eq:GC2}
\end{align}
for all $a,b=1,\dots,B\nu_i$.

In addition we adopt the Gutzwiller approximation, i.e., in Wick expansions with respect to the Slater determinant $\ket{\Psi_0}$ we neglect all contributions that vanish in the limit of infinite coordination number.


\section{Functional reformulation underpinning the unification of ghost-GA, ghost-DMET, and DMFT}
\label{sec:lagrange_formulation}

In this section we introduce a functional reformulation of the ghost-GA variational problem as a QE theory.
This formulation is the central structural result of the present work: it provides the bridge between the static ghost-DMET perspective of Ref.~\cite{gDMET} and the dynamical DMFT functional framework.

The section is organized as follows:
\begin{itemize}
\item In Sec.~\ref{sec:Lagrange-functional} we introduce the Lagrange function at the center of this work.
\item In Sec.~\ref{sec:lagrange_equations} we derive the corresponding stationarity conditions.
\item In Sec.~\ref{sec:observables} we summarize how physical observables and Green's functions are evaluated from the variational parameters.
\item In Sec.~\ref{sec:Asymptotic-Sigma} we discuss the asymptotic behavior of the self-energy and the associated spectral-weight relation.
\item In Sec.~\ref{sec:Gauge} we discuss the gauge structure of the Lagrange function and show explicitly that all physical quantities are invariant under the corresponding transformations.
\item In Sec.~\ref{sec:dmft_like_algorithm} we show how the stationarity conditions implied by this formulation organize into a DMFT-like iterative structure.
\end{itemize}

\begin{figure}[t]
  \centering
  \includegraphics[width=\columnwidth]{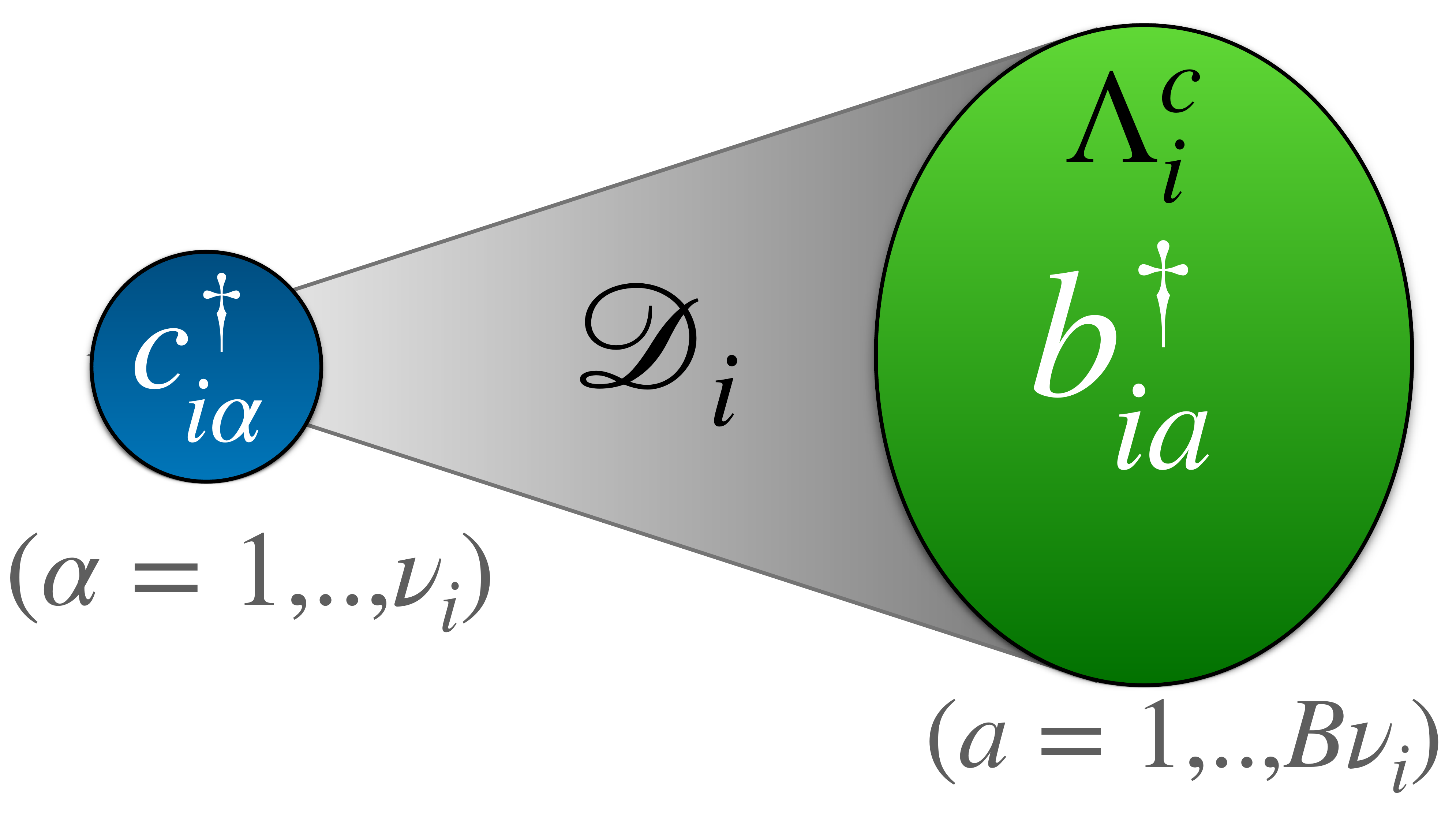} 
  \caption{Schematic representation of the correlated embedding Hamiltonian $\hat{H}^{i}_{\mathrm{emb}}[\D_i,\Lambda_i^c]$ [Eq.~\eqref{eq:Hemb}]. The physical local degrees of freedom of fragment $i$ (governed by $\hat{H}^{i}_{\mathrm{loc}}$) are hybridized with $B\nu_i$ auxiliary bath modes $\{b_{ia}\}$ through the matrix $\D_i$, while the quadratic bath term is parametrized by $\Lambda_i^c$.}
  \label{Figure2}
\end{figure}

\subsection{Functional formulation of ghost-GA}
\label{sec:Lagrange-functional}

Our reformulation of the ghost-GA equations is encoded in the following Lagrange function:
\begin{align}
\mathcal{L} 
= \mathcal{L}^{\text{qp}} 
+ \mathcal{L}^{\text{emb}} 
- \mathcal{L}^{0, \text{emb}} 
\,,
\label{eq:L_lagrange}
\end{align}
where: 
\begin{align}
&\mathcal{L}^{\text{qp}}_{\mathcal{R}, \Lambda} =
\langle \Psi_0 | \hat{H}_{\mathrm{qp}}
| \Psi_0 \rangle
+ E \left( 1 - \langle \Psi_0 | \Psi_0 \rangle \right)
\\
&\mathcal{L}^{\text{emb}}_{\{\mathcal{D}_i,\Lambda_i^c\}} =
\sum_{i=1}^{\N}
\Big[
\langle \Phi_i | \hat{H}^{i}_{\mathrm{emb}} 
| \Phi_i \rangle
+ E^c_i \left( 1 - \langle \Phi_i | \Phi_i \rangle \right)
\Big]
\\
&\mathcal{L}^{0, \text{emb}}_{\{\mathcal{D}_i\}, \{\Lambda_i^c\}, \{\mathcal{R}_i\}, \{ \Lambda_i \} }\ =
\sum_{i=1}^{\N}
\Big[
\langle \Phi^0_i | \hat{H}^{i}_{0,\mathrm{emb}}
| \Phi^0_i \rangle \\
& \qquad \qquad \quad \qquad \qquad \qquad \qquad + E^{0c}_i \left( 1 - \langle \Phi^0_i | \Phi^0_i \rangle \right)
\Big] \nonumber
\,.
\end{align}
Here the ghost-GA variational parameters are
$\R_i,\D_i \in \mathbb{C}^{B\nu_i\times\nu_i}$ and
$\Lambda_i,\Lambda_i^c \in \mathbb{C}^{B\nu_i\times B\nu_i}$,
with $\Lambda_i=\Lambda_i^\dagger$ and $\Lambda_i^c=(\Lambda_i^c)^\dagger$, which parametrize the following auxiliary Hamiltonians:
\begin{align}
&\hat{H}_{\mathrm{qp}}[\R,\Lambda]
=
\sum_{i,j=1}^{\N}
\sum_{a=1}^{B\nu_i}\sum_{b=1}^{B\nu_j}
\big[\R_i\,t_{ij}\,\R_j^\dagger\big]_{ab}\,\fc_{ia}\fa_{jb}
\label{eq:Hqp}
\\
& \quad +
\sum_{i=1}^{\N}\sum_{a,b=1}^{B\nu_i}
\big[\Lambda_i\big]_{ab}\,\fc_{ia}\fa_{ib},
\nonumber
\\
& \hat{H}^{i}_{\mathrm{emb}}[\D_i,\Lambda_i^c]
=
\hat{H}^{i}_{\mathrm{loc}}[\cc_{i\alpha},\ca_{i\alpha}] +
\sum_{a,b=1}^{B\nu_i}\big[\Lambda_i^c\big]_{ab}\,\bba_{ib}\bbc_{ia}
\label{eq:Hemb}
\\
& \quad +
\sum_{a=1}^{B\nu_i}\sum_{\alpha=1}^{\nu_i}
\Big(\big[\D_i\big]_{a\alpha}\cc_{i\alpha}\bba_{ia}+\text{H.c.}\Big)
\,,
\nonumber
\\
& \hat{H}^{i}_{0,\mathrm{emb}}[\D_i,\Lambda_i^c;\R_i,\Lambda_i]
=
\sum_{a,b=1}^{B\nu_i}\big[\Lambda_i\big]_{ab}\,\fc_{ia}\fa_{ib}
\label{eq:H0emb} \\
& \quad +
\sum_{a,b=1}^{B\nu_i}
\Big(\big[\D_i\R_i^{T}\big]_{ab}\,\fc_{ib}\bba_{ia}+\text{H.c.}\Big)
\nonumber
+
\sum_{a,b=1}^{B\nu_i}\big[\Lambda_i^c\big]_{ab}\,\bba_{ib}\bbc_{ia}
\,.
\end{align}
In Eqs.~\eqref{eq:Hemb}--\eqref{eq:H0emb}, $\bba_{ia}$ ($a=1,\dots,B\nu_i$) denote auxiliary bath fermionic modes associated with fragment $i$.
We also introduced the following block matrices:
\begin{align}
\R&=
\begin{pmatrix}
\R_1 & \mathbf{0} & \dots & \mathbf{0}\\
\mathbf{0} & \R_2 & \dots & \mathbf{0}\\
\vdots & \vdots & \ddots & \vdots\\
\mathbf{0} & \mathbf{0} & \dots & \R_{\N}
\end{pmatrix}
,
\quad
\Lambda=
\begin{pmatrix}
\Lambda_1 & \mathbf{0} & \dots & \mathbf{0}\\
\mathbf{0} & \Lambda_2 & \dots & \vdots\\
\vdots & \vdots & \ddots & \vdots\\
\mathbf{0} & \dots & \dots & \Lambda_{\N}
\end{pmatrix}.
\label{eq:block_R_Lambda}
\end{align}

As in Refs.~\cite{Our-PRX,Ghost-GA,ALM_g-GA,gRISB,gDMET}, the Lagrange function involves the so-called quasiparticle Hamiltonian $\hat{H}_{\mathrm{qp}}$, which is quadratic and acts on the ghost fermions. 
It also involves, for each fragment $i$, an embedding Hamiltonian (EH) $\hat{H}^{i}_{\mathrm{emb}}$, consisting of the physical local Hamiltonian $\hat{H}^{i}_{\mathrm{loc}}$ coupled to bath modes through the hybridization matrix $\D_i$, with $\Lambda_i^c$ parametrizing the quadratic bath term, depicted in Fig.~\ref{Figure2}.

In addition, the present Lagrange construction involves a new quadratic auxiliary EH $\hat{H}^{i}_{0,\mathrm{emb}}$ for each fragment $i$, featuring a coupling between local ghost degrees of freedom and the bath modes.
This new embedding Hamiltonian is the key element behind both the formal and algorithmic developments presented in this work.


A central point of the present formulation is that the same Lagrange construction already contains the dynamical objects entering the correspondence with DMFT.
In particular, as shown in Sec.~\ref{sec:finiteT_functional_BK}, the present Lagrange construction admits a finite-temperature generalization, which can be equivalently rewritten in the following form:
\begin{align}
&\Lag_\beta[\{\D_i\},\{\Lambda_i^c\},\{\R_i\},\{\Lambda_i\}]
\nonumber\\
&\quad
=
\Omega_{\rm qp}[\R,\Lambda]
+\sum_{i=1}^{\N}\Omega_{\rm emb}^i[\D_i,\Lambda_i^c]
\nonumber\\&\quad
-\sum_{i=1}^{\N}\Omega_{0,{\rm emb}}^i[\D_i,\Lambda_i^c;\R_i,\Lambda_i]
\nonumber\\
&\quad
\equiv
-\,T\sum_{n}e^{i\omega_n0^+}\Tr
\ln\!\big(\io\mathbf 1-h_0-\Sigma(\io)\big)
\nonumber\\
&\quad
+T\sum_{i=1}^{\N}\sum_{n}e^{i\omega_n0^+}\Tr
\ln\!\big(\io\mathbf 1\!-\!\eps_i\!-\!\Delta_i(\io)\!-\!\Sigma_i(\io)\big)
\nonumber\\
&\quad
+\sum_{i=1}^{\N}\Omega_{\rm imp}^i[\Delta_i(\io)]
\,,
\label{eq:Lbeta_preview}
\end{align}
where $\beta=T^{-1}$, $T$ is the temperature, $\omega_n=(2n+1)\pi T$ are the fermionic Matsubara frequencies, $\Sigma(\io)$ is the fragment-diagonal matrix with local blocks $\Sigma_i(\io)$, and the dependence on the static matrices $(\R_i,\Lambda_i,\D_i,\Lambda_i^c)$ enters through
\begin{align}
\Sigma_i(z)
&=
z\mathbf{1}_{\nu_i}
-\Big[\R_i^\dagger\big(z\mathbf{1}_{B\nu_i}-\Lambda_i\big)^{-1}\R_i\Big]^{-1}
-\eps_i,
\label{eq:Sigma_preview}
\\
\Delta_i(z)
&=
\D_i^{T}\big(z\mathbf{1}_{B\nu_i}+\Lambda_i^c\big)^{-1}\D_i^{*}
\,.
\label{eq:Delta_preview}
\end{align}
Here the thermodynamic-potential contributions are defined as:
\begin{align}
\Omega_{\rm qp}[\R,\Lambda]
&=
-\frac{1}{\beta}\ln\!\Big(\Tr e^{-\beta\,\hat H_{\rm qp}}\Big),
\\
\Omega_{\rm emb}^i[\D_i,\Lambda_i^c]
&=
-\frac{1}{\beta}\ln\!\Big(\Tr e^{-\beta\,\hat H_{\rm emb}^i}\Big),
\\
\Omega_{0,{\rm emb}}^i[\D_i,\Lambda_i^c;\R_i,\Lambda_i]
&=
-\frac{1}{\beta}\ln\!\Big(\Tr e^{-\beta\,\hat H_{0,{\rm emb}}^i}\Big),
\\
\Omega_{\rm bath}^i[\Lambda_i^c]
&=
-\frac{1}{\beta}\ln\!\Big(\Tr e^{-\beta\,\hat H_{\rm bath}^i[\Lambda_i^c]}\Big),
\\
\Omega_{\rm imp}^i[\Delta_i(\io)]
&=
\Omega_{\rm emb}^i[\D_i,\Lambda_i^c]
-
\Omega_{\rm bath}^i[\Lambda_i^c],
\\
\hat H_{\rm bath}^i[\Lambda_i^c]
&=
\sum_{a,b=1}^{B\nu_i}
[\Lambda_i^c]_{ab}\,\bba_{ib}\bbc_{ia}
\,.
\end{align}
This makes explicit the resemblance with the DMFT functional structure~\cite{LDA+U+DMFT}. From this perspective, the ghost-GA variational parameters can be viewed as encoding a specific finite-pole parametrization of the local self-energy and hybridization function, while at the same time defining, from the ghost-DMET perspective~\cite{gDMET}, the auxiliary reference systems entering the density-matrix matching conditions.

We will show that, while the stationary point of this functional can be formulated in terms of static density-matrix matching, it recovers DMFT exactly in the limit $B\rightarrow\infty$, for any temperature $T$.


\subsection{Zero-temperature stationarity equations}
\label{sec:lagrange_equations}

In this subsection we focus on the zero-temperature case, which is also the setting in which the connection with the ghost-DMET perspective developed below is most transparent. The corresponding finite-temperature generalization is presented later in Sec.~\ref{subsec:finiteT_extension}.

The Lagrange function $\Lag$, defined above in Eq.~\eqref{eq:L_lagrange}, has to be extremized with respect to $\ket{\Psi_0}$, $E$, $\ket{\Phi^0}$, $E^{0c}$, $\ket{\Phi}$, $E^c$, $\D$, $\Lambda^c$, $\R$ and $\Lambda$.

It is convenient to introduce the matrix:
\begin{equation}
h^*[\R,\Lambda]=\R\,t\,\R^\dagger+ \Lambda
\,,
\label{eq:hstar_compact}
\end{equation}
where $t$ is defined in Eq.~\eqref{eq:block_t} and 
$\R,\Lambda$ are defined in Eq.~\eqref{eq:block_R_Lambda}.
The matrix $h^*$ can also be written in block form as follows:
\begin{align}
h^*[\R,\Lambda] =
\begin{pmatrix}
\Lambda_1  &\R_1t_{12}\R^\dagger_2& \dots & \R_{1}t_{1 \N}\R^\dagger_{\N} \\
\R_2t_{21}\R^\dagger_1 & \Lambda_2 & \dots & \vdots \\
\vdots  & \vdots  & \ddots & \vdots \\
\R_{\N}t_{\N 1}\R^\dagger_{1} & \dots& \dots & \Lambda_{\N}
\end{pmatrix}
\,.
\label{SM-h*}
\end{align}

Having established the notation above, the stationarity conditions of $\Lag$ are the following:
\begin{widetext}
\begin{align}
\hat{H}_{\mathrm{qp}}[\R,\Lambda]\ket{\Psi_0}
&=E_0\ket{\Psi_0}
\,,
\label{e1}
\\
\hat{H}^{i}_{\mathrm{emb}}[\D_i,\Lambda_i^c]\ket{\Phi_i}
&=E^{c}_i\ket{\Phi_i}
\,,
\label{e2}
\\
\hat{H}^{i}_{0,\mathrm{emb}}[\D_i,\Lambda_i^c;\R_i,\Lambda_i]\ket{\Phi^0_i}
&=E^{0c}_i\ket{\Phi^0_i}
\,,
\label{e3}
\\
\Av{{\Phi}^0_i}{\fc_{ia}\fa_{ib}}
&=
\frac{\partial}{\partial [\Lambda_i]_{ab}}
\Av{\Psi_0}{\hat{H}_{\mathrm{qp}}[\R,\Lambda]}
=
\Av{\Psi_0}{\fc_{ia}\fa_{ib}}
\,,
\label{e4}
\\
\Av{\Phi^0_i}{
\fc_{ia}\left(\sum_{b=1}^{B\nu_i}[\D_i]_{b\alpha}\bba_{ib}\right)}
&=
\frac{\partial}{\partial [\R_i]_{a\alpha}}
\Av{\Psi_0}{\hat{H}_{\mathrm{qp}}[\R,\Lambda]}
=\Av{\Psi_0}{\fc_{ia}\left(\sum_{j=1}^{\N}\sum_{\beta=1}^{\nu_j}
[t_{ij}]_{\alpha\beta}\sum_{b=1}^{B\nu_j}[\R_j^\dagger]_{\beta b}\fa_{jb}\right)}
\,,
\label{e5}
\\
\Av{\Phi^0_i}{\bba_{ib}\bbc_{ia}}
&=
\Av{\Phi_i}{\bba_{ib}\bbc_{ia}}
\,,
\label{e6}
\\
\begin{split}
\Av{\Phi^0_i}{
\left(\sum_{a=1}^{B\nu_i}[\R_i]_{a\alpha}\fc_{ia}\right)\bba_{ib}}
&=
\Av{\Phi_i}{\cc_{i\alpha}\bba_{ib}}
\,.
\end{split}
\label{e7}
\end{align}
\end{widetext}

Here the expectation values with respect to $\ket{\Psi_0}$ can be calculated recognizing that the corresponding single-particle density matrix is the transpose of $n_F(h^*[\R,\Lambda])$, where $n_F$ is the zero-temperature Fermi function and $n_F(h^*)$ is defined in terms of the spectral decomposition $h^*=U\,\varepsilon\,U^\dagger$, where $\varepsilon$ is diagonal and we denote its diagonal entries by $\varepsilon_n$, as follows:
\begin{equation}
n_F(h^*)=U\,n_F(\varepsilon)\,U^\dagger
\,,
\qquad
[n_F(\varepsilon)]_{nn}=\theta(-\varepsilon_n)
\,,
\label{eq:nF_def}
\end{equation}
where $\theta$ is the Heaviside step function.

In addition, as shown in Refs.~\cite{Ghost-GA,ALM_g-GA}, within the particle-number conserving (normal-state) variational setup of Sec.~\ref{sec:gga_setup} (in particular, for the standard choice $B$ odd and $m_i=\frac{B-1}{2}\nu_i$ in Eq.~\eqref{eq:mi_rule}), the ground-state eigenvalue problems in Eqs.~\eqref{e2} and \eqref{e3} are to be understood in the half-filled sector of the corresponding impurity$+$bath Hilbert spaces, i.e.: 
\begin{align}
\left[
\sum_{\alpha=1}^{\nu_i}\cc_{i\alpha}\ca_{i\alpha}
+
\sum_{a=1}^{B\nu_i}\bbc_{ia}\bba_{ia}
\right]
\,\ket{\Phi_i}
&=\frac{(B+1)\nu_i}{2}\,\ket{\Phi_i}
\,,
\label{eq:half_fill_Hemb}
\\
\left[
\sum_{a=1}^{B\nu_i}\fc_{ia}\fa_{ia}
+
\sum_{a=1}^{B\nu_i}\bbc_{ia}\bba_{ia}
\right]
\,\ket{\Phi_i^0}
&=
B\nu_i\,\ket{\Phi_i^0}
\,.
\label{eq:half_fill_H0emb}
\end{align}

\subsection{Physical observables and Green's functions}
\label{sec:observables}

Once the Lagrange problem defined by Eqs.~\eqref{eq:L_lagrange} and \eqref{e1}--\eqref{e7} is converged, physical observables can be evaluated directly in terms of the variational parameters.

\subsubsection{Ground-state expectation values}

Within the ghost-GA, ground-state expectation values are computed as follows:
\begin{itemize} 
\item As shown in Refs.~\cite{Our-PRX,Ghost-GA,ALM_g-GA}, for any local operator $\hat{O}^{i}_{\mathrm{loc}}[\cc_{i\alpha},\ca_{i\alpha}]$
acting on fragment $i$, one has
\begin{align}
\langle\Psi_G|\hat{O}^{i}_{\mathrm{loc}}|\Psi_G\rangle
&=
\langle\Phi_i|\hat{O}^{i}_{\mathrm{loc}}|\Phi_i\rangle
\,.
\label{eq:local_obs}
\end{align}
In particular,
\begin{align}
\langle\Psi_G|\hat{H}^{i}_{\mathrm{loc}}|\Psi_G\rangle
&=
\langle\Phi_i|\hat{H}^{i}_{\mathrm{loc}}|\Phi_i\rangle
\,.
\label{eq:local_energy}
\end{align}

\item For $i\neq j$, the ghost-GA approximation yields
\begin{align}
\begin{split}
\langle\Psi_G|\cc_{i\alpha}\ca_{j\beta}|\Psi_G\rangle
&=
\sum_{a=1}^{B\nu_i}\sum_{b=1}^{B\nu_j}
\big[\R_i^\dagger\big]_{\alpha a}
\langle\Psi_0|\fc_{ia}\fa_{jb}|\Psi_0\rangle
\big[\R_j\big]_{b\beta}
\,.
\end{split}
\label{eq:nonlocal_1body}
\end{align}

\item The total energy is
\begin{align}
\begin{split}
\E
&=
\sum_{i=1}^{\N}\langle\Phi_i|\hat{H}^{i}_{\mathrm{loc}}|\Phi_i\rangle
\\
&\quad+
\sum_{\substack{i,j=1\\ i\neq j}}^{\N}
\sum_{\alpha=1}^{\nu_i}\sum_{\beta=1}^{\nu_j}
[t_{ij}]_{\alpha\beta}\,
\langle\Psi_G|\cc_{i\alpha}\ca_{j\beta}|\Psi_G\rangle
\,.
\end{split}
\label{eq:total_energy}
\end{align}

\end{itemize}

\begin{figure}[t]
\centering
\includegraphics[width=\columnwidth]{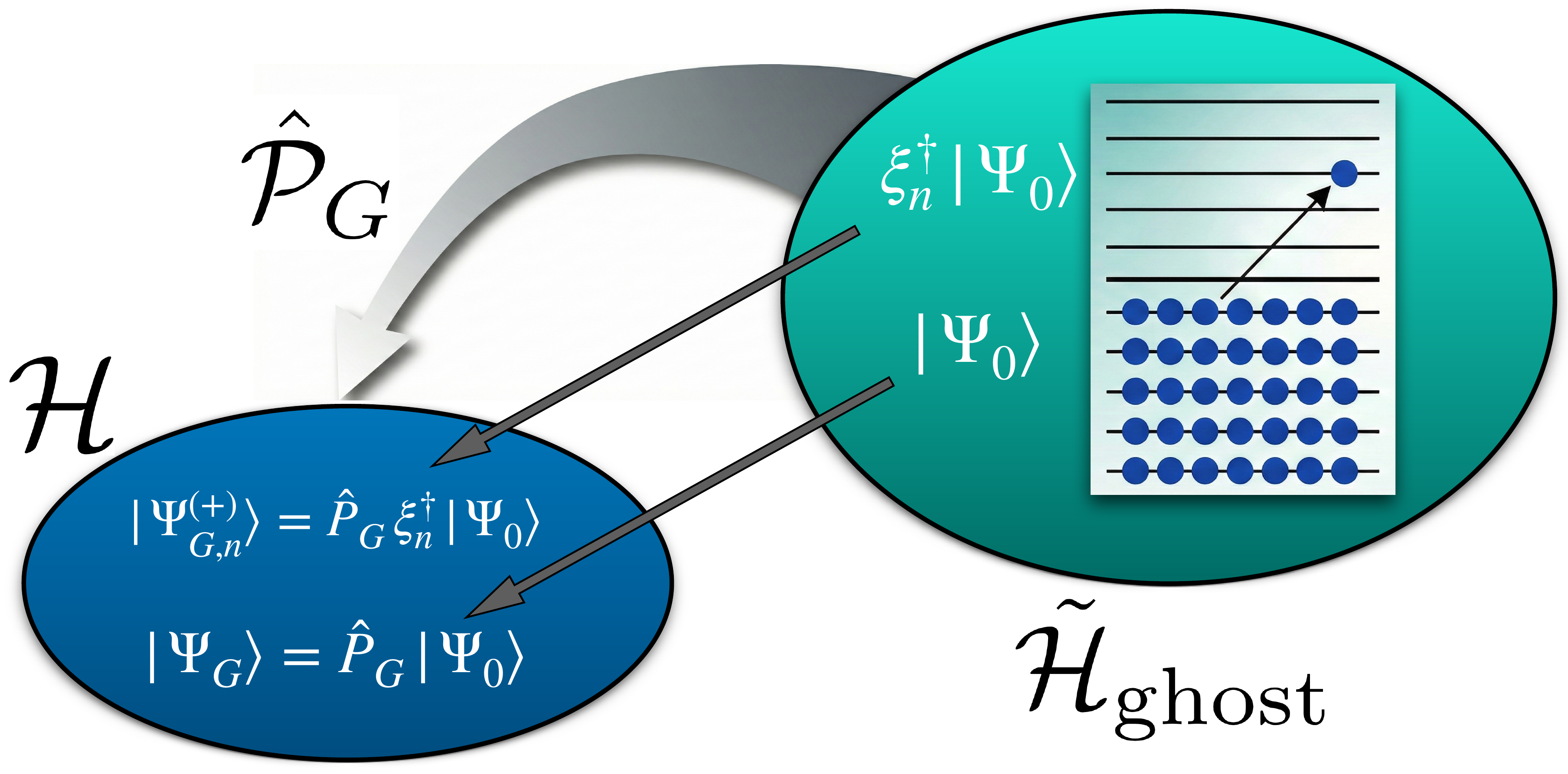}
\caption{Schematic representation of the ghost-GA excitation construction. A quasiparticle excitation $\xi_n^\dagger|\Psi_0\rangle$ of the auxiliary reference state $|\Psi_0\rangle$ in the ghost Hilbert space $\tilde{H}_{\text{ghost}}$ is mapped into the physical Hilbert space $\tilde{H}$ by the same variationally-optimized local map $\hat{\mathcal P}_G$ that defines the ground-state $\ket{\Psi_G}$, yielding the corresponding physical excited state $\ket{\Psi^{(+)}_{G,n}}$.}
\label{Figure3}
\end{figure}

\subsubsection{Single-particle Green's function}

Single-particle spectra in ghost-GA are obtained by constructing explicit particle and hole excitations as stationary points of the same variational energy functional that defines the ground state~\cite{Ghost-GA,ALM_g-GA}.
Once the ground-state stationary point is found, the quadratic quasiparticle Hamiltonian $\hat{H}_{\mathrm{qp}}[\R,\Lambda]$ [Eq.~\eqref{eq:Hqp}] defines single-particle eigenmodes $\{\xi_n,\xi_n^\dagger\}$ acting on the auxiliary Slater determinant $\ket{\Psi_0}$.
In the thermodynamic limit, the corresponding $N\!\pm\!1$ states can be taken as stationary (variational) solutions obtained by adding or removing one such quasiparticle mode and applying the same optimized local map $\hat{\mathcal P}_G$~\cite{Ghost-GA,ALM_g-GA}, namely:
\begin{align}
\ket{\Psi^{(+)}_{G,n}}&=\hat{\mathcal P}_G\,\xi_n^\dagger\,\ket{\Psi_0}
\,,
&
\ket{\Psi^{(-)}_{G,n}}&=\hat{\mathcal P}_G\,\xi_n\,\ket{\Psi_0}
\,,
\label{eq:qp_excited_states}
\end{align}
see Fig.~\ref{Figure3}.
Computing the one-particle Green's function within this variational excitation manifold yields the following closed expression for $G(z)$ in terms of the converged variational parameters:
\begin{align}
G(z)
&=
\R^\dagger\big[z\mathbf{1}-h^*[\R,\Lambda]\big]^{-1}\R
\,,
\label{eq:G_def}
\end{align}
where $h^*[\R,\Lambda]$ is defined in Eqs.~\eqref{eq:hstar_compact} and \eqref{SM-h*}, $\R$ is defined in Eq.~\eqref{eq:block_R_Lambda}, and $\mathbf{1}$ denotes the identity matrix.

It is convenient to write $G(z)$ in Dyson form:
\begin{align}
G(z)
&=
\big[z\mathbf{1}-h_0-\Sigma(z)\big]^{-1},
\label{eq:G_dyson}
\end{align}
where $h_0$ is the one-body matrix introduced in Eq.~\eqref{eq:h0_block}. As shown below, the resulting self-energy is local in the fragment index, i.e.:
\begin{align}
\Sigma(z)
&=
\begin{pmatrix}
\Sigma_{1}(z) & \mathbf{0} & \dots & \mathbf{0}\\
\mathbf{0} & \Sigma_{2}(z) & \dots & \vdots\\
\vdots & \vdots & \ddots & \vdots\\
\mathbf{0} & \dots & \dots & \Sigma_{\N}(z)
\end{pmatrix}
\,,
\label{eq:Sigma_block}
\end{align}
with the following local blocks:
\begin{align}
\Sigma_i(z)
&=
z\mathbf{1}_{\nu_i}
-\Big[\R_i^\dagger\big(z\mathbf{1}_{B\nu_i}-\Lambda_i\big)^{-1}\R_i\Big]^{-1}
-\eps_i
\,.
\label{eq:Sigma_local}
\end{align}
Here $\eps_i$ denotes the diagonal block of $h_0$ on fragment $i$ [Eq.~\eqref{eq:h0_block}], and $\Lambda_i$ is the corresponding diagonal block of $\Lambda$ [Eq.~\eqref{eq:block_R_Lambda}].

\subsubsection{Local self-energy from the quasiparticle Hamiltonian}
\label{sec:self_energy_from_Hqp}

Here we derive the local self-energy formula in Eq.~\eqref{eq:Sigma_local}.
We start from the physical Green's function induced by the quadratic quasiparticle Hamiltonian
$\hat{H}_{\mathrm{qp}}[\R,\Lambda]$ [Eq.~\eqref{eq:Hqp}] written in Eq.~(\ref{eq:G_def}), see Ref.~\cite{Ghost-GA}, namely:
\begin{align}
G(z)
&=
\R^\dagger\Big[z\mathbf{1}-\Lambda-\R\,t\,\R^\dagger\Big]^{-1}\R
\,.
\label{eq:G_from_Hqp}
\end{align}

\begin{itemize}

\item We compute $G(z)^{-1}$ by applying Eq.~\eqref{eq:woodbury_projected_inverse}
(Appendix~\ref{sec:woodbury_projected_inverses}) with
\begin{align}
A=z\mathbf{1}-\Lambda,
\qquad
U=\R,
\qquad
C=t
\,.
\label{eq:woodbury_subs_Hqp}
\end{align}
This gives
\begin{align}
G(z)^{-1}
&=
\Big[\R^\dagger\big(z\mathbf{1}-\Lambda\big)^{-1}\R\Big]^{-1}
-t
\,.
\label{eq:Ginv_Hqp}
\end{align}

\item Comparing Eq.~\eqref{eq:Ginv_Hqp} with the Dyson form
\begin{align}
G(z)
&=
\big[z\mathbf{1}-h_0-\Sigma(z)\big]^{-1},
\label{eq:G_dyson_recall}
\end{align}
[Eq.~\eqref{eq:G_dyson}], and using $h_0=\eps+t$ [Eq.~\eqref{eq:eps_def}], we have
\begin{align}
G(z)^{-1}
&=
z\mathbf{1}-\eps-t-\Sigma(z)
\,.
\label{eq:Ginv_dyson}
\end{align}
Equating Eqs.~\eqref{eq:Ginv_Hqp} and \eqref{eq:Ginv_dyson} and canceling $-t$ yields
\begin{align}
\Sigma(z)
&=
z\mathbf{1}
-\Big[\R^\dagger\big(z\mathbf{1}-\Lambda\big)^{-1}\R\Big]^{-1}
-\eps
\,.
\label{eq:Sigma_global}
\end{align}

\item Since $\R$ and $\Lambda$ are block diagonal in the fragment index
[Eq.~\eqref{eq:block_R_Lambda}], the matrix
$\R^\dagger\big(z\mathbf{1}-\Lambda\big)^{-1}\R$ is also block diagonal, hence
$\Sigma(z)$ is fragment diagonal. Taking the $i$-th diagonal block of
Eq.~\eqref{eq:Sigma_global} yields Eq.~\eqref{eq:Sigma_local}.

\end{itemize}

\subsection{Asymptotic behavior of $\Sigma_i(z)$ and spectral-weight}
\label{sec:Asymptotic-Sigma}

Starting from Eq.~\eqref{eq:Sigma_local}, we extract the leading behavior of $\Sigma_i(z)$ at large $|z|$.
Using
\begin{align}
\big(z\mathbf{1}_{B\nu_i}-\Lambda_i\big)^{-1}
&=
\frac{1}{z}\mathbf{1}_{B\nu_i}
+\frac{1}{z^2}\Lambda_i
+\mathcal{O}(z^{-3})
\,,
\end{align}
we obtain
\begin{align}
\R_i^\dagger\big(z\mathbf{1}_{B\nu_i}-\Lambda_i\big)^{-1}\R_i
&=
\frac{1}{z}\,\R_i^\dagger\R_i
\nonumber\\
&+\frac{1}{z^2}\,\R_i^\dagger\Lambda_i\R_i
+\mathcal{O}(z^{-3})
\,.
\label{eq:leading}
\end{align}
Substituting Eq.~\eqref{eq:leading} into Eq.~\eqref{eq:Sigma_local} yields
\begin{align}
\Sigma_i(z)
&=
z\Big[\mathbf{1}_{\nu_i}-(\R_i^\dagger\R_i)^{-1}\Big]
\label{eq:Sigma_linear_term}
\\
&\quad
+(\R_i^\dagger\R_i)^{-1}(\R_i^\dagger\Lambda_i\R_i)(\R_i^\dagger\R_i)^{-1}
-\eps_i
+\mathcal{O}(z^{-1})
\,.
\nonumber
\end{align}

For later convenience we note that, since
\begin{align}
G_i(z)
&=
\int_{-\infty}^{+\infty}d\omega\,\frac{A_i(\omega)}{z-\omega}
\,,
\end{align}
from Eq.~\eqref{eq:G_dyson}, together with Eq.~\eqref{eq:Sigma_linear_term},
it follows that, in ghost-GA:
\begin{align}
\int_{-\infty}^{+\infty}d\omega\,A_i(\omega)
&=
\R_i^\dagger\R_i
\,.
\label{eq:sumrule_RdagR}
\end{align}
Therefore, the unphysical linear term in the asymptotic expansion of the self-energy vanishes in the limit where   
$\R_i^\dagger\R_i=\mathbf{1}_{\nu_i}$, recovering the correct spectral sum rule.

\subsection{Gauge invariance}
\label{sec:Gauge}

A remarkable property of the Lagrange function $\Lag$ in Eq.~\eqref{eq:L_lagrange} is its invariance under a group of unitary gauge
transformations in the $B\nu_i$-dimensional ghost ($f$) and bath ($b$) single-particle spaces of each fragment $i$.

Let $\theta_i,\phi_i\in\mathbb{C}^{B\nu_i\times B\nu_i}$ be Hermitian matrices and define
\begin{align}
u_i(\theta_i)=e^{i\theta_i},
\qquad
u_i^\dagger(\theta_i)u_i(\theta_i)=\mathbf{1}
\,,
\label{eq:gauge_ui}
\\
v_i(\phi_i)=e^{i\phi_i},
\qquad
v_i^\dagger(\phi_i)v_i(\phi_i)=\mathbf{1}
\,.
\label{eq:gauge_vi}
\end{align}
We also introduce the corresponding second-quantized unitary operators acting on the ghost and bath Fock spaces:
\begin{align}
\hat{U}_i(\theta_i)
&=
\exp\!\left(
i\sum_{a,b=1}^{B\nu_i}[\theta_i]_{ab}\,\fc_{ia}\fa_{ib}
\right)
\,,
\label{eq:gauge_Ui}
\\
\hat{U}(\theta)
&=
\prod_{i=1}^{\N}\hat{U}_i(\theta_i)
\,,
\label{eq:gauge_U}
\\
\hat{V}_i(\phi_i)
&=
\exp\!\left(
i\sum_{a,b=1}^{B\nu_i}[\phi_i]_{ab}\,\bbc_{ia}\bba_{ib}
\right)
\,.
\label{eq:gauge_Vi}
\end{align}
Since $\theta_i=\theta_i^\dagger$ and $\phi_i=\phi_i^\dagger$, the operators $\hat{U}_i$, $\hat{U}$, and $\hat{V}_i$ are unitary.

The gauge transformation is defined by
\begin{align}
\ket{\Psi_0} &\rightarrow \hat{U}^\dagger(\theta)\ket{\Psi_0},
\label{eq:gauge_Psi0}
\\
\ket{\Phi_i} &\rightarrow \hat{V}_i^\dagger(\phi_i)\ket{\Phi_i}
\,,
\label{eq:gauge_Phi}
\\
\ket{\Phi_i^0} &\rightarrow \hat{U}_i^\dagger(\theta_i)\,\hat{V}_i^\dagger(\phi_i)\ket{\Phi_i^0}
\,,
\label{eq:gauge_Phi0}
\end{align}
together with the parameter transformations
\begin{align}
\R_i &\rightarrow u_i^\dagger(\theta_i)\R_i
\,,
\label{eq:gauge_R}
\\
\Lambda_i &\rightarrow u_i^\dagger(\theta_i)\Lambda_i\,u_i(\theta_i)
\,,
\label{eq:gauge_L}
\\
\D_i &\rightarrow v_i^{T}(\phi_i)\D_i
\,,
\label{eq:gauge_D}
\\
\Lambda_i^c &\rightarrow v_i^\dagger(\phi_i)\Lambda_i^c\,v_i(\phi_i)
\,.
\label{eq:gauge_Lc}
\end{align}
The invariance of the Lagrange function is verified by noting that the hybridization matrix entering $\hat{H}^{i}_{0,\mathrm{emb}}$ transforms covariantly:
from Eqs.~\eqref{eq:gauge_R} and \eqref{eq:gauge_D} one has
\begin{align}
\D_i\R_i^{T}
\ \longrightarrow\
v_i^{T}\,(\D_i\R_i^{T})\,u_i^{*}
\,,
\label{eq:gauge_DRt}
\end{align}
which is compensated by the rotations of the $f$ and $b$ operators implied by
Eqs.~\eqref{eq:gauge_Psi0} and \eqref{eq:gauge_Phi0}.

Note also that all physical observables are invariant under the combined gauge group defined in
Eqs.~\eqref{eq:gauge_Psi0}--\eqref{eq:gauge_Lc}.
In particular, the local expectation values in Eq.~\eqref{eq:local_obs}, the non-local one-body correlators in
Eq.~\eqref{eq:nonlocal_1body}, and the total energy in Eq.~\eqref{eq:total_energy} are invariant.
Moreover, the local self-energy blocks
\begin{equation}
\Sigma_i(z)
=
z\mathbf{1}_{\nu_i}
-\Big[\R_i^\dagger\big(z\mathbf{1}_{B\nu_i}-\Lambda_i\big)^{-1}\R_i\Big]^{-1}
-\eps_i
\label{eq:Sigma_local-rep}
\end{equation}
are invariant, since the combination
$\R_i^\dagger\big(z\mathbf{1}_{B\nu_i}-\Lambda_i\big)^{-1}\R_i$ is unchanged under
Eqs.~\eqref{eq:gauge_R} and \eqref{eq:gauge_L}.
Similarly, the hybridization function of the correlated embedding Hamiltonian
\begin{equation}
\Delta_i(z)=\D_i^{T}\big(z\mathbf{1}+\Lambda_i^c\big)^{-1}\D_i^{*}
\label{eq:hyb_def}
\end{equation}
is invariant under Eqs.~\eqref{eq:gauge_D} and \eqref{eq:gauge_Lc}.

\subsection{DMFT-like algorithmic structure}
\label{sec:dmft_like_algorithm}

The stationarity conditions \eqref{e1}--\eqref{e7} admit a DMFT-like interpretation and suggest a fixed-point iteration between two sets of quantities. The pair $(\R_i,\Lambda_i)$ parametrizes a fragment-diagonal self-energy through Eq.~\eqref{eq:Sigma_local}. The pair $(\D_i,\Lambda_i^c)$ specifies the bath sector of the embedding Hamiltonians $\hat{H}^{i}_{\mathrm{emb}}$ and $\hat{H}^{i}_{0,\mathrm{emb}}$ [Eqs.~\eqref{eq:Hemb}--\eqref{eq:H0emb}], and therefore plays the role of Weiss-field parameters.

A practical implementation can be organized as the following four-step cycle:
\begin{enumerate}
\item Starting from a guess for $(\R_i,\Lambda_i)$, solve the quasiparticle problem \eqref{e1} for $\ket{\Psi_0}$ and evaluate the one-body quantities entering \eqref{e4}--\eqref{e5}.

This mirrors the DMFT step in which, given a local self-energy $\Sigma(z)$ (here encoded by $(\R_i,\Lambda_i)$ through Eq.~\eqref{eq:Sigma_local}), one constructs the lattice Green's function by the Dyson equation \eqref{eq:G_dyson}.

\item With $(\R_i,\Lambda_i)$ fixed, determine $(\D_i,\Lambda_i^c)$ by solving the auxiliary quadratic embedding problem \eqref{e3} for $\ket{\Phi_i^0}$ (in the half-filled impurity$+$bath sector specified in Eq.~\eqref{eq:half_fill_H0emb}) and enforcing the consistency relations \eqref{e4} and \eqref{e5}.

This corresponds to the subsequent DMFT Weiss-field update, i.e., determining the impurity hybridization functions $\Delta_i(z)$ (here encoded by $(\D_i,\Lambda_i^c)$ through Eq.~\eqref{eq:hyb_def}) consistently with the lattice information computed in the previous step.

\item With $(\D_i,\Lambda_i^c)$ fixed, solve the interacting embedding problems \eqref{e2} for $\ket{\Phi_i}$ (in the half-filled impurity$+$bath sector specified in Eq.~\eqref{eq:half_fill_Hemb}).

This mirrors the impurity step of $T=0$ DMFT;
with a key difference: since the self-consistency constraints are expressed in terms of equal-time expectation values, only the ground state of each finite-bath embedding Hamiltonian $\hat{H}^{i}_{\mathrm{emb}}[\D_i,\Lambda_i^c]$ is required.

\item With $(\D_i,\Lambda_i^c)$ fixed, update $(\R_i,\Lambda_i)$ by solving \eqref{e3} for $\ket{\Phi_i^0}$ and enforcing the matching relations \eqref{e6} and \eqref{e7}. The updated $(\R_i,\Lambda_i)$ define the updated self-energy through Eq.~\eqref{eq:Sigma_local}.

This mirrors the DMFT self-energy update: the impurity information obtained in step~3 is fed back into the lattice through a new local self-energy, here parametrized by $(\R_i,\Lambda_i)$ via Eq.~\eqref{eq:Sigma_local}, which is then used again in step~1.
\end{enumerate}

The cycle is iterated until convergence of physical quantities, such as the total energy \eqref{eq:total_energy}, the local observables \eqref{eq:local_obs}--\eqref{eq:local_energy}, and the local self-energy blocks $\Sigma_i(z)$ in Eq.~\eqref{eq:Sigma_local}.
Accordingly, convergence is assessed in terms of gauge-invariant physical quantities, rather than in terms of the parameters $(\R_i,\Lambda_i,\D_i,\Lambda_i^c)$ themselves.

This algorithmic structure is used in Sec.~\ref{sec:equivalence_gga_dmft_Binfty} as the basis for establishing the $B\rightarrow\infty$ correspondence with the DMFT fixed point at zero temperature.

\subsubsection{Analytic hybridization update from the Schmidt--DMET embedding of $\hat{H}_{\mathrm{qp}}$ at zero temperature}
\label{sec:analytic_hybridization_update}

In this section we show that, at temperature $T=0$, step~2 of the DMFT-like cycle in Sec.~\ref{sec:dmft_like_algorithm} can be carried out analytically.
The key point is that, once $(\R,\Lambda)$ are fixed, $\ket{\Psi_0}$ is the Slater-determinant ground state of the quadratic Hamiltonian $\hat{H}_{\mathrm{qp}}[\R,\Lambda]$ [Eq.~\eqref{e1}], so that the corresponding embedding problem can be constructed explicitly from the Schmidt decomposition of $\hat{H}_{\mathrm{qp}}$ as in Appendix~\ref{app:schmidt_theorem}.
This is precisely the mechanism underlying the ghost-DMET formulation introduced in Ref.~\cite{gDMET}, here recovered directly from the Lagrange-functional perspective of Sec.~\ref{sec:lagrange_formulation}.

The starting observation is that Eqs.~\eqref{e4} and \eqref{e5} can be rewritten in the following derivative form:
\begin{widetext}
\begin{align}
\frac{\partial}{\partial [\Lambda_i]_{ab}}
\Av{\Psi_0}{\hat{H}_{\mathrm{qp}}[\R,\Lambda]}
&=
\frac{\partial}{\partial [\Lambda_i]_{ab}}
\Av{\Phi^0_i}{\hat{H}^{i}_{0,\mathrm{emb}}[\D_i,\Lambda_i^c;\R_i,\Lambda_i]}
\,,
\label{eq:derivative_match_Lambda}
\\
\frac{\partial}{\partial [\R_i]_{a\alpha}}
\Av{\Psi_0}{\hat{H}_{\mathrm{qp}}[\R,\Lambda]}
&=
\frac{\partial}{\partial [\R_i]_{a\alpha}}
\Av{\Phi^0_i}{\hat{H}^{i}_{0,\mathrm{emb}}[\D_i,\Lambda_i^c;\R_i,\Lambda_i]}
\,.
\label{eq:derivative_match_R}
\end{align}
\end{widetext}
Equations~\eqref{eq:derivative_match_Lambda}--\eqref{eq:derivative_match_R} are exactly of the type discussed in Appendix~\ref{app:schmidt_theorem}, where expectation values of operator derivatives are preserved by the Slater-determinant embedding construction; see in particular the identities
\eqref{eq:app_derivative_identity_AA} and \eqref{eq:app_derivative_identity}.

To make this correspondence explicit, we consider the bipartition of the quasiparticle Hamiltonian
$\hat{H}_{\mathrm{qp}}[\R,\Lambda]$ [Eq.~\eqref{eq:Hqp}] into the subsystem
$A=\{\,\fa_{ia}\,\}_{a=1}^{B\nu_i}$ and its complement
$B=\{\,\fa_{jc}\,\}_{j\neq i}$.
For fixed $(\R,\Lambda)$, $\ket{\Psi_0}$ is the Slater-determinant ground state of $\hat{H}_{\mathrm{qp}}[\R,\Lambda]$ [Eq.~\eqref{e1}], and Appendix~\ref{app:schmidt_theorem} constructs an auxiliary bath and a quadratic embedding Hamiltonian $\hat{H}^{\mathrm{emb}}_A$ on the active modes such that the $A$--$A$ one-body density matrix and the $A$--$B$ coupling derivatives of $\hat{H}_{\mathrm{qp}}$ are reproduced by construction.
Identifying this $\hat{H}^{\mathrm{emb}}_A$ with $\hat{H}^{i}_{0,\mathrm{emb}}[\D_i,\Lambda_i^c;\R_i,\Lambda_i]$ [Eq.~\eqref{eq:H0emb}] therefore yields an explicit realization of step~2: the parameters $(\D_i,\Lambda_i^c)$ are obtained directly from the Slater-determinant embedding data of $\hat{H}_{\mathrm{qp}}[\R,\Lambda]$, and Eqs.~\eqref{e4}--\eqref{e5} are satisfied identically.

Denoting the local single-particle density matrix of $\ket{\Psi_0}$ by
\begin{equation}
[\Delta_i]_{ab}=\langle\Psi_0|\fc_{ia}\fa_{ib}|\Psi_0\rangle
\,,
\end{equation}
and identifying the matrix $\rho_A^{0}$ of Appendix~\ref{app:schmidt_theorem} with the local ghost density matrix $\Delta_i$ defined above,
the specialization of the embedding construction in Appendix~\ref{app:schmidt_theorem} to the present $A/B$ partition yields an explicit analytic update for the hybridization parameters.
In particular, using the bath-orbital definition Eq.~\eqref{eq:app_B} (specialized to the present $A/B$ partition), together with the algebraic simplification of the bath block in Eq.~\eqref{eq:Bdag_hBB_B_final}, one obtains:
\begin{widetext}
\begin{align}
\big[\D_i^T\big]_{\alpha a}
&=
\sum_{b=1}^{B\nu_i}
\Bigg[
\sum_{\substack{j=1\\ j\neq i}}^{\N}
\sum_{\beta=1}^{\nu_j}
\sum_{c=1}^{B\nu_j}
[t_{ij}]_{\alpha\beta}\,
[\R_j^\dagger]_{\beta c}\,
\langle\Psi_0|\fc_{ib}\fa_{jc}|\Psi_0\rangle
\Bigg]\,
\Big[\Delta_i^T(\mathbf{1}-\Delta_i^T)\Big]^{-\frac{1}{2}}_{ba}
\,,
\label{eq:DiT_realspace}
\\
\Lambda^c_i
&=
-\frac{1}{2}\left[
\sqrt{\frac{\Delta^T_i}{\mathbf{1}-\Delta^T_i}}\,
\Lambda_i\,
\sqrt{\frac{\mathbf{1}-\Delta^T_i}{\Delta^T_i}}
+\text{H.c.}
\right]
-
\left[
\frac{\tfrac{\mathbf{1}}{2}-\Delta^T_i}{\sqrt{\Delta^T_i(\mathbf{1}-\Delta^T_i)}}\;
\R_i\,\D^T_i
+\text{H.c.}
\right]
\,.
\label{eq:Lambdaci_analytic_update}
\end{align}
\end{widetext}

In summary, Eqs.~\eqref{eq:DiT_realspace} and \eqref{eq:Lambdaci_analytic_update} provide an analytic implementation of the hybridization determination step, i.e., the Weiss-field update encoded by $(\D_i,\Lambda_i^c)$ [Eq.~\eqref{eq:hyb_def}], directly from the one-body correlators of the quasiparticle Slater determinant $\ket{\Psi_0}$.

\subsubsection{Numerical density-matrix fit for the self-energy update}
\label{sec:numerical_self_energy_update}

The update of $(\R_i,\Lambda_i)$ in step~4 of Sec.~\ref{sec:dmft_like_algorithm} can be formulated in complete analogy with the hybridization update discussed in Sec.~\ref{sec:analytic_hybridization_update}.
Also in this case, for fixed bath parameters $(\D_i,\Lambda_i^c)$, one is faced with a one-body density-matrix fit problem: determine the quadratic parameters $(\R_i,\Lambda_i)$ entering $\hat{H}^{i}_{0,\mathrm{emb}}[\D_i,\Lambda_i^c;\R_i,\Lambda_i]$ [Eq.~\eqref{eq:H0emb}] such that the corresponding ground state $\ket{\Phi_i^0}$ reproduces the one-body information extracted from the interacting embedding ground state $\ket{\Phi_i}$ [Eq.~\eqref{e2}], as required by Eqs.~\eqref{e6}--\eqref{e7}.
Note that, analogously to Sec.~\ref{sec:analytic_hybridization_update}, the matching conditions \eqref{e6}--\eqref{e7} can be formally rewritten in the following derivative form:
\begin{widetext}
\begin{align}
\frac{\partial}{\partial [\Lambda_i^c]_{ab}}
\Av{\Phi^0_i}{\hat{H}^{i}_{0,\mathrm{emb}}[\D_i,\Lambda_i^c;\R_i,\Lambda_i]}
&=
\frac{\partial}{\partial [\Lambda_i^c]_{ab}}
\Av{\Phi_i}{\hat{H}^{i}_{\mathrm{emb}}[\D_i,\Lambda_i^c]},
\label{eq:derivative_match_Lambdac}
\\
\frac{\partial}{\partial [\D_i]_{b\alpha}}
\Av{\Phi^0_i}{\hat{H}^{i}_{0,\mathrm{emb}}[\D_i,\Lambda_i^c;\R_i,\Lambda_i]}
&=
\frac{\partial}{\partial [\D_i]_{b\alpha}}
\Av{\Phi_i}{\hat{H}^{i}_{\mathrm{emb}}[\D_i,\Lambda_i^c]}
\,.
\label{eq:derivative_match_D}
\end{align}
\end{widetext}

However, while the hybridization update discussed in Sec.~\ref{sec:analytic_hybridization_update} admits a closed-form solution because $\ket{\Psi_0}$ is a Slater determinant, so that the corresponding density-matrix fit can be realized analytically through the Schmidt--DMET embedding construction of Appendix~\ref{app:schmidt_theorem}, here the target state $\ket{\Phi_i}$ is the ground state of the interacting Hamiltonian $\hat{H}^{i}_{\mathrm{emb}}[\D_i,\Lambda_i^c]$.
Therefore, the quadratic auxiliary state $\ket{\Phi_i^0}$ cannot be constructed analytically in general, and the determination of $(\R_i,\Lambda_i)$ must be performed numerically by enforcing the one-body matching conditions [Eqs.~\eqref{e6} and \eqref{e7}] (equivalently, Eqs.~\eqref{eq:derivative_match_Lambdac} and \eqref{eq:derivative_match_D}).

In practice, this amounts to a numerical one-body density-matrix fit that updates the local self-energy parameters $(\R_i,\Lambda_i)$, and hence $\Sigma_i(z)$ through Eq.~\eqref{eq:Sigma_local}, by matching the relevant reduced one-body information between the interacting embedding problem and its quadratic auxiliary counterpart (at fixed $\D_i,\Lambda_i^c$).

In this context, we mention that advanced strategies from the DMET literature, such as projected schemes~\cite{DMET-fit-1} or semidefinite programming~\cite{DMET-fit-2}, could be directly adapted here to further enhance numerical stability.


\section{Derivation of the QE Lagrange function from the standard ghost-GA formulation}
\label{sec:derivation_new_lagrange}

In Sec.~\ref{sec:lagrange_formulation} we introduced the Lagrange function~\eqref{eq:L_lagrange}, involving the additional quadratic embedding Hamiltonian
$\hat{H}^{i}_{0,\mathrm{emb}}$ defined in Eq.~\eqref{eq:H0emb}.
The purpose of this section is to show that Eq.~\eqref{eq:L_lagrange} can be derived directly from the ghost-GA Lagrange formulation previously derived in Refs.~\cite{Ghost-GA,ALM_g-GA}, and is therefore an equivalent reformulation of the variational problem outlined in Sec.~\ref{subsec:gGA}.

The standard embedding formulation of ghost-GA is encoded in the following Lagrange function~\cite{Ghost-GA,ALM_g-GA}:
\begin{align}
\Lag
&=
\langle \Psi_0 | \hat{H}_{\mathrm{qp}}[\R,\Lambda] | \Psi_0 \rangle
+ E \left( 1 - \langle \Psi_0 | \Psi_0 \rangle \right)
\nonumber\\
&\quad
+ \sum_{i=1}^{\N}
\Big[
\langle \Phi_i | \hat{H}^{i}_{\mathrm{emb}}[\D_i,\Lambda_i^c] | \Phi_i \rangle
+ E^c_i \left( 1 - \langle \Phi_i | \Phi_i \rangle \right)
\Big]
\nonumber\\
&\quad
-\sum_{i=1}^{\N}\,\mathcal{F}_{\R_i,\Lambda_i,\D_i,\Lambda_i^c}^i[\Delta_i]
\,,
\label{eq:L_old}
\end{align}
where $\hat{H}_{\mathrm{qp}}$ and $\hat{H}^{i}_{\mathrm{emb}}$ are the same operators defined in
Eqs.~\eqref{eq:Hqp} and \eqref{eq:Hemb}, and 
\begin{align}
&\mathcal{F}_{\R_i,\Lambda_i,\D_i,\Lambda_i^c}^i[\Delta_i]
=
\sum_{a,b=1}^{B\nu_i}\Big(\big[\Lambda_i\big]_{ab}+\big[\Lambda_i^c\big]_{ab}\Big)\,[\Delta_i]_{ab}
\nonumber\\
&\quad
+\sum_{a,b=1}^{B\nu_i}\sum_{\alpha=1}^{\nu_i}
\Big(
[\D_i]_{a\alpha}\,[\R_i]_{b\alpha}\,
\big[\Delta_i(\mathbf{1}-\Delta_i)\big]^{\tfrac{1}{2}}_{ba}
+\text{c.c.}
\Big)
\label{eq:F_old}
\end{align}
encodes the matching conditions between the quasiparticle sector and the embedding sector.
The ghost-GA solution is obtained extremizing Eq.~\eqref{eq:L_old} with respect to 
$\ket{\Psi_0}$, $E$, $\{\ket{\Phi_i}\}$, $\{E_i^c\}$, and the matrices $\{\R_i,\Lambda_i,\D_i,\Lambda_i^c,\Delta_i\}$.

In Sec.~\ref{sec:direct_equivalence_to_gga} we show that the standard ghost-GA functional Eq.~\eqref{eq:L_old}
can be recast into the QE form Eq.~\eqref{eq:L_lagrange} by introducing the auxiliary quadratic embedding problem.
Conversely, Sec.~\ref{sec:inverse_equivalence_to_gga} proves that any stationary point of Eq.~\eqref{eq:L_lagrange}
induces, up to gauge, a stationary point of Eq.~\eqref{eq:L_old}.
Together, these results establish the equivalence of the two Lagrange formulations.

\subsection{Purification of $\Delta_i$ and recovery of the QE functional}
\label{sec:direct_equivalence_to_gga}

In this subsection we show how the standard ghost-GA functional in Eq.~\eqref{eq:L_old} can be recast into the QE form Eq.~\eqref{eq:L_lagrange}. 
The key step is to represent $\Delta_i$ through the auxiliary quadratic state $\ket{\Phi_i^0}$. To this end, we
note that extremizing Eq.~\eqref{eq:L_old} with respect to $\Lambda_i$ implies that:
\begin{equation}
    [{\Delta}_i]_{ab}=\Av{\Psi_0}{\fc_{ia} \fa_{ib}}
    \,.
\end{equation}
At the saddle point, $\Delta_i$ can also be expressed as follows:
\begin{align}
    \Av{{\Phi}^0_i}{\fc_{ia}\fa_{ib}} & = [{\Delta}_i]_{ab}
    \,,
    \label{SC-imp-proof}
    \\
    \Av{{\Phi}^0_i}{\fc_{ia}\bba_{ib}} & = [{\Delta}_i(\mathbf{1}-{\Delta}_i)]^{\frac{1}{2}}_{ab}
    \,,
    \label{SC-hybr-proof}
    \\
    \Av{{\Phi}^0_i}{\bba_{ib}\bbc_{ia}} & = [{\Delta}_i]_{ab}
    \,,
    \label{SC-bath-proof}
\end{align}
where $|\Phi_i^0\rangle \equiv \ket{{\Phi}^0_i[\R_i,\Lambda_i,\D_i,\Lambda^c_i]}$ is the normalized ground state of $\hat{H}^{i}_{0,\mathrm{emb}}[\D_i,\Lambda_i^c;\R_i,\Lambda_i]$, as defined in Eq.~\eqref{eq:H0emb}.

Conceptually, Eqs.~\eqref{SC-imp-proof}--\eqref{SC-bath-proof} can be viewed as encoding $\Delta_i$ into a specific purification fixed by the saddle-point data $(\R_i,\Lambda_i,\D_i,\Lambda_i^c)$.

Once Eqs.~\eqref{SC-imp-proof}--\eqref{SC-bath-proof} are established, the equivalence with Eq.~\eqref{eq:L_lagrange} follows because $\mathcal{F}^i_{\R_i,\Lambda_i,\D_i,\Lambda_i^c}[\Delta_i]$ in Eq.~\eqref{eq:L_old} can be replaced by $\langle \Phi_i^0|\hat{H}^{i}_{0,\mathrm{emb}}[\D_i,\Lambda_i^c;\R_i,\Lambda_i]|\Phi_i^0\rangle$ (with $E_i^{0c}$ enforcing normalization), which is exactly what Eq.~\eqref{eq:L_lagrange} implements.

\subsubsection{Standard ghost-GA Lagrange equations}
\label{sec:old_lagrange_equations}

Following Refs.~\cite{Ghost-GA,ALM_g-GA}, we represent the matrices $\Delta_i$, $\Lambda_i$, and $\Lambda^c_i$ as expansions in terms of an orthonormal basis of Hermitian matrices, denoted $\left[h_i\right]_s$ (with respect to the canonical scalar product $(A, B) = \Tr \left[A^{\dagger}B\right]$):
\begin{align}
    \label{coeffDelta}
    \Delta_i =& \sum_{s=1}^{(B{\nu}_i)^2} \left[d_i\right]_s \left[h^T_i\right]_s 
    \,,
    \\
    \label{coeffL}
    \Lambda_i =& \sum_{s=1}^{(B{\nu}_i)^2} \left[l_i\right]_s \left[h_i\right]_s  
    \,,
    \\
    \label{coeffLc}
    \Lambda^c_i =& \sum_{s=1}^{(B{\nu}_i)^2} \left[l^c_i\right]_s \left[h_i\right]_s \,,
\end{align}
where $\left[d_i\right]_s$, $\left[l_i\right]_s$, and $\left[l^c_i\right]_s$ are real-valued coefficients.

Having established this notation, the saddle-point of the ghost-GA Lagrange function in Eq.~\eqref{eq:L_old} is given by the following equations:
\begin{widetext}
    \begin{align}
    \hat{H}_{\text{qp}}[\R,\Lambda]\ket{\Psi_0} &= E_0\ket{\Psi_0}
    \,,
    \label{SM-Hqp-summary}
    \\
    \hat{H}^{i}_{\mathrm{emb}}[\D_i,\Lambda_i^c]\ket{\Phi_i}
    &=E^{c}_i\ket{\Phi_i}
    \,,
    \label{SM-Hemb-summary}
    \\
    [\Delta_i]_{ab} &= 
    \frac{\partial}{\partial [\Lambda_i]_{ab}}
    \Av{\Psi_0}{\hat{H}_{\mathrm{qp}}[\R,\Lambda]}
    =\Av{\Psi_0}{\fc_{ia}\fa_{ib}}
    \,,
    \label{SM-Delta-summary}
    \\     
    \sum_{c=1}^{B{\nu}_i} \left[\D_i\right]_{c\alpha}\left[\Delta_i\left(\mathbf{1}- \Delta_i\right)\right]_{ac}^{\tfrac{1}{2}} &= 
    \frac{\partial}{\partial [\R_i]_{a\alpha}}
    \Av{\Psi_0}{\hat{H}_{\mathrm{qp}}[\R,\Lambda]}
    =\Av{\Psi_0}{\fc_{ia}\left(\sum_{j=1}^{\N}\sum_{\beta=1}^{\nu_j}
    [t_{ij}]_{\alpha\beta}\sum_{b=1}^{B\nu_j}[\R_j^\dagger]_{\beta b}\fa_{jb}\right)}
    \,,
    \label{detD}
    \\
    [l^c_i]_{s} 
    &= -
    [l_i]_{s}
    -\sum_{c,b=1}^{B{\nu}_i}\sum_{\alpha=1}^{\nu_i}\frac{\partial}{\partial \left[d_i\right]_s} \left(\left[\Delta_i\left(\mathbf{1}-\Delta_i\right)\right]^{\tfrac{1}{2}}_{cb}\left[\D_i\right]_{b\alpha}\left[\R_i\right]_{c\alpha} + \mathrm{c.c.}\right)
    \,,
    \label{detLc}
    \\
    \left[\Delta_i\right]_{ab} &= \bra{\Phi_i}\bba_{ib}\bbc_{ia}\ket{\Phi_i} 
    \,,
    \label{detF2}
    \\
    \sum_{a=1}^{B\nu_i}
    \left[\R_i\right]_{a\alpha}
    \left[\Delta_i\left(\mathbf{1}-\Delta_i\right)\right]_{ab}^{\tfrac{1}{2}} &= 
    \bra{\Phi_i}\cc_{i\alpha}\bba_{ib}\ket{\Phi_i}
    \,.
    \label{detF1}
\end{align}
\end{widetext}

As shown in Refs.~\cite{Ghost-GA,ALM_g-GA}, within the particle-number conserving variational ansatz of Sec.~\ref{sec:gga_setup} the ground-state EH eigenproblem Eq.~\eqref{SM-Hemb-summary} is understood in the half-filled impurity$+$bath sector, as specified by Eq.~\eqref{eq:half_fill_Hemb}.
The physical observables can be evaluated directly in terms of the variational parameters as follows:
\begin{itemize}
\item For any local operator $\hat{O}^{i}_{\mathrm{loc}}[\cc_{i\alpha},\ca_{i\alpha}]$
acting on fragment $i$ one has:
\begin{align}
\langle\Psi_G|\hat{O}^{i}_{\mathrm{loc}}|\Psi_G\rangle
&=
\langle\Phi_i|\hat{O}^{i}_{\mathrm{loc}}|\Phi_i\rangle
\,.
\label{eq:local_obs-copy}
\end{align}

\item For $i\neq j$, the ghost-GA approximation yields
\begin{align}
\begin{split}
\langle\Psi_G|\cc_{i\alpha}\ca_{j\beta}|\Psi_G\rangle
&=
\sum_{a=1}^{B\nu_i}\sum_{b=1}^{B\nu_j}
\big[\R_i^\dagger\big]_{\alpha a}
\langle\Psi_0|\fc_{ia}\fa_{jb}|\Psi_0\rangle
\big[\R_j\big]_{b\beta}
\,.
\end{split}
\label{eq:nonlocal_1body-copy}
\end{align}
\end{itemize}

\subsubsection{Identification of $\hat{H}^{i}_{0,\mathrm{emb}}$ as the GA EH of $\hat{H}_{\mathrm{qp}}$}
\label{sec:proof_SC}

Here we prove Eqs.~\eqref{SC-imp-proof}--\eqref{SC-bath-proof} at the saddle point of the standard ghost-GA Lagrange problem \eqref{eq:L_old}.

For this purpose, it is useful to consider the quasiparticle Hamiltonian
$\hat{H}_{\mathrm{qp}}[\R,\Lambda]$ [Eq.~\eqref{eq:Hqp}], treating $\R$ and $\Lambda$ as fixed parameters, and to apply to it the standard GA construction recalled above. We denote by a superscript $0$ the GA variables associated with this auxiliary problem, in order to distinguish them from the corresponding variables of the correlated physical problem. Thus, we consider the GA Lagrange problem for $\hat{H}_{\mathrm{qp}}[\R,\Lambda]$, extremized with respect to $\ket{\Psi_0^0}$, $E_0^0$, $\{\ket{\Phi_i^0}\}$, $\{E_i^{0c}\}$, and the matrices $\{\R_i^0,\Lambda_i^0,\D_i^0,\Lambda_i^{0c},\Delta_i^0\}$. Note that, since in this auxiliary problem the local degrees of freedom of fragment $i$ are the $B\nu_i$ ghost modes, the matrices $\R_i^0$, $\Lambda_i^0$, $\D_i^0$, $\Lambda_i^{0c}$, and $\Delta_i^0$ are all of size $B\nu_i\times B\nu_i$.

Since $\hat{H}_{\mathrm{qp}}[\R,\Lambda]$ is one-body, we already know that its GA solution $\ket{\Psi_0}$ is exactly reproduced by the GA, setting the Gutzwiller projector as the identity operator.
Therefore, the corresponding GA equations can be solved setting:
$\ket{\Psi^0_0}=\ket{\Psi_0}$, $E^0_0=E_0$,
$\R^0_i=\mathbf{1}_{B\nu_i}$
and $\Lambda^0_i=\Lambda_i$.
Therefore, specializing Eqs.~\eqref{SM-Hqp-summary}-\eqref{detF1} to this problem, we obtain the following equations:
\begin{widetext}
    \begin{align}
    \hat{H}_{\text{qp}}[\R,\Lambda]\ket{\Psi_0} &= E_0\ket{\Psi_0}
    \,,
    \label{SM-Hqp-summary0}
    \\
    \hat{H}^{i}_{\mathrm{emb}}[\D^0_i,\Lambda_i^{0c}]\ket{\Phi^0_i}
    &=E^{0c}_i\ket{\Phi^0_i}
    \,,
    \label{SM-Hemb-summary0}
    \\
    [\Delta^0_i]_{ab} &= 
    \Av{\Psi_0}{\fc_{ia}\fa_{ib}}
    \,,
    \label{SM-Delta-summary0}
    \\     
    \sum_{c=1}^{B{\nu}_i} \left[\D^0_i\right]_{cA}\left[\Delta^0_i\left(\mathbf{1}- \Delta^0_i\right)\right]_{ac}^{\tfrac{1}{2}} &= 
    \sum_{j=1}^{\N}
    \sum_{b=1}^{B\nu_j}
    \left[
    \sum_{\alpha=1}^{\nu_i}\sum_{\beta=1}^{\nu_j}[\R_i]_{A\alpha}
    [t_{ij}]_{\alpha\beta}[\R_j^\dagger]_{\beta b}
    \right]
    \Av{\Psi_0}{\fc_{ia}\fa_{jb}}
    \nonumber\\
    &=
    \sum_{\alpha=1}^{\nu_i}[\R_i]_{A\alpha}
    \Av{\Psi_0}{\fc_{ia}\left(\sum_{j=1}^{\N}\sum_{\beta=1}^{\nu_j}
    [t_{ij}]_{\alpha\beta}\sum_{b=1}^{B\nu_j}[\R_j^\dagger]_{\beta b}\fa_{jb}\right)}
    \,,
    \label{detD0}
    \\
    [l^{0c}_i]_{s} 
    &= -
    [l_i]_{s}
    -\sum_{c,b=1}^{B{\nu}_i}\sum_{A=1}^{B\nu_i}\frac{\partial}{\partial \left[d^{0}_i\right]_s} \left(\left[\Delta^0_i\left(\mathbf{1}-\Delta^0_i\right)\right]^{\tfrac{1}{2}}_{cb}\left[\D^0_i\right]_{bA}\left[\mathbf{1}_{B\nu_i}\right]_{cA} + \mathrm{c.c.}\right) 
    \,,
    \label{detLc0}
    \\
    \left[\Delta^0_i\right]_{ab} &= \bra{\Phi^0_i}\bba_{ib}\bbc_{ia}\ket{\Phi^0_i} 
    \,,
    \label{detF20}
    \\
    \left[\Delta^0_i\left(\mathbf{1}-\Delta^0_i\right)\right]_{ab}^{\tfrac{1}{2}} &= 
    \bra{\Phi^0_i}\fc_{ia}\bba_{ib}\ket{\Phi^0_i}
    \,,
    \label{detF10}
\end{align}
\end{widetext}
where 
\begin{align}
    \hat{H}^{i}_{\mathrm{emb}}[\D^0_i,\Lambda_i^{0c}]
&=
\sum_{a,b=1}^{B\nu_i}\big[\Lambda_i\big]_{ab}\,\fc_{ia}\fa_{ib} +
\sum_{a,b=1}^{B\nu_i}\big[\Lambda_i^{0c}\big]_{ab}\,\bba_{ib}\bbc_{ia}
\nonumber\\
&+
\sum_{a,b=1}^{B\nu_i}
\Big([\D^0_i]_{ab}\,\fc_{ib}\bba_{ia}+\text{H.c.}\Big)
\end{align}
and we introduced the following expansions:
\begin{align}
    \label{coeffDelta0}
    \Delta^0_i =& \sum_{s=1}^{(B{\nu}_i)^2} \left[d^0_i\right]_s \left[h^T_i\right]_s 
    \,,
    \\
    \label{coeffLc0}
    \Lambda^{0c}_i =& \sum_{s=1}^{(B{\nu}_i)^2} \left[l^{0c}_i\right]_s \left[h_i\right]_s \,,
\end{align}
where $\left[d^0_i\right]_s$ and $\left[l^{0c}_i\right]_s$ are real-valued coefficients.

Note that, since this auxiliary GA solution is realized with the Gutzwiller projector equal to the identity, particle number is conserved locally and the selection rule Eq.~\eqref{eq:mi_rule} reduces here to $m_i=0$ (ordinary GA, no ghost extension). 
Therefore, as shown in Refs.~\cite{Ghost-GA,ALM_g-GA}, this fixes the EH eigenproblem [Eq.~\eqref{SM-Hemb-summary0}] to be solved in the half-filled $f{+}b$ sector, as specified by Eq.~\eqref{eq:half_fill_H0emb}.

By comparing Eqs.~\eqref{SM-Hqp-summary0}-\eqref{detF10}
with Eqs.~\eqref{SM-Hqp-summary}-\eqref{detF1}
it follows that, at the saddle point of the correlated Lagrange function [Eq.~\eqref{eq:L_old}], the following conditions must hold:
\begin{align}
    [\Delta_i^0]_{ab}&=[\Delta_i]_{ab}
    \label{s1}
    \,,
    \\
    [\D^0_i]_{ab}&=[\D_i\R_i^T]_{ab}
    \label{s2}
    \,,
    \\
    [\Lambda_i^{0c}]_{ab}&=[\Lambda_i^{c}]_{ab}
    \label{s3}
    \,.
\end{align}
In particular, this implies that
\begin{equation}
    \hat{H}^{i}_{\mathrm{emb}}[\D^0_i,\Lambda_i^{0c}]=
    \hat{H}^{i}_{0,\mathrm{emb}}[\D_i,\Lambda_i^c;\R_i,\Lambda_i]
    \,,
    \label{s4}
\end{equation}
with $\hat{H}^{i}_{0,\mathrm{emb}}[\D_i,\Lambda_i^c;\R_i,\Lambda_i]$
defined in Eq.~\eqref{eq:H0emb}.

We also use the same GA/QE property recalled in Eq.~\eqref{eq:local_obs-copy}, here applied to the auxiliary GA problem defined by $\hat{H}_{\mathrm{qp}}[\R,\Lambda]$: the embedding state $\ket{\Phi_i^0}$ reproduces the expectation values of all local operators of that problem (built from the local $f$ and $b$ modes).

In summary:
\begin{itemize}
    \item Eq.~\eqref{SC-imp-proof} follows from Eqs.~\eqref{s4}, \eqref{SM-Delta-summary0}, and by applying the local-observable identity mentioned above to $\hat{O}^{i}_{\mathrm{loc}}=\fc_{ia}\fa_{ib}$;

    \item Eq.~\eqref{SC-hybr-proof} follows from Eqs.~\eqref{s4}, \eqref{s1}
    and \eqref{detF10};

    \item Eq.~\eqref{SC-bath-proof} follows from Eqs.~\eqref{s4}, \eqref{s1} and \eqref{detF20}.
\end{itemize}

\subsection{Converse implication: recovery of the standard ghost-GA functional}
\label{sec:inverse_equivalence_to_gga}

In Sec.~\ref{sec:direct_equivalence_to_gga} we showed that the QE Lagrange function [\eqref{eq:L_lagrange}] can be obtained from the standard ghost-GA Lagrange formulation~\eqref{eq:L_old}--\eqref{eq:F_old} by introducing the auxiliary quadratic problem $\hat H_{0,{\rm emb}}^i$ and the state $\ket{\Phi_i^0}$.
Here we prove the inverse statement: any stationary point of Eq.~\eqref{eq:L_lagrange} induces, up to gauge transformation, a stationary point of Eqs.~\eqref{eq:L_old}--\eqref{eq:F_old}.
This shows that introducing $\hat H_{0,{\rm emb}}^i$ does not create additional physical stationary points, but only uncovers an enlarged gauge-group structure, discussed in detail in Sec.~\ref{sec:gauge_structure_two_Lagrangians}.

Let
$\ket{\Psi_0},\{\ket{\Phi_i}\},\{\ket{\Phi_i^0}\},\{\R_i,\Lambda_i\},\{\D_i,\Lambda_i^c\}$
realize a stationary point of the QE functional~\eqref{eq:L_lagrange}, and
define:
\begin{equation}
[\Delta_i]_{ab} =
\Av{{\Phi}^0_i}{\fc_{ia}\fa_{ib}}
=
\Av{\Psi_0}{\fc_{ia}\fa_{ib}}
\,.
\label{eq:inverse_equiv_Delta_def}
\end{equation}
Since $\hat H_{0,{\rm emb}}^i$ is one-body and its ground state $\ket{\Phi_i^0}$ is a Slater determinant, the fermionic Schmidt theorem of Appendix~\ref{app:schmidt_theorem} can be applied to the bipartition:
\begin{equation}
\mathcal{H}^{(i)}_{0,{\rm emb}}=\mathcal{H}^{(i)}_{f}\otimes \mathcal{H}^{(i)}_{b}
\,.
\end{equation}
Therefore, there exists a unitary transformation of the form $
v_i(\phi_i)=e^{i\phi_i}$ such that
\begin{align}
\D_i &\rightarrow v_i^{T}(\phi_i)\D_i
\,,
\label{eq:gauge_D-repeated}
\\
\Lambda_i^c &\rightarrow v_i^\dagger(\phi_i)\Lambda_i^c\,v_i(\phi_i)
\,,
\label{eq:gauge_Lc-repeated}
\\
\ket{\Phi_i^0} &\rightarrow 
\exp\!\left(
i\sum_{a,b=1}^{B\nu_i}[\phi_i]_{ab}\,\bbc_{ia}\bba_{ib}
\right)
\ket{\Phi_i^0}
\,,
\label{eq:gauge_Phi0-repeated}
\end{align}
acting exclusively on the bath space, such that:
\begin{align}
\big\langle \Phi_i^0 \big| \fc_{ia}\fa_{ib} \big| \Phi_i^0 \big\rangle
&=
[\Delta_i]_{ab}
\,,
\label{eq:inverse_equiv_SC_imp}
\\
\big\langle \Phi_i^0 \big| \fc_{ia}\bba_{ib} \big| \Phi_i^0 \big\rangle
&=
\big[\Delta_i(\mathbf{1}-\Delta_i)\big]^{1/2}_{ab}
\,,
\label{eq:inverse_equiv_SC_hybr}
\\
\big\langle \Phi_i^0 \big| \bba_{ib}\,\bbc_{ia} \big| \Phi_i^0 \big\rangle
&=
[\Delta_i]_{ab}
\,.
\label{eq:inverse_equiv_SC_bath}
\end{align}
Since $\Lag$ in Eq.~\eqref{eq:L_lagrange} is invariant under the bath-space unitary rotation
\eqref{eq:gauge_D-repeated}--\eqref{eq:gauge_Phi0-repeated} (with $\ket{\Psi_0}$, $\{\ket{\Phi_i}\}$, and $\{\R_i,\Lambda_i\}$ unchanged),
the transformed set of variables
$\ket{\Psi_0},\{\ket{\Phi_i}\},\{\ket{\Phi_i^0}\},\{\R_i,\Lambda_i\},\{\D_i,\Lambda_i^c\}$
also realizes a (physically equivalent) stationary point.

Using Eq.~\eqref{eq:H0emb} together with
Eqs.~\eqref{eq:inverse_equiv_SC_imp}--\eqref{eq:inverse_equiv_SC_bath}, we can evaluate 
$\langle \Phi_i^0|\hat H_{0,{\rm emb}}^i|\Phi_i^0\rangle$ explicitly in terms of $\Delta_i$:
\begin{align}
&\big\langle \Phi_i^0 \big|\hat H^{i}_{0,{\rm emb}}[\D_i,\Lambda_i^c;\R_i,\Lambda_i]\big| \Phi_i^0 \big\rangle
=
\mathcal{F}_{\R_i,\Lambda_i,\D_i,\Lambda_i^c}^i[\Delta_i]
\,.
\label{eq:H0emb_expect_equals_Fold}
\end{align}

Substituting Eq.~\eqref{eq:H0emb_expect_equals_Fold} into Eq.~\eqref{eq:L_lagrange} (with the normalization of
$\ket{\Phi_i^0}$ enforced by $E_i^{0c}$) shows that, after the bath-gauge choice above, the QE Lagrange function
reduces exactly to the standard ghost-GA Lagrange function in Eq.~\eqref{eq:L_old}, with $\Delta_i$ understood as the
quasiparticle density matrix, see  Eq.~\eqref{eq:inverse_equiv_Delta_def}.

Finally, starting from the standard formulation~\eqref{eq:L_old}, one may eliminate the matrices $\Delta_i$ as independent
variables by substituting the stationarity condition with respect to $\Lambda_i$,
$[\Delta_i]_{ab}=\langle\Psi_0|\fc_{ia}\fa_{ib}|\Psi_0\rangle$ (see Eq.~\eqref{SM-Delta-summary}).
This produces the same reduced functional obtained above from Eq.~\eqref{eq:L_lagrange} after eliminating
the auxiliary sector $\{\ket{\Phi_i^0}\}$, proving that the QE formulation does not introduce additional physical stationary
points beyond those related by the bath gauge transformations.

\subsection{Enlarged gauge structure of the new Lagrange formulation}
\label{sec:gauge_structure_two_Lagrangians}

As shown in Sec.~\ref{sec:Gauge}, Eq.~\eqref{eq:L_lagrange} is invariant under independent local unitary rotations
of the ghost and bath single-particle spaces,
$\{u_i(\theta_i)\}$ and $\{v_i(\phi_i)\}$, implemented by
Eqs.~\eqref{eq:gauge_Psi0}--\eqref{eq:gauge_Lc}; the mixed hybridization matrix in $\hat H^{i}_{0,{\rm emb}}$
transforms covariantly as in Eq.~\eqref{eq:gauge_DRt}.

By contrast, the standard ghost-GA functional in Eqs.~\eqref{eq:L_old}--\eqref{eq:F_old} is invariant only under the
diagonal subgroup in which the two rotations are locked,
\begin{equation}
v_i(\phi_i)=u_i(\theta_i)
\qquad (i=1,\dots,\N),
\label{eq:diagonal_gauge_constraint}
\end{equation}
together with $\Delta_i\rightarrow u_i^\dagger(\theta_i)\,\Delta_i\,u_i(\theta_i)$.
This restriction follows directly from Eq.~\eqref{eq:F_old}, where $\D_i$ and $\R_i$ appear only through contractions
with $\Delta_i$ and $\big[\Delta_i(\mathbf{1}-\Delta_i)\big]^{1/2}$, which transform by conjugation with the same $u_i$.
Therefore, Eq.~\eqref{eq:L_old} implicitly fixes the relative $f/b$ gauge.

In summary, Eq.~\eqref{eq:L_lagrange} exposes the
full gauge-invariance group under which all physical observables are unchanged, see Eqs.~\eqref{eq:gauge_Psi0}--\eqref{eq:gauge_Lc}. 
This enlarged gauge structure of Eq.~\eqref{eq:L_lagrange} is essential for the relation with DMFT established in Sec.~\ref{sec:equivalence_gga_dmft_Binfty}, since it allows one to treat the
self-energy sector $(\R_i,\Lambda_i)$ and the Weiss-field sector $(\D_i,\Lambda_i^c)$ as independent, gauge-covariant
parametrizations of the two dynamical objects entering the DMFT self-consistency map.

\section{Proof of equivalence between ghost-GA and DMFT at $B\to\infty$}
\label{sec:equivalence_gga_dmft_Binfty}

In Sec.~\ref{sec:lagrange_formulation} we showed that the ghost-GA solution can be obtained extremizing a Lagrange function with respect to a set of matrices denoted as $\R_i,\D_i,\Lambda_i,\Lambda_i^c$.
These variational parameters are fully encoded, up to a gauge transformation, into a list of hybridization functions $\Delta_i(z)$ given by Eq.~\eqref{eq:hyb_def}:
\begin{equation}
\Delta_i(z)=\D_i^{T}\big(z\mathbf{1}+\Lambda_i^c\big)^{-1}\D_i^{*}
\,.
\label{eq:hyb_def-copy}
\end{equation}
and a local self-energy, represented as in Eq.~\eqref{eq:Sigma_block}:
\begin{align}
\Sigma(z)
&=
\begin{pmatrix}
\Sigma_{1}(z) & \mathbf{0} & \dots & \mathbf{0}\\
\mathbf{0} & \Sigma_{2}(z) & \dots & \vdots\\
\vdots & \vdots & \ddots & \vdots\\
\mathbf{0} & \dots & \dots & \Sigma_{\N}(z)
\end{pmatrix}\,,
\label{eq:DMFT_Sigma_local}
\end{align}
with $\Sigma_i(z)$ given by Eq.~\eqref{eq:Sigma_local}:
\begin{equation}
\Sigma_i(z)
=
z\mathbf{1}_{\nu_i}
-\Big[\R_i^\dagger\big(z\mathbf{1}_{B\nu_i}-\Lambda_i\big)^{-1}\R_i\Big]^{-1}
-\eps_i
\label{eq:Sigma_local-copy}
\,.
\end{equation}
Furthermore, in Sec.~\ref{sec:dmft_like_algorithm} we showed that the ghost-GA Lagrange equations can be solved with an algorithmic structure that closely resembles DMFT.

In this section we prove that ghost-GA solution reduces to DMFT in the limit $B\to\infty$.
This is accomplished as follows: 
\begin{itemize}

\item First, in Sec.~\ref{sec:parametrization} we note that, as pointed out in previous work, for each fragment $i=1,\dots,\N$, any DMFT hybridization function $\Delta_i(z)$ and any DMFT self-energy $\Sigma_i(z)$ can be reproduced exactly by the ghost-GA parametrizations \eqref{eq:hyb_def-copy} and \eqref{eq:Sigma_local-copy} in the limit $B\to\infty$.

\item Second, in the subsequent subsections we prove explicitly that, in the limit where the so-obtained representations of $\Sigma_i(z)$ and $\Delta_i(z)$
become exact, the resulting parameters $\R_i,\D_i,\Lambda_i,\Lambda_i^c$ satisfy the ghost-GA stationarity conditions~\eqref{e1}--\eqref{e7}, 
implying that the ghost-GA fixed point coincides with the DMFT fixed point as $B\to\infty$.

\end{itemize}

\subsection{Ghost-GA parametrization of DMFT hybridization function and self-energy}
\label{sec:parametrization}

Fix a fragment $i$. At a DMFT fixed point, the local objects $\Delta_i(z)$ and $\Sigma_i(z)$ are causal matrix-valued functions (in particular, they admit a representation in terms of a noninteracting bath and of a causal local self-energy). In this subsection we explain why the ghost-GA parametrizations \eqref{eq:hyb_def-copy} and \eqref{eq:Sigma_local-copy} are general enough to reproduce such functions in the limit $B\to\infty$.

\subsubsection{Hybridization function}

The parametrization \eqref{eq:hyb_def-copy},
\begin{align}
\Delta_i(z)=\D_i^{T}\big(z\mathbf{1}+\Lambda_i^c\big)^{-1}\D_i^{*}
\,,
\end{align}
is the hybridization function generated by a quadratic bath with $B\nu_i$ fermionic modes and one-body bath matrix $-\Lambda_i^c$. Diagonalizing $\Lambda_i^c$ with a unitary transformation and expanding the inverse shows that $\Delta_i(z)$ is a rational matrix function with at most $B\nu_i$ poles on the real axis. Increasing $B$ increases the number of available bath poles and therefore the resolution of the representation. In particular, the DMFT Weiss field for fragment $i$ is, by construction, the hybridization function of an impurity model with a (possibly infinitely large) noninteracting bath; discretizing that bath with $B\nu_i$ modes yields a sequence of rational hybridization functions of the form \eqref{eq:hyb_def-copy} converging to the target $\Delta_i(z)$ as $B\to\infty$.

\subsubsection{Self-energy}

The parametrization \eqref{eq:Sigma_local-copy},
\begin{align}
\Sigma_i(z)
&=
z\mathbf{1}_{\nu_i}
-\Big[\R_i^\dagger\big(z\mathbf{1}_{B\nu_i}-\Lambda_i\big)^{-1}\R_i\Big]^{-1}
-\eps_i
\,,
\end{align}
is invariant under the local gauge transformation \eqref{eq:gauge_R}--\eqref{eq:gauge_L}. Using this freedom, choose a unitary $u_i$ (e.g.\ from an SVD of $\R_i$) such that, in the transformed basis,
\begin{align}
u_i^\dagger \R_i
&=
\begin{pmatrix}
R_{0,i}\\
\mathbf{0}
\end{pmatrix},
\quad
u_i^\dagger \Lambda_i\,u_i
=
\begin{pmatrix}
\lambda_{0,i} & \lambda_{1,i}\\
\lambda_{1,i}^\dagger & \lambda_{2,i}
\end{pmatrix}
\,,
\label{eq:gauge_pole_form}
\end{align}
where $R_{0,i}\in\mathbb{C}^{\nu_i\times\nu_i}$ is invertible and $\lambda_{2,i}=\lambda_{2,i}^\dagger$ has size $(B\nu_i-\nu_i)\times(B\nu_i-\nu_i)$. Using an additional gauge rotation acting only within the $(B\nu_i-\nu_i)$-dimensional subspace corresponding to the lower block in Eq.~\eqref{eq:gauge_pole_form}, we can further assume that $\lambda_{2,i}$ is diagonal. Let $M_i=B\nu_i-\nu_i$.

Applying the block-inverse identity Eq.~\eqref{eq:block_inverse_M} (Appendix~\ref{sec:woodbury_projected_inverses})
to the block form \eqref{eq:gauge_pole_form} yields the following expression for the projected resolvent
entering \eqref{eq:Sigma_local-copy}:
\begin{align}
\label{eq:proj_resolvent_schur}
&\R_i^\dagger\big(z\mathbf{1}_{B\nu_i}-\Lambda_i\big)^{-1}\R_i
=
\\&\qquad
R_{0,i}^\dagger
\Big[
z\mathbf{1}_{\nu_i}-\lambda_{0,i}
-\lambda_{1,i}\big(z\mathbf{1}_{M_i}-\lambda_{2,i}\big)^{-1}\lambda_{1,i}^\dagger
\Big]^{-1}
R_{0,i}
\,.
\nonumber
\end{align}
Substituting Eq.~\eqref{eq:proj_resolvent_schur} into \eqref{eq:Sigma_local-copy} yields
\begin{align}
\Sigma_i(z)
&=
z\Big[\mathbf{1}_{\nu_i}-(R_{0,i}^\dagger R_{0,i})^{-1}\Big]
-\eps_i
+R_{0,i}^{-1}\lambda_{0,i}(R_{0,i}^\dagger)^{-1}
\nonumber\\
&\quad
+R_{0,i}^{-1}\lambda_{1,i}\big(z\mathbf{1}_{M_i}-\lambda_{2,i}\big)^{-1}
\lambda_{1,i}^\dagger(R_{0,i}^\dagger)^{-1}.
\label{eq:Sigma_rational_form}
\end{align}
Since $\lambda_{2,i}$ is diagonal, Eq.~\eqref{eq:Sigma_rational_form} can be written explicitly as
\begin{align}
\big[\Sigma_i(z)\big]_{\alpha\beta}
&=
z\Big[\mathbf{1}_{\nu_i}-(R_{0,i}^\dagger R_{0,i})^{-1}\Big]_{\alpha\beta}
-\big[\eps_i\big]_{\alpha\beta}
\nonumber\\
&\quad
+\sum_{a,b=1}^{\nu_i}\big[R_{0,i}^{-1}\big]_{\alpha a}\,
\big[\lambda_{0,i}\big]_{ab}\,
\big[(R_{0,i}^\dagger)^{-1}\big]_{b\beta}
\nonumber\\
&\quad
+\sum_{a,b=1}^{\nu_i}\sum_{c=1}^{M_i}\big[R_{0,i}^{-1}\big]_{\alpha a}\,
\big[\lambda_{1,i}\big]_{ac}\,
\nonumber\\&\qquad\quad
\times \frac{1}{z-\big[\lambda_{2,i}\big]_{cc}}\,
\big[\lambda_{1,i}^\dagger\big]_{cb}\,
\big[(R_{0,i}^\dagger)^{-1}\big]_{b\beta}
\,.
\label{eq:Sigma_pole_expansion}
\end{align}
Equation~\eqref{eq:Sigma_pole_expansion} exhibits $\Sigma_i(z)$ as a pole expansion whose pole locations are the diagonal entries $[\lambda_{2,i}]_{cc}$. 
Moreover, each pole has a positive-semidefinite residue by construction, consistent with a causal representation.

In summary, increasing $B$ (hence increasing $M_i=B\nu_i-\nu_i$) increases the number of available poles and allows one to approximate an arbitrary target causal $\Sigma_i(z)$ with arbitrary accuracy. In the limit $B\to\infty$ the representation can be made exact.

\subsubsection{Isometry from spectral sum rule}

Once the representation of $\Sigma_i(z)$ is exact, the high-frequency behavior of the corresponding Green's function satisfies the correct spectral sum rule. Therefore, as discussed in Sec.~\ref{sec:Asymptotic-Sigma}, the ghost-GA parameters satisfy
\begin{align}
\R_i^\dagger\R_i=\mathbf{1}_{\nu_i}
\,,
\label{eq:isometry_from-fit}
\end{align}
consistently with Eq.~\eqref{eq:sumrule_RdagR}.

In particular, in the gauge of Eq.~\eqref{eq:gauge_pole_form} this implies
\begin{align}
R_{0,i}^\dagger R_{0,i}=\mathbf{1}_{\nu_i}
\,,
\label{eq:R0_isometry_from_fit}
\end{align}
i.e., $R_{0,i}$ is unitary. Therefore Eq.~\eqref{eq:Sigma_rational_form} reduces to:
\begin{align}
& \big[\Sigma_i(z)\big]_{\alpha\beta}
=
-\big[\eps_i\big]_{\alpha\beta}
+\sum_{a,b=1}^{\nu_i}\big[R_{0,i}^\dagger\big]_{\alpha a}\,
\big[\lambda_{0,i}\big]_{ab}\,
\big[R_{0,i}\big]_{b\beta}
\nonumber\\
&
+\sum_{a,b=1}^{\nu_i}\sum_{c=1}^{M_i}\big[R_{0,i}^\dagger\big]_{\alpha a}\,
\big[\lambda_{1,i}\big]_{ac}
\frac{1}{z-\big[\lambda_{2,i}\big]_{cc}}\,
\big[\lambda_{1,i}^\dagger\big]_{cb}\,
\big[R_{0,i}\big]_{b\beta}
\,.
\label{eq:Sigma_pole_expansion_isometric}
\end{align}

\subsection{Setup: from a DMFT fixed point to a ghost-GA realization}
\label{sec:dmft_to_gga_setup}

In this section we assume a fixed-point solution of zero-temperature DMFT for the lattice Hamiltonian \eqref{eq:H_def}.
For each fragment $i=1,\dots,\N$, we denote by $\Delta_i(z)$ the corresponding Weiss (hybridization) function and by $\Sigma_i(z)$ the corresponding local self-energy.
The associated impurity Green's function is
\begin{align}
G_i(z)
&=
\big[z\mathbf{1}_{\nu_i}-\eps_i-\Delta_i(z)-\Sigma_i(z)\big]^{-1}
\,.
\label{eq:dmft_impurity_dyson}
\end{align}
At the DMFT fixed point, $G_i(z)$ coincides with the $i$-th diagonal block of the lattice Green's function obtained from the same local self-energy.
Introducing the fragment-diagonal matrix
\begin{align}
\Sigma(z)
&=
\begin{pmatrix}
\Sigma_{1}(z) & \mathbf{0} & \dots & \mathbf{0}\\
\mathbf{0} & \Sigma_{2}(z) & \dots & \vdots\\
\vdots & \vdots & \ddots & \vdots\\
\mathbf{0} & \dots & \dots & \Sigma_{\N}(z)
\end{pmatrix},
\label{eq:dmft_Sigma_block}
\end{align}
the lattice Green's function is
\begin{align}
G(z)
&=
\big[z\mathbf{1}-h_0-\Sigma(z)\big]^{-1}
\nonumber\\
&=
\begin{pmatrix}
G_{1}(z) & G_{12}(z) & \dots & G_{1\N}(z)\\
G_{21}(z) & G_{2}(z) & \dots & \vdots\\
\vdots & \vdots & \ddots & \vdots\\
G_{\N 1}(z) & \dots & \dots & G_{\N}(z)
\end{pmatrix}
\,,
\label{eq:dmft_lattice_dyson_blocks}
\end{align}
where $\mathbf{1}$ denotes the identity matrix in the full physical one-particle space and $h_0$ is the one-body matrix introduced in Eq.~\eqref{eq:h0_block}.

We assume that $B$ is chosen sufficiently large so that $\Delta_i(z)$ and $\Sigma_i(z)$ can be represented (to the desired accuracy, and exactly in the limit $B\to\infty$) by the ghost-GA parametrizations discussed in Sec.~\ref{sec:parametrization}, namely:
\begin{align}
\Delta_i(z)
&=\D_i^{T}\big(z\mathbf{1}+\Lambda_i^c\big)^{-1}\D_i^{*}
\,,
\label{eq:setup_Delta_fit}
\\
\Sigma_i(z)
&=
z\mathbf{1}_{\nu_i}
-\Big[\R_i^\dagger\big(z\mathbf{1}_{B\nu_i}-\Lambda_i\big)^{-1}\R_i\Big]^{-1}
-\eps_i
\,.
\label{eq:setup_Sigma_fit}
\end{align}
In the remainder of this section we show that, under the standing assumptions above, the matrices
$(\R_i,\Lambda_i,\D_i,\Lambda_i^c)$ obtained in this way satisfy the ghost-GA stationarity conditions \eqref{e1}--\eqref{e7}.

In order to prove our claim, we consider the three operators defined in Sec.~\ref{sec:lagrange_formulation}:
the quasiparticle Hamiltonian $\hat{H}_{\mathrm{qp}}[\R,\Lambda]$ in Eq.~\eqref{eq:Hqp},
the interacting embedding Hamiltonian $\hat{H}^{i}_{\mathrm{emb}}[\D_i,\Lambda_i^c]$ in Eq.~\eqref{eq:Hemb},
and the auxiliary quadratic embedding Hamiltonian $\hat{H}^{i}_{0,\mathrm{emb}}[\D_i,\Lambda_i^c;\R_i,\Lambda_i]$ in Eq.~\eqref{eq:H0emb}.

\begin{figure}
\centering
\includegraphics[width=\columnwidth]{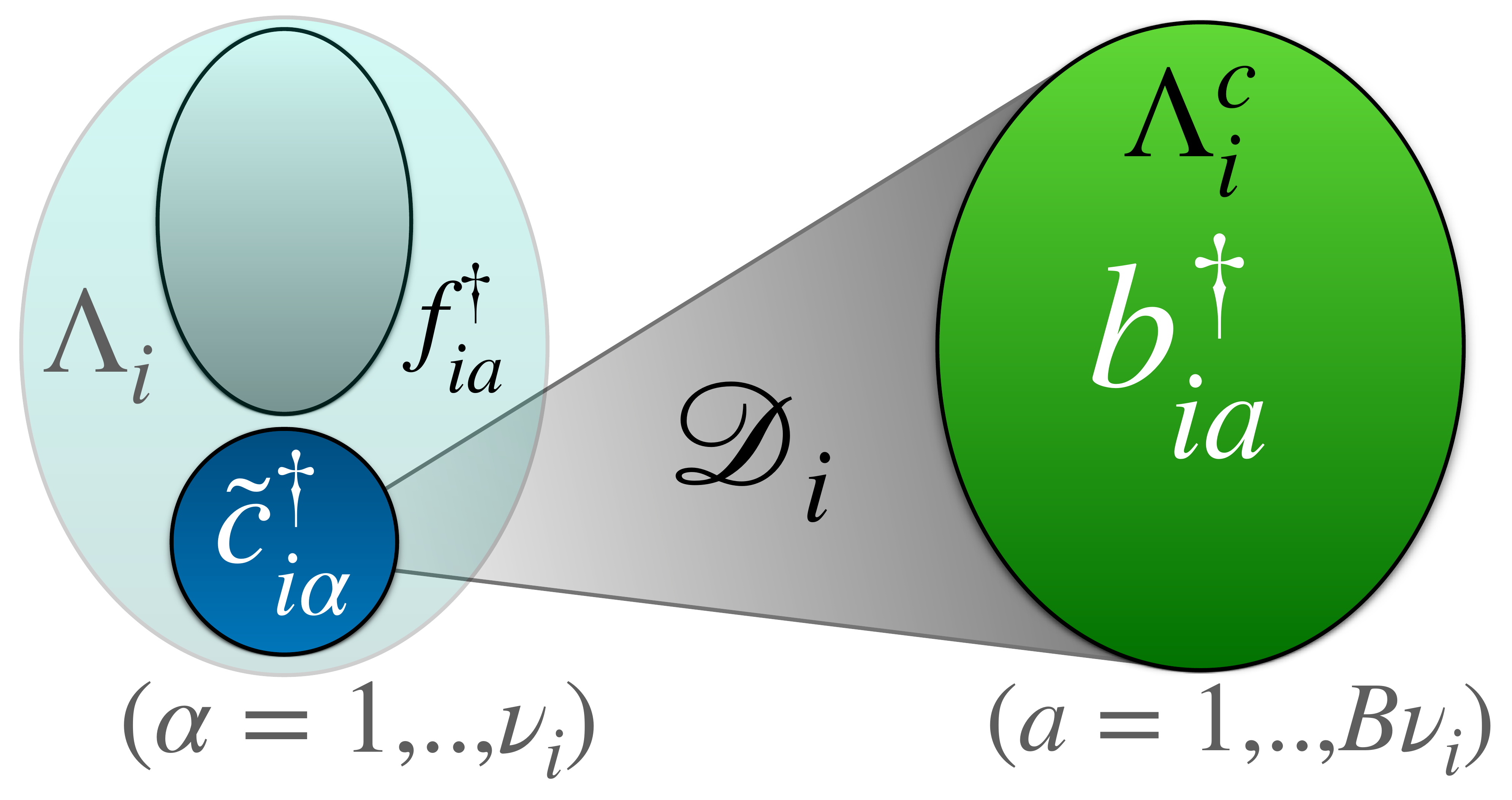}
\caption{Schematic representation of the auxiliary quadratic embedding Hamiltonian $\hat{H}^{i}_{0,\mathrm{emb}}[\D_i,\Lambda_i^c;\R_i,\Lambda_i]$ for a fixed fragment $i$. The impurity modes $\{\tilde c_{i\alpha}\}_{\alpha=1,\dots,\nu_i}$ are embedded within the local $f$ sector and hybridize with the bath modes $\{b_{ia}\}_{a=1,\dots,B\nu_i}$ through $\D_i$. The bath one-body term is controlled by $\Lambda_i^c$ [Eq.~\eqref{eq:setup_Delta_fit}], while the local quadratic term $\Lambda_i$ acts in the full local $f$ space, mixing $\tilde c_{i\alpha}$ with the remaining local degrees of freedom (complement within the $f$ sector).}
\label{Figure4}
\end{figure}

\begin{figure}
\centering
\includegraphics[width=\columnwidth]{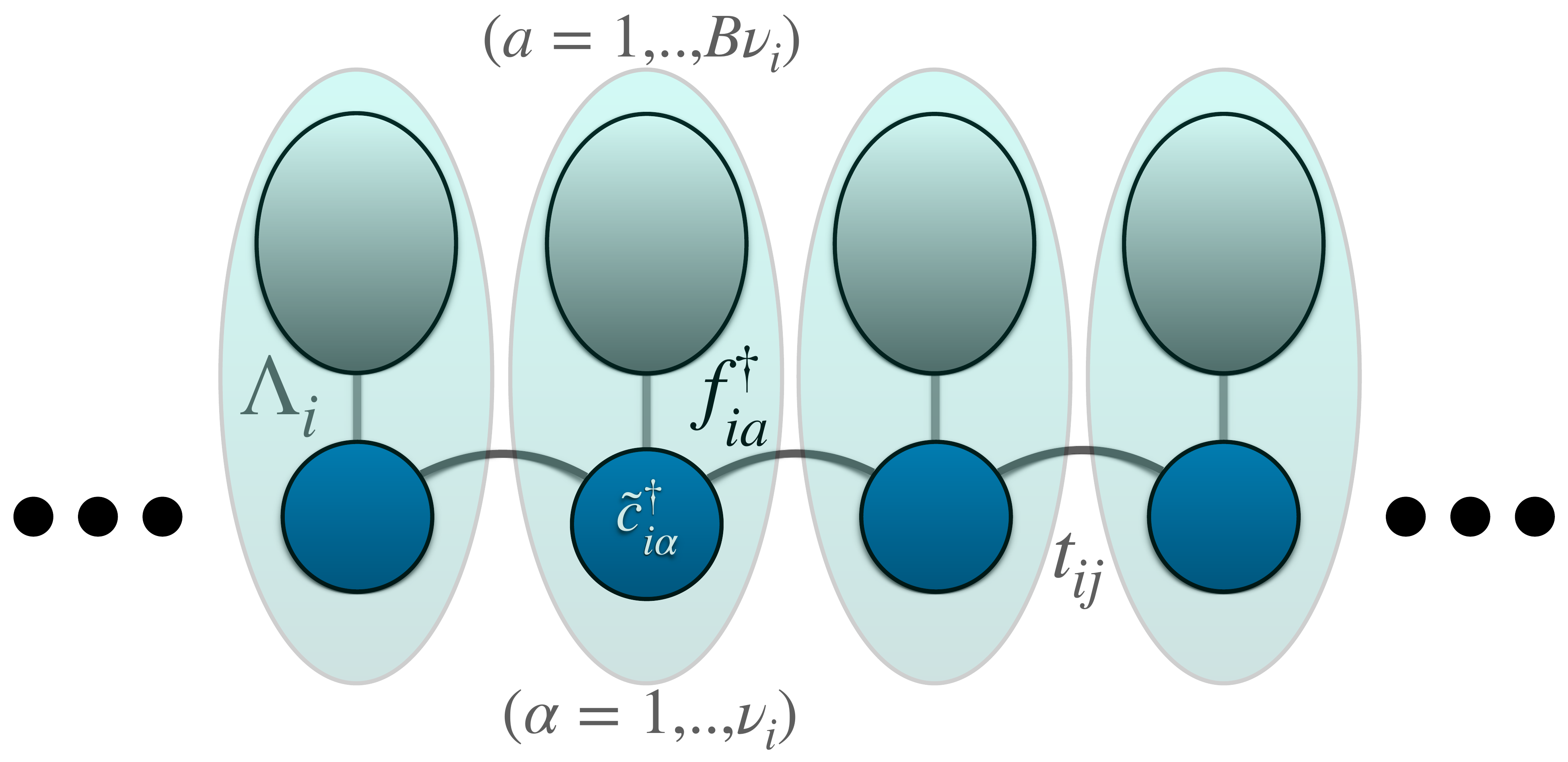}
\caption{Schematic representation of the quasiparticle Hamiltonian $\hat{H}_{\mathrm{qp}}[\R,\Lambda]$. Inter-fragment hopping is restricted to the $\{\tilde c_{i\alpha}\}$ subspace and is governed by the physical hopping matrices $t_{ij}$, whereas the local quadratic term $\Lambda_i$ acts on the full local $f$ sector and therefore couples $\tilde c_{i\alpha}$ to the remaining local degrees of freedom within $\{f_{ia}\}$.}
\label{Figure5}
\end{figure}

\subsection{Role of isometry relation in $B\rightarrow\infty$ limit}

A key observation, at the core of our proof, is that in the $B\rightarrow\infty$ limit the correct spectral sum rule implies (Sec.~\ref{sec:Asymptotic-Sigma}):
\begin{align}
\R_i^\dagger\R_i=\mathbf{1}_{\nu_i}
\,.
\label{eq:setup_isometry}
\end{align}
Equation~\eqref{eq:setup_isometry} allows us to define, for each fragment $i$, canonical impurity operators within the local $f$ sector:
\begin{align}
\tilde c_{i\alpha}
=
\sum_{a=1}^{B\nu_i}[\R_i^\dagger]_{\alpha a}\,\fa_{ia}\,,
\qquad
(\alpha=1,\dots,\nu_i)
\,.
\label{eq:setup_ctilde_def}
\end{align}
In fact, from this relation it follows that:
\begin{align}
\{\tilde{c}_{i\alpha},\tilde{c}_{i\beta}^\dagger\}
&=
\sum_{a=1}^{B\nu_i}[\R_i^\dagger]_{\alpha a}[\R_i]_{a\beta}
=
[\R_i^\dagger\R_i]_{\alpha\beta}
=
\delta_{\alpha\beta}
\,.
\label{eq:setup_ctilde_canonical}
\end{align}

This has two key consequences:
\begin{enumerate}
    \item The auxiliary quadratic embedding Hamiltonian $\hat{H}^{i}_{0,\mathrm{emb}}$, previously introduced in Eq.~\eqref{eq:H0emb}, can be represented as:
    \begin{align}
    & \hat{H}^{i}_{0,\mathrm{emb}}[\D_i,\Lambda_i^c;\R_i,\Lambda_i]
    =
    \sum_{a,b=1}^{B\nu_i}\big[\Lambda_i\big]_{ab}\,\fc_{ia}\fa_{ib}
    \label{eq:H0emb-2}\\
    & \quad +
    \sum_{a=1}^{B\nu_i}\sum_{\alpha=1}^{\nu_i}
    \Big(\big[\D_i\big]_{a\alpha}\tilde{c}^\dagger_{i\alpha}\bba_{ia}+\text{H.c.}\Big)
    +
    \sum_{a,b=1}^{B\nu_i}\big[\Lambda_i^c\big]_{ab}\,\bba_{ib}\bbc_{ia}
    \,.
    \nonumber
    \end{align}
    Equation~\eqref{eq:H0emb-2} makes explicit that the bath modes $\{b_{ia}\}$ hybridize directly with the impurity operators $\{\tilde c_{i\alpha}\}$, while the remaining local degrees of freedom within the $f$ space enter through the quadratic term governed by $\Lambda_i$,
    see Fig.~\ref{Figure4}.

    \item The quasiparticle Hamiltonian $\hat{H}_{\mathrm{qp}}[\R,\Lambda]$, previously introduced in Eq.~\eqref{eq:Hqp}, can be represented as:
    \begin{align}
    \hat{H}_{\mathrm{qp}}[\R,\Lambda]
    &=
    \sum_{\substack{i,j=1\\ i\neq j}}^{\N}
    \sum_{\alpha=1}^{\nu_i}\sum_{\beta=1}^{\nu_j}
    [t_{ij}]_{\alpha\beta}\,\tilde c_{i\alpha}^\dagger \tilde c_{j\beta}
    \nonumber\\
    &\quad+
    \sum_{i=1}^{\N}\sum_{a,b=1}^{B\nu_i}
    [\Lambda_i]_{ab}\,\fc_{ia}\fa_{ib}
    \,.
    \label{eq:Hqp_ctilde}
    \end{align}
    This form makes explicit that inter-fragment hopping acts only within the $\{\tilde c_{i\alpha}\}$ subspace, while the local quadratic term governed by $\Lambda_i$ acts in the full local $f$ space. Hence it couples $\{\tilde c_{i\alpha}\}$ to the remaining local degrees of freedom within $\{f_{ia}\}$,
    see Fig.~\ref{Figure5}.

\end{enumerate}

Within this setting, the ghost-GA can be interpreted through the lens of the ghost-DMET QE scheme of Ref.~\cite{gDMET}:
(I) On one hand, the ghost-GA stationarity conditions in Eqs.~\eqref{e6}--\eqref{e7} can be written as
\begin{align}
\Av{\Phi_i^0}{\bba_{ib}\bbc_{ia}}
&=
\Av{\Phi_i}{\bba_{ib}\bbc_{ia}},
\\
\Av{\Phi_i^0}{\tilde{c}^\dagger_{i\alpha}\,\bba_{ib}}
&=
\Av{\Phi_i}{\cc_{i\alpha}\,\bba_{ib}}.
\end{align}
Therefore, $\hat{H}^{i}_{0,\mathrm{emb}}$ acts as an interface that matches the impurity environment of $\hat{H}^{i}_{\mathrm{emb}}$ at the level of equal-time impurity--bath correlators, with the physical impurity operators $\{c_{i\alpha}\}$ replaced by the canonical modes 
$\{\tilde{c}_{i\alpha}\}$.
(II) On the other hand, as discussed in Sec.~\ref{sec:analytic_hybridization_update}, the stationarity conditions in Eqs.~\eqref{e4}--\eqref{e5} encode the one-body matching between $\hat{H}^{i}_{0,\mathrm{emb}}$ and $\hat{H}_{\mathrm{qp}}$, implied by the Schmidt embedding construction.

In the following subsections we exploit this structure to complete the DMFT correspondence proof: starting from a DMFT fixed point and its ghost-GA representation in the $B\to\infty$ limit, we show explicitly that the resulting parameters $(\R_i,\Lambda_i,\D_i,\Lambda_i^c)$ satisfy the stationarity conditions \eqref{e4}--\eqref{e7}. Specifically, Sec.~\ref{sec:weiss_from_H0emb} derives the impurity representation implied by $\hat{H}^{i}_{0,\mathrm{emb}}$ and establishes the equal-time identities that yield \eqref{e6}--\eqref{e7}, while Sec.~\ref{sec:e4_e5_proof} relates the corresponding Weiss field to the lattice (cavity) construction and thereby verifies \eqref{e4}--\eqref{e5}.


\subsection{Weiss field from $\hat{H}^{i}_{0,\mathrm{emb}}[\D_i,\Lambda_i^c;\R_i,\Lambda_i]$}
\label{sec:weiss_from_H0emb}

Let us focus on a fixed fragment $i$ and consider the auxiliary quadratic embedding Hamiltonian
$\hat{H}^{i}_{0,\mathrm{emb}}[\D_i,\Lambda_i^c;\R_i,\Lambda_i]$
previously introduced in Eq.~\eqref{eq:H0emb-2}.
Our  goal is to prove that the Green's function of the
degrees of freedom $\{\tilde{c}_{i\alpha}\}$, as defined in Eq.~\eqref{eq:setup_ctilde_def}, is given by the following equation:
\begin{align}
G_{\tilde c,i}(z)
&=
\big[
z\mathbf{1}_{\nu_i}-\eps_i-\Delta_i(z)-\Sigma_i(z)
\big]^{-1}
\,,
\label{eq:Dyson_ctilde}
\end{align}
with $\Sigma_i(z)$ given by Eq.~\eqref{eq:Sigma_local} and $\Delta_i(z)$ given by Eq.~\eqref{eq:hyb_def}:
\begin{align}
\Delta_i(z)
&=
\D_i^{T}\big(z\mathbf{1}_{B\nu_i}+\Lambda_i^c\big)^{-1}\D_i^{*}
\,,
\label{eq:Delta_def_local-2}
\\
\Sigma_i(z)
&=
z\mathbf{1}_{\nu_i}
-\Big[\R_i^\dagger\big(z\mathbf{1}_{B\nu_i}-\Lambda_i\big)^{-1}\R_i\Big]^{-1}
-\eps_i
\label{eq:Sigma_local-2}
\,.
\end{align}

To prove Eq.~\eqref{eq:Dyson_ctilde}, let us consider the
single-particle resolvent of $\hat H^{i}_{0,\mathrm{emb}}$, which can be expressed in the following block-matrix form:
\begin{align}
\mathcal{G}_i(z)
=
\Bigg[
z\mathbf{1}_{2B\nu_i}-
\begin{pmatrix}
\Lambda_i & W_i\\
W_i^\dagger & -\Lambda_i^c
\end{pmatrix}
\Bigg]^{-1}
\,,
\label{eq:G0_full}
\end{align}
where we introduced the $B\nu_i\times B\nu_i$ matrix
\begin{align}
W_i=\R_i\,\D_i^{T}
\,.
\label{eq:Wi_def}
\end{align}

\begin{itemize}

\item We denote by $G_{f,i}(z)$ the $(f,f)$ block of $\mathcal{G}_i(z)$ in Eq.~\eqref{eq:G0_full}.
Applying the block-inverse formula Eq.~\eqref{eq:block_inverse_M} (Appendix~\ref{sec:woodbury_projected_inverses}) to the matrix
\begin{align}
z\mathbf{1}_{2B\nu_i}-\begin{pmatrix}
\Lambda_i & W_i\\
W_i^\dagger & -\Lambda_i^c
\end{pmatrix}
=
\begin{pmatrix}
A & U\\
U^\dagger & C
\end{pmatrix},
\end{align}
with the identifications
\begin{align}
A=z\mathbf{1}_{B\nu_i}-\Lambda_i,
\quad
U=-W_i,
\quad
C=z\mathbf{1}_{B\nu_i}+\Lambda_i^c
\,,
\end{align}
we obtain:
\begin{align}
G_{f,i}(z)
&=
\Big[
A-U\,C^{-1}U^\dagger
\Big]^{-1}
\nonumber\\
&=
\Big[
z\mathbf{1}_{B\nu_i}-\Lambda_i
-
\R_i\,\Delta_i(z)\,\R_i^\dagger
\Big]^{-1}.
\label{eq:G0_f_block}
\end{align}

\item By definition \eqref{eq:setup_ctilde_def}, the $\tilde c$-Green's function induced by $\hat H^{i}_{0,\mathrm{emb}}$ is the projection of $G_{f,i}(z)$:
\begin{align}
G_{\tilde c,i}(z)
&=
\R_i^\dagger\,G_{f,i}(z)\,\R_i
\nonumber\\
&=
\R_i^\dagger\Big[z\mathbf{1}_{B\nu_i}-\Lambda_i-\R_i\Delta_i(z)\R_i^\dagger\Big]^{-1}\R_i
\,.
\label{eq:G0_ctilde_def}
\end{align}

We now compute $\big[G_{\tilde c,i}(z)\big]^{-1}$ explicitly, applying Eq.~\eqref{eq:woodbury_projected_inverse} (Appendix~\ref{sec:woodbury_projected_inverses}) to Eq.~\eqref{eq:G0_ctilde_def} with
\begin{align}
A=z\mathbf{1}_{B\nu_i}-\Lambda_i
\,,
\quad U=\R_i
\,,
\quad C=\Delta_i(z)
\,.
\label{eq:woodbury_subs}
\end{align}

Using these definitions we obtain:
\begin{align}
\big[G_{\tilde c,i}(z)\big]^{-1}
&=
\Big[\R_i^\dagger\big(z\mathbf{1}_{B\nu_i}-\Lambda_i\big)^{-1}\R_i\Big]^{-1}
-\Delta_i(z)
\,.
\label{eq:G0_ctilde_inverse_explicit}
\end{align}

\item Recalling Eq.~\eqref{eq:Sigma_local-2},
one rewrites
\begin{align}
\Big[\R_i^\dagger\big(z\mathbf{1}_{B\nu_i}-\Lambda_i\big)^{-1}\R_i\Big]^{-1}
&=
z\mathbf{1}_{\nu_i}-\eps_i-\Sigma_i(z)
\,.
\label{eq:Sigma_rearranged}
\end{align}
Substituting Eq.~\eqref{eq:Sigma_rearranged} into Eq.~\eqref{eq:G0_ctilde_inverse_explicit} yields the DMFT-like Dyson form of Eq.~\eqref{eq:Dyson_ctilde}.

\end{itemize}

In summary, we proved Equation~\eqref{eq:Dyson_ctilde}, which shows that $\hat{H}^{i}_{0,\mathrm{emb}}[\D_i,\Lambda_i^c;\R_i,\Lambda_i]$ provides a quadratic realization of an impurity problem for the modes $\tilde c_{i\alpha}$, where the bath parameters $(\D_i,\Lambda_i^c)$ generate the Weiss field $\Delta_i(z)$ in Eq.~\eqref{eq:Delta_def_local-2}, while the parameters $(\R_i,\Lambda_i)$ generate the local self-energy $\Sigma_i(z)$ in Eq.~\eqref{eq:Sigma_local-2}.

As a consequence, the full single-particle Green's function matrix of the impurity--bath subsystem is also reproduced.
Indeed, let $\mathcal{G}^{(cb)}_i(z)$ denote the full $(c_i,b_i)$ Green's function matrix of the correlated embedding Hamiltonian $\hat{H}^{i}_{\mathrm{emb}}[\D_i,\Lambda_i^c]$ in Eq.~\eqref{eq:Hemb}, and let $\mathcal{G}^{(\tilde c b)}_i(z)$ denote the full $(\tilde c_i,b_i)$ Green's function matrix of the auxiliary quadratic embedding Hamiltonian $\hat{H}^{i}_{0,\mathrm{emb}}[\D_i,\Lambda_i^c;\R_i,\Lambda_i]$.
Then, in both cases, the inverse Green's function matrix is
\begin{align}
\big[\mathcal{G}^{(cb)}_i(z)\big]^{-1}
&=
\begin{pmatrix}
z\mathbf{1}_{\nu_i}-\eps_i-\Sigma_i(z) & -\D_i^{T}\\
-\D_i^{*} & z\mathbf{1}_{B\nu_i}+\Lambda_i^c
\end{pmatrix},
\nonumber\\
\big[\mathcal{G}^{(\tilde c b)}_i(z)\big]^{-1}
&=
\begin{pmatrix}
z\mathbf{1}_{\nu_i}-\eps_i-\Sigma_i(z) & -\D_i^{T}\\
-\D_i^{*} & z\mathbf{1}_{B\nu_i}+\Lambda_i^c
\end{pmatrix}.
\label{eq:G_ctilde_b_inverse}
\end{align}
Consequently, $\mathcal{G}^{(cb)}_i(z)=\mathcal{G}^{(\tilde c b)}_i(z)$ as matrix-valued functions, even though the impurity operators are different in the two realizations (physical $c_{i\alpha}$ versus $\tilde c_{i\alpha}$).
Therefore the corresponding equal-time one-body correlators coincide.
In particular, we have:
\begin{align}
\Av{\Phi^0_i}{\bbc_{ia}\bba_{ib}}
&=
\Av{\Phi_i}{\bbc_{ia}\bba_{ib}}
\,,
\label{e6-ctilde}
\\
\Av{\Phi^0_i}{\tilde c_{i\alpha}^\dagger\,\bba_{ib}}
&=
\Av{\Phi_i}{\cc_{i\alpha}\bba_{ib}}
\,,
\label{e7-ctilde}
\end{align}
where $\ket{\Phi_i}$ denotes the ground state of $\hat{H}^{i}_{\mathrm{emb}}[\D_i,\Lambda_i^c]$ and $\ket{\Phi_i^0}$ the ground state of $\hat{H}^{i}_{0,\mathrm{emb}}[\D_i,\Lambda_i^c;\R_i,\Lambda_i]$.

\subsection{Lattice Weiss field from the quasiparticle cavity construction}
\label{sec:e4_e5_proof}

In this subsection we prove Eqs.~\eqref{e4} and \eqref{e5}. The key step is to identify the lattice Weiss field from the quasiparticle cavity construction and show that, at the DMFT fixed point, it coincides with the impurity Weiss field $\Delta_i(z)$.

We consider the single-particle resolvent of the quadratic quasiparticle Hamiltonian
$\hat{H}_{\mathrm{qp}}[\R,\Lambda]$ [Eq.~\eqref{eq:Hqp}],
\begin{align}
\mathcal{G}^{\mathrm{qp}}(z)
&=
\big[z\mathbf{1}-h^*[\R,\Lambda]\big]^{-1}
\nonumber\\
&=\big[(z\mathbf{1}-\Lambda)-\R\,t\,\R^\dagger\big]^{-1}
\nonumber\\
&=
\begin{pmatrix}
\mathcal{G}^{\mathrm{qp}}_{1}(z) & \mathcal{G}^{\mathrm{qp}}_{12}(z) & \dots & \mathcal{G}^{\mathrm{qp}}_{1\N}(z)\\
\mathcal{G}^{\mathrm{qp}}_{21}(z) & \mathcal{G}^{\mathrm{qp}}_{2}(z) & \dots & \vdots\\
\vdots & \vdots & \ddots & \vdots\\
\mathcal{G}^{\mathrm{qp}}_{\N 1}(z) & \dots & \dots & \mathcal{G}^{\mathrm{qp}}_{\N}(z)
\end{pmatrix},
\label{eq:Gqp_def_blocks}
\end{align}
where $h^*[\R,\Lambda]=\R\,t\,\R^\dagger+\Lambda$ [Eq.~\eqref{eq:hstar_compact}],
$\mathcal{G}^{\mathrm{qp}}_{j}(z)$ denotes the $j$-th diagonal block (size $B\nu_j\times B\nu_j$),
and $\mathcal{G}^{\mathrm{qp}}_{jk}(z)$ denotes the $(j,k)$ off-diagonal block.

We now project the quasiparticle resolvent $\mathcal{G}^{\mathrm{qp}}(z)$ onto the $\tilde c$ subspace
defined by $\R$ [Eq.~\eqref{eq:block_R_Lambda}] and define
\begin{align}
G(z)
&=
\R^\dagger\,\mathcal{G}^{\mathrm{qp}}(z)\,\R
\nonumber\\
&=
\begin{pmatrix}
G_{1}(z) & G_{12}(z) & \dots & G_{1\N}(z)\\
G_{21}(z) & G_{2}(z) & \dots & \vdots\\
\vdots & \vdots & \ddots & \vdots\\
G_{\N 1}(z) & \dots & \dots & G_{\N}(z)
\end{pmatrix},
\label{eq:Gi_from_qp}
\end{align}
where $G_i(z)$ denotes the $i$-th diagonal block (of size $\nu_i\times\nu_i$).

Using Eq.~\eqref{eq:woodbury_projected_inverse} (Appendix~\ref{sec:woodbury_projected_inverses}) directly on
$\mathcal{G}^{\mathrm{qp}}(z)$ with
$A=z\mathbf{1}-\Lambda$, $U=\R$, and $C=t$ yields
\begin{align}
G(z)^{-1}
&=
\Big[\R^\dagger(z\mathbf{1}-\Lambda)^{-1}\R\Big]^{-1}-t
\,.
\label{eq:G_dyson_step1}
\end{align}
Finally, invoking Eq.~\eqref{eq:Sigma_local-2} (and $h_0=\eps+t$, Eq.~\eqref{eq:eps_def}) to rewrite
$\big[\R^\dagger(z\mathbf{1}-\Lambda)^{-1}\R\big]^{-1}=z\mathbf{1}-\eps-\Sigma(z)$, we obtain the lattice Dyson form
\begin{align}
G(z)^{-1}
&=
z\mathbf{1}-h_0-\Sigma(z)
\,,
\label{eq:G_dyson_latt_global}
\end{align}
where $\Sigma(z)$ has the local (fragment-diagonal) structure defined in Eq.~\eqref{eq:Sigma_block},
with diagonal blocks $\Sigma_i(z)$ given by Eq.~\eqref{eq:Sigma_local-2}.

\subsubsection*{Lattice Weiss field}

We define the lattice Weiss (hybridization) field $\Delta_i^{\mathrm{latt}}(z)$ by the local Dyson relation:
\begin{align}
G_i(z)^{-1}
&=
z\mathbf{1}_{\nu_i}-\eps_i-\Delta^{\mathrm{latt}}_i(z)
-\Sigma_i(z)
\,.
\label{eq:Gi_dyson_latt}
\end{align}
We now compute $\Delta^{\mathrm{latt}}_i(z)$ explicitly in terms of the quasiparticle cavity construction.

\begin{itemize}
\item Let $\bar{h}^{\,*}_i[\R,\Lambda]$ denote the block matrix obtained from $h^*[\R,\Lambda]$ by removing
the block row and column corresponding to fragment $i$, and define the corresponding cavity resolvent
\begin{align}
\bar{\mathcal{G}}^{\mathrm{qp}}_i(z)
&=
\big[z\mathbf{1}-\bar{h}^{\,*}_i[\R,\Lambda]\big]^{-1},
\label{eq:Gqp_cavity_def}
\end{align}
with blocks $\bar{\mathcal{G}}^{\mathrm{qp}}_{i;jk}(z)$ for $j,k\neq i$.
To obtain the local ghost block $\mathcal{G}^{\mathrm{qp}}_{i}(z)$, we apply the block-inverse identity
Eq.~\eqref{eq:block_inverse_M} (Appendix~\ref{sec:woodbury_projected_inverses}) to the matrix
$z\mathbf{1}-h^*[\R,\Lambda]$ decomposed in the $(i)$ block versus its complement. 
This yields:
\begin{align}
\mathcal{G}^{\mathrm{qp}}_{i}(z)
&=
\Big[z\mathbf{1}_{B\nu_i}-\Lambda_i-\R_i\,X_i(z)\,\R_i^\dagger\Big]^{-1}
\,,
\label{eq:Gqp_local_block-X}
\end{align}
where we introduced the $\nu_i\times \nu_i$ matrix
\begin{align}
X_i(z)
&=
\sum_{\substack{j,k=1\\ j\neq i,\,k\neq i}}^{\N}
t_{ij}\,\R_j^\dagger\,\bar{\mathcal{G}}^{\mathrm{qp}}_{i;jk}(z)\,\R_k\,t_{ki}
\,.
\label{eq:Delta_latt_def-X}
\end{align}

\item Projecting Eq.~\eqref{eq:Gqp_local_block-X} onto the $\{\tilde c_{i\alpha}\}$ subspace gives
$G_i(z)=\R_i^\dagger\,\mathcal{G}^{\mathrm{qp}}_{i}(z)\,\R_i$.
Applying Eq.~\eqref{eq:woodbury_projected_inverse} (Appendix~\ref{sec:woodbury_projected_inverses}) to this expression with
$A=z\mathbf{1}_{B\nu_i}-\Lambda_i$, $U=\R_i$, and $C=X_i(z)$, and using Eq.~\eqref{eq:Sigma_local-2} to rewrite
\begin{equation}
\big[\R_i^\dagger(z\mathbf{1}_{B\nu_i}-\Lambda_i)^{-1}\R_i\big]^{-1}=z\mathbf{1}_{\nu_i}-\eps_i-\Sigma_i(z)
\,, 
\end{equation}
we obtain
\begin{align}
G_i(z)^{-1}
&=
z\mathbf{1}_{\nu_i}-\eps_i-\Sigma_i(z)-X_i(z)
\,.
\label{eq:Gi_dyson_latt_with_X}
\end{align}

\item Comparing Eq.~\eqref{eq:Gi_dyson_latt_with_X} with the defining relation \eqref{eq:Gi_dyson_latt} shows that
$X_i(z)=\Delta_i^{\mathrm{latt}}(z)$.

\end{itemize}

In summary, the cavity expression for the
lattice Weiss field is:
\begin{equation}
\Delta^{\mathrm{latt}}_i(z)
=
\sum_{\substack{j,k=1\\ j\neq i,\,k\neq i}}^{\N}
t_{ij}\,\R_j^\dagger\,\bar{\mathcal{G}}^{\mathrm{qp}}_{i;jk}(z)\,\R_k\,t_{ki}
\,.
\label{eq:Delta_latt_def}
\end{equation}

At the DMFT fixed point assumed in Sec.~\ref{sec:dmft_to_gga_setup}, the local lattice Green's function block $G_i(z)$ satisfies the impurity Dyson equation \eqref{eq:dmft_impurity_dyson} with the same pair $\big(\Sigma_i(z),\Delta_i(z)\big)$.
Since $\Delta_i^{\mathrm{latt}}(z)$ is defined by Eq.~\eqref{eq:Gi_dyson_latt}, comparing \eqref{eq:Gi_dyson_latt} with \eqref{eq:dmft_impurity_dyson} yields:
\begin{align}
\Delta^{\mathrm{latt}}_i(z)
&=\Delta_i(z)
\label{eq:Delta_latt_equals_Delta}
\end{align}
and, in turn:
\begin{align}
\mathcal{G}^{\mathrm{qp}}_{i}(z)
&=
\Big[z\mathbf{1}_{B\nu_i}-\Lambda_i-\R_i\,\Delta^{\mathrm{latt}}_i(z)\,\R_i^\dagger\Big]^{-1}
\nonumber\\
&=
\Big[z\mathbf{1}_{B\nu_i}-\Lambda_i-\R_i\,\Delta_i(z)\,\R_i^\dagger\Big]^{-1}
\,.
\label{eq:Gqp_local_block}
\end{align}


\subsubsection*{Local ghost density matching}

We first prove Eq.~\eqref{e4}, i.e., the equality of the local ghost density matrices of $\hat{H}^{i}_{0,\mathrm{emb}}$ and $\hat{H}_{\mathrm{qp}}$.
From Sec.~\ref{sec:weiss_from_H0emb} the $(f,f)$ block of the resolvent of
$\hat{H}^{i}_{0,\mathrm{emb}}[\D_i,\Lambda_i^c;\R_i,\Lambda_i]$ is
\begin{align}
G_{f,i}(z)
&=
\Big[z\mathbf{1}_{B\nu_i}-\Lambda_i-\R_i\,\Delta_i(z)\,\R_i^\dagger\Big]^{-1}
\,.
\label{eq:Gf_0emb}
\end{align}
Using Eq.~\eqref{eq:Delta_latt_equals_Delta} together with Eq.~\eqref{eq:Gqp_local_block} we obtain
\begin{align}
G_{f,i}(z)=\mathcal{G}^{\mathrm{qp}}_{i}(z)
\,.
\label{eq:Gf_match}
\end{align}
Taking the equal-time limit of Eq.~\eqref{eq:Gf_match} yields
\begin{align}
\Av{\Phi^0_i}{\fc_{ia}\fa_{ib}}
&=
\Av{\Psi_0}{\fc_{ia}\fa_{ib}}
\,,
\end{align}
which is Eq.~\eqref{e4}.

\subsubsection*{Hybridization matching}

We next prove Eq.~\eqref{e5}, which expresses the matching of the mixed impurity--bath correlator with the corresponding lattice quantity entering the hybridization update.
Let $G_{bf,i}(z)$ denote the $(b,f)$ block of the resolvent $\mathcal{G}_i(z)$ in Eq.~\eqref{eq:G0_full}.
Using the block inverse formula Eq.~\eqref{eq:block_inverse_M}
(Appendix~\ref{sec:woodbury_projected_inverses}) with the same identifications as in
Sec.~\ref{sec:weiss_from_H0emb} gives
\begin{align}
G_{bf,i}(z)
&=
\big(z\mathbf{1}_{B\nu_i}+\Lambda_i^c\big)^{-1}\D_i^{*}\R_i^\dagger\,G_{f,i}(z)
\,,
\label{eq:Gbf_0emb}
\end{align}
hence
\begin{align}
\D_i^{T}G_{bf,i}(z)
&=
\Delta_i(z)\,\R_i^\dagger\,G_{f,i}(z)
\,.
\label{eq:DGbf_0emb}
\end{align}

On the lattice side, block inversion of $z\mathbf{1}-h^*[\R,\Lambda]$ in the same $(i)$ versus $(j\neq i)$ structure yields, for $j\neq i$,
\begin{align}
\mathcal{G}^{\mathrm{qp}}_{ji}(z)
&=
\sum_{\substack{k=1\\ k\neq i}}^{\N}
\bar{\mathcal{G}}^{\mathrm{qp}}_{i;jk}(z)\,\R_k\,t_{ki}\,\R_i^\dagger\,\mathcal{G}^{\mathrm{qp}}_{i}(z)
\,.
\label{eq:Gqp_offdiag}
\end{align}
Multiplying Eq.~\eqref{eq:Gqp_offdiag} by $t_{ij}\R_j^\dagger$ and summing over $j\neq i$ gives
\begin{align}
\sum_{\substack{j=1\\ j\neq i}}^{\N}
t_{ij}\,\R_j^\dagger\,\mathcal{G}^{\mathrm{qp}}_{ji}(z)
&=
\Delta^{\mathrm{latt}}_i(z)\,\R_i^\dagger\,\mathcal{G}^{\mathrm{qp}}_{i}(z)
\,.
\label{eq:Mqp_identity}
\end{align}
Using Eqs.~\eqref{eq:Delta_latt_equals_Delta} and \eqref{eq:Gf_match}, and comparing
Eqs.~\eqref{eq:DGbf_0emb} and \eqref{eq:Mqp_identity}, we obtain
\begin{align}
\D_i^{T}G_{bf,i}(z)
&=
\sum_{\substack{j=1\\ j\neq i}}^{\N}
t_{ij}\,\R_j^\dagger\,\mathcal{G}^{\mathrm{qp}}_{ji}(z)
\,.
\label{eq:e5_freq_identity}
\end{align}
Taking the equal-time limit of Eq.~\eqref{eq:e5_freq_identity} yields Eq.~\eqref{e5}, i.e.:
\begin{widetext}
\begin{align}
\Av{\Phi^0_i}{
\fc_{ia}\left(\sum_{b=1}^{B\nu_i}[\D_i]_{b\alpha}\bba_{ib}\right)}
&=
\sum_{j=1}^{\N}\sum_{\beta=1}^{\nu_j}
[t_{ij}]_{\alpha\beta}\,
\Av{\Psi_0}{\fc_{ia}\left(\sum_{b=1}^{B\nu_j}[\R_j^\dagger]_{\beta b}\fa_{jb}\right)}
\,.
\end{align}
\end{widetext}

In summary, the infinite-$B$ ghost-GA embedding construction reproduces the DMFT fixed point exactly.
The central mechanism is that the DMFT objects entering the self-consistency map (local propagators and the hybridization function) are represented through a sequence of finite-dimensional auxiliary Hamiltonians whose parameters are updated self-consistently.
In the $B\to\infty$ limit the auxiliary bath becomes complete, and the resulting update map coincides with the standard DMFT update, yielding the same local self-energy and hybridization function.

A conceptual insight emerging from the proof above is that, while at finite $B$ the hybridization function $\Delta_i(z)$ parametrizing the EH \eqref{eq:Hemb} via Eq.~\eqref{eq:hyb_def} is generally different from the lattice Weiss field $\Delta_i^{\mathrm{latt}}(z)$ defined by the cavity construction [Eq.~\eqref{eq:Delta_latt_def}] and implied by the lattice resolvent \eqref{eq:G_dyson}, they become equal in the limit $B\to\infty$.

Note that this finite-$B$ ghost-GA framework contrasts with how finite-bath approximations are introduced in standard DMFT implementations, which are based on discretizing the Weiss field, replacing the continuous hybridization function by a finite-bath representation.
In the present scheme, instead, $\Delta_i(z)$ enters only via the ghost-GA stationarity conditions that determine the self-energy, while
the local Green's function is computed from \eqref{eq:G_dyson} with $\Sigma_i(z)$ given by \eqref{eq:Sigma_local}.
Systematic comparisons~\cite{TH1,TH2} demonstrated that the ghost-GA variational optimization yields a faster convergence of ground-state observables as a function of bath size than standard finite-bath DMFT, even though, as emphasized earlier, only the ground state of the EH is required.

\subsection{Unified perspective: ghost-GA, ghost-DMET, and DMFT}
\label{sec:unified_perspective}

A key conceptual consequence of this section is that DMFT can be viewed as implementable through ground-state information of the interacting auxiliary problem.
At $T=0$, the quantities required to close the self-consistency loop are obtained from ground-state expectation values of the embedding Hamiltonian, rather than from explicitly computing frequency-dependent correlation functions or excited-state spectra.
In this sense, the proof provides a principled route to formulate DMFT as a ground-state-based embedding algorithm, while retaining full equivalence in the complete-bath limit.

The equivalence established in this work also clarifies the relationship between the present variational framework and density matrix embedding theory (DMET)~\cite{DMET,DMET-spectral}, specifically in light of the ghost-DMET formulation proposed in Ref.~\cite{gDMET}.
In fact, as shown in Ref.~\cite{gDMET}, the stationarity conditions of the ghost-GA energy functional can be derived from a QE perspective by constructing the auxiliary quasiparticle Hamiltonian $\hat{H}_{\mathrm{qp}}[\R,\Lambda]$ and requiring self-consistency between the impurity and the bath.
However, a crucial formal step in interpreting the ghost-GA equations as a DMET-like embedding relies on the condition that the embedding matrix $\R_i$ acts as an isometry from the physical space to the auxiliary space, i.e.,
$\R_i^\dagger \R_i \approx \mathbf{1}_{\nu_i}$.
In the finite-$B$ ghost-GA framework, this condition is generally satisfied only approximately (although often to high accuracy~\cite{Ghost-GA,ALM_g-GA,TH1,TH2}), and the deviation $\mathbf{1}_{\nu_i} - \R_i^\dagger \R_i$ represents the spectral weight ``missing'' from the finite-bath description.
The present work reveals that in the limit $B\to\infty$, the ghost-GA parameters must reproduce the exact high-frequency behavior of the local Green's function [Eq.~\eqref{eq:sumrule_RdagR}], which enforces the exact sum rule:
\begin{align}
\lim_{B\to\infty} \R_i^\dagger \R_i = \mathbf{1}_{\nu_i}
\,.
\label{eq:R_isometry_exact}
\end{align}

In summary, we showed that the $B\to\infty$ limit not only recovers the exact DMFT self-energy, but also rigorously enforces the isometry condition required to interpret the theory from a DMET perspective.

\section{Finite-$T$ ghost-GA functional and DMFT functional structure}
\label{sec:finiteT_functional_BK}

In this section we derive a finite-temperature grand-potential functional for the ghost-GA, and show that it admits a fully gauge-invariant reformulation in terms of the dynamical Weiss field and self-energy. In the infinite-bath limit, this dynamical functional coincides with the standard DMFT functional~\cite{LDA+U+DMFT}. As a consequence, the DMFT stationarity conditions can be enforced within the ghost-GA functional through stationarity with respect to static auxiliary parameters, providing a purely static route to the finite-temperature DMFT fixed point.

The section is structured as follows: In Sec.~\ref{subsec:finiteT_extension} we introduce the finite-temperature grand-potential functional in terms of auxiliary quasiparticle and embedding Hamiltonians and derive the corresponding stationarity conditions as thermal expectation-value matching relations. In Sec.~\ref{subsec:finiteT_tracelog} we rewrite the same functional in a gauge-invariant form in terms of the induced dynamical Weiss field and self-energy, making contact with DMFT functional formulations. In Sec.~\ref{subsec:finiteT_tracelog_proof} we provide the algebraic derivation connecting these two representations. 
Finally, in Sec.~\ref{sec:dmft_stationary_Binfty} we take the infinite-bath limit and identify the resulting dynamical functional with the DMFT functional~\cite{LDA+U+DMFT}, thereby establishing a static-parameter formulation of the DMFT stationarity conditions in terms of auxiliary problems.

\subsection{Finite-temperature extension}
\label{subsec:finiteT_extension}

Although the discussion and implementation above focus on $T=0$, the same construction admits a direct finite-temperature generalization.
To make this explicit, we define the following finite-$T$ grand-potential Lagrange function:
\begin{align}
\mathcal{L}_\beta[\R,\Lambda,\D,\Lambda^c]
&=
\Omega_{\rm qp}[\R,\Lambda]
+\sum_{i=1}^{\N}\Omega_{\rm emb}^i[\D_i,\Lambda_i^c]
\nonumber\\&
-\sum_{i=1}^{\N}\Omega_{0,{\rm emb}}^i[\D_i,\Lambda_i^c;\R_i,\Lambda_i]
\,,
\label{eq:Lbeta_def_req}
\end{align}
where $\beta= 1/T$ and each term is defined as:
\begin{align}
\Omega_{\rm qp}
&=
-\frac{1}{\beta}\ln\!\Big(\Tr e^{-\beta\,\hat H_{\rm qp}[\R,\Lambda]}\Big)
\,,
\label{Omega-qp}
\\
\Omega_{\rm emb}^i
&=
-\frac{1}{\beta}\ln\!\Big(\Tr e^{-\beta\,\hat H_{\rm emb}^i[\D_i,\Lambda_i^c]}\Big)
\,,
\label{Omega-Hemb}
\\
\Omega_{0,{\rm emb}}^i
&=
-\frac{1}{\beta}\ln\!\Big(\Tr e^{-\beta\,\hat H_{0,{\rm emb}}^i[\D_i,\Lambda_i^c;\R_i,\Lambda_i]}\Big)
\,.
\label{Omega-Hemb0}
\end{align}
Note that $\mathcal{L}_\beta$ reduces to the ghost-GA Lagrange function in Eq.~\eqref{eq:L_lagrange} in the limit $\beta\to\infty$.

Extremizing $\mathcal{L}_\beta$ yields the following thermal generalization of the stationarity conditions in Eqs.~\eqref{e4}--\eqref{e7}:
\begin{widetext}
\begin{align}
\left\langle \fc_{ia}\fa_{ib}\right\rangle_{\beta,{\rm qp}}
&=
\left\langle \fc_{ia}\fa_{ib}\right\rangle_{\beta,0{\rm emb},i}
\,,
\label{eq:e4_beta_explicit}
\\
\left\langle
\fc_{ia}\left(\sum_{b=1}^{B\nu_i}[\D_i]_{b\alpha}\bba_{ib}\right)
\right\rangle_{\beta,0{\rm emb},i}
&=
\left\langle
\fc_{ia}\left(\sum_{j=1}^{\N}\sum_{\beta=1}^{\nu_j}
[t_{ij}]_{\alpha\beta}\sum_{b=1}^{B\nu_j}[\R_j^\dagger]_{\beta b}\fa_{jb}\right)
\right\rangle_{\beta,{\rm qp}}
\,,
\label{eq:e5_beta_explicit}
\\
\left\langle \bba_{ib}\bbc_{ia}\right\rangle_{\beta,0{\rm emb},i}
&=
\left\langle \bba_{ib}\bbc_{ia}\right\rangle_{\beta,{\rm emb},i}
\,,
\label{eq:e6_beta_explicit}
\\
\left\langle
\left(\sum_{a=1}^{B\nu_i}[\R_i]_{a\alpha}\fc_{ia}\right)\bba_{ib}
\right\rangle_{\beta,0{\rm emb},i}
&=
\left\langle \cc_{i\alpha}\bba_{ib}\right\rangle_{\beta,{\rm emb},i}
\,,
\label{eq:e7_beta_explicit}
\end{align}
\end{widetext}
where the thermal averages in each auxiliary problem are defined as follows:
\begin{align}
\langle \hat O\rangle_{\beta,{\rm qp}}
&=
\frac{\Tr\!\left(e^{-\beta \hat H_{\rm qp}}\hat O\right)}{\Tr\!\left(e^{-\beta \hat H_{\rm qp}}\right)}
\,,
\label{eq:thermal_averages_explicit1}
\\
\langle \hat O\rangle_{\beta,{\rm emb},i}
&=
\frac{\Tr\!\left(e^{-\beta \hat H_{\rm emb}^i}\hat O\right)}{\Tr\!\left(e^{-\beta \hat H_{\rm emb}^i}\right)}
\,,
\label{eq:thermal_averages_explicit2}
\\
\langle \hat O\rangle_{\beta,0{\rm emb},i}
&=
\frac{\Tr\!\left(e^{-\beta \hat H_{0,{\rm emb}}^i}\hat O\right)}{\Tr\!\left(e^{-\beta \hat H_{0,{\rm emb}}^i}\right)}
\,.
\label{eq:thermal_averages_explicit3}
\end{align}

Remarkably, the proof of Sec.~\ref{sec:equivalence_gga_dmft_Binfty} relies only on algebraic resolvent identities and on the self-consistency structure of the auxiliary construction, and therefore carries over to finite temperature upon working on the Matsubara axis ($z=i\omega_n$) and replacing ground-state expectation values with thermal averages [Eqs.~\eqref{eq:thermal_averages_explicit1}--\eqref{eq:thermal_averages_explicit3}].
Consequently, the stationary point of $\mathcal{L}_\beta$ converges to the finite-temperature DMFT fixed point in the limit $B\to\infty$ for any fixed $\beta$, 
and not only in the  limit $\beta\to\infty$, where $\mathcal{L}_\beta$ reduces to the ghost-GA Lagrange function in Eq.~\eqref{eq:L_lagrange} and Eqs.~\eqref{eq:e4_beta_explicit}--\eqref{eq:e7_beta_explicit}
reduce to Eqs.~\eqref{e4}--\eqref{e7}.

In the following subsection we show numerically that this convergence is not only an asymptotic statement about the strict $B\to\infty$ limit: for the cases considered, the finite-temperature ghost-GA results at $B=3$ are already very close to DMFT.

\subsection{Gauge-invariant dynamical form of the finite-$T$ functional}
\label{subsec:finiteT_tracelog}

In this subsection we rewrite Eq.~\eqref{eq:Lbeta_def_req} in a gauge-invariant form in terms of the induced self-energy and hybridization functions, thereby making contact with the DMFT functional~\cite{LDA+U+DMFT}.

We express the finite-$T$ functional Eq.~\eqref{eq:Lbeta_def_req} in terms of the local self-energy and hybridization functions induced by the following ghost-GA parametrizations:
\begin{align}
\Sigma_i(\io)
&=
\io\mathbf 1_{\nu_i}
-\Big[\R_i^\dagger(\io\mathbf 1_{B\nu_i}-\Lambda_i)^{-1}\R_i\Big]^{-1}
-\eps_i,
\label{eq:Sigmai_req}
\\
\Delta_i(\io)
&=
\D_i^{T}\big(\io\mathbf 1_{B\nu_i}+\Lambda_i^c\big)^{-1}\D_i^{*}
\,.
\label{eq:Deltai_req}
\end{align}

As we are going to show, Eq.~\eqref{eq:Lbeta_def_req} can be written in the following equivalent form:
\begin{align}
&\mathcal{L}_\beta[\{\Sigma_i(i\omega_n)\},\{\Delta_i(i\omega_n)\}]
=
\nonumber\\
&\quad
-\,T\sum_{n}e^{i\omega_n0^+}\Tr
\ln\!\big(\io\mathbf 1-h_0-\Sigma(\io)\big)
\nonumber\\
&\quad
+T\sum_{i=1}^{\N}\sum_{n}e^{i\omega_n0^+}\Tr
\ln\!\big(\io\mathbf 1 \!-\!  \eps_i \!-\! \Delta_i(\io) \!-\!  \Sigma_i(\io)\big)
\nonumber\\
&\quad
+\sum_{i=1}^{\N}\Omega_{\rm imp}^i[\Delta_i(\io)]
\,,
\label{eq:Lbeta_simplified_Omega}
\end{align}
where $\Sigma_i(\io)$ and $\Delta_i(\io)$ are given by Eqs.~\eqref{eq:Sigmai_req} and \eqref{eq:Deltai_req},
\begin{align}
\Omega_{\rm imp}^i[\Delta_i(\io)]
&=
\Omega_{\rm emb}^i[\D_i,\Lambda_i^c]-\Omega_{\rm bath}^i[\Lambda_i^c]
\,,
\label{eq:Omega_imp_def}
\\
\Omega_{\rm bath}^i[\Lambda_i^c]
&=
\Tr(\Lambda_i^c)
\label{eq:Omega_bath_def}
\\&
-\,T\sum_{n}e^{i\omega_n0^+}\Tr_{B\nu_i}\ln\!\big(\io\mathbf 1+\Lambda_i^c\big)
\nonumber
\,,
\end{align}
and $\Omega_{{\rm emb}}^i$
is defined in Eq.~\eqref{Omega-Hemb},
so that $\Omega_{\rm imp}^i$ depends on $(\D_i,\Lambda_i^c)$ only through the hybridization function
$\Delta_i(\io)$ in Eq.~\eqref{eq:Deltai_req}.

Equation~\eqref{eq:Lbeta_simplified_Omega} is directly comparable to the DMFT functional~\cite{LDA+U+DMFT} written in terms of the dynamical objects $\Sigma(\io)$ and $\Delta_i(\io)$.
The key distinction is that here $\Sigma(\io)$ and $\Delta_i(\io)$ are not treated as arbitrary dynamical variables, but are restricted to the finite-dimensional parametrizations in Eqs.~\eqref{eq:Sigmai_req}--\eqref{eq:Deltai_req}, in terms of the ghost-GA variational parameters $(\R,\Lambda,\D,\Lambda^c)$.

A central practical feature of this specific parametrization is that extremizing $\mathcal{L}_\beta$ with respect to the underlying matrices $(\R,\Lambda,\D,\Lambda^c)$ yields self-consistency equations that can be enforced entirely through \emph{static} auxiliary problems (thermal expectation values in $\hat H_{\rm qp}$, $\hat H_{\rm emb}^i$, and $\hat H_{0,{\rm emb}}^i$), rather than by treating $\Sigma(\io)$ and $\Delta_i(\io)$ as unconstrained dynamical variables.
This finite-$T$ functional viewpoint is not tied to the $T=0$ ghost-GA formulation, although it can be motivated from that perspective in the limit $\beta\to\infty$.

\subsection{Derivation of the gauge-invariant dynamical form of the finite-$T$ functional}
\label{subsec:finiteT_tracelog_proof}

In this subsection we prove that the finite-$T$ grand-potential Lagrange function in Eq.~\eqref{eq:Lbeta_def_req}
is equivalent to the gauge-invariant form in Eq.~\eqref{eq:Lbeta_simplified_Omega}.
Starting from Eq.~\eqref{eq:Lbeta_def_req} together with the definitions
Eqs.~\eqref{Omega-qp}--\eqref{Omega-Hemb0}, we rewrite the three contributions
$\Omega_{\rm qp}$, $\Omega_{\rm emb}^i$, and $\Omega_{0,{\rm emb}}^i$ one by one.

Throughout the derivation we use the following elementary identities:
\begin{align}
\Tr\ln M &= \ln\det M,
\label{eq:Trln_lndet}
\\
\det\!\big(A-U C U^\dagger\big)
&=
\det(A)\,
\det\!\big(\mathbf 1-C\,U^\dagger A^{-1}U\big),
\label{eq:det_lemma}
\\
\det
\begin{pmatrix}
A & B\\
C & D
\end{pmatrix}
&=
\det(D)\,
\det\!\big(A-BD^{-1}C\big),
\label{eq:Schur_det}
\end{align}
valid whenever the indicated inverses exist.

\subsubsection{Quasiparticle contribution}

Since $\hat H_{\rm qp}$ is quadratic, Eq.~\eqref{Omega-qp} can be written in the Matsubara trace--log form
\begin{align}
\Omega_{\rm qp}
&=
-\,T\sum_{n}e^{i\omega_n0^+}\Tr
\ln\!\big(\io\mathbf 1-h^*[\R,\Lambda]\big)
\,,
\label{eq:Omega_qp_trlog}
\end{align}
where $h^*[\R,\Lambda]=\R\,t\,\R^\dagger+\Lambda$ [Eq.~\eqref{eq:hstar_compact}].
Using Eq.~\eqref{eq:det_lemma} with $A=\io\mathbf 1-\Lambda$, $U=\R$, and $C=t$, we obtain
\begin{align}
&\Tr\ln\!\big(\io\mathbf 1-h^*[\R,\Lambda]\big)
=
\Tr\ln(\io\mathbf 1-\Lambda)
\nonumber\\&\qquad\qquad
+\Tr\ln\!\Big(\mathbf 1-t\,\R^\dagger(\io\mathbf 1-\Lambda)^{-1}\R\Big)
\,.
\label{eq:Trln_qp_step1}
\end{align}
From the definition \eqref{eq:Sigmai_req} (and $h_0=\eps+t$, Eq.~\eqref{eq:eps_def}) one has
\begin{align}
\R^\dagger(\io\mathbf 1-\Lambda)^{-1}\R
&=
\big[\io\mathbf 1-\eps-\Sigma(\io)\big]^{-1}
\,,
\label{eq:RLR_qp}
\end{align}
hence
\begin{align}
&\Tr\ln\!\Big(\mathbf 1-t\,\R^\dagger(\io\mathbf 1-\Lambda)^{-1}\R\Big)
=
\nonumber\\&\qquad
\Tr\ln\!\big(\io\mathbf 1-h_0-\Sigma(\io)\big)
\nonumber\\&\qquad
-\Tr\ln\!\big(\io\mathbf 1-\eps-\Sigma(\io)\big)
\,.
\label{eq:Trln_qp_step2}
\end{align}
Substituting Eq.~\eqref{eq:Trln_qp_step2} into Eq.~\eqref{eq:Trln_qp_step1} yields
\begin{align}
\Tr\ln\!\big(\io\mathbf 1-h^*[\R,\Lambda]\big)
&=
\Tr\ln\!\big(\io\mathbf 1-h_0-\Sigma(\io)\big)
\nonumber\\&
+\Tr\ln(\io\mathbf 1-\Lambda)
\nonumber\\&
-\Tr\ln\!\big(\io\mathbf 1-\eps-\Sigma(\io)\big)
\,.
\label{eq:Trln_qp_FT}
\end{align}
Inserting Eq.~\eqref{eq:Trln_qp_FT} into Eq.~\eqref{eq:Omega_qp_trlog} yields the corresponding expression for $\Omega_{\rm qp}$.

\subsubsection{Quadratic embedding contribution}

Since $\hat H_{0,{\rm emb}}^i$ is quadratic, Eq.~\eqref{Omega-Hemb0} can be written as
\begin{align}
\Omega_{0,{\rm emb}}^i
&=
\Tr(\Lambda_i^c)
-\,T\sum_{n}e^{i\omega_n0^+}\Tr
\ln\!\big(\io\mathbf 1-h_i^{0,{\rm emb}}\big)
\,,
\label{eq:Omega_0emb_trlog}
\end{align}
where the one-body matrix associated with $\hat H_{0,{\rm emb}}^i$ [Eq.~\eqref{eq:H0emb}] is
\begin{align}
h_i^{0,{\rm emb}}
&=
\begin{pmatrix}
\Lambda_i & W_i\\
W_i^\dagger & -\Lambda_i^c
\end{pmatrix},
\qquad
W_i=\R_i\D_i^{T}
\,.
\label{eq:h0emb_onebody_FT}
\end{align}
Using Eq.~\eqref{eq:Schur_det} with $D=\io\mathbf 1+\Lambda_i^c$ yields
\begin{align}
&\Tr\ln\!\big(\io\mathbf 1-h_i^{0,{\rm emb}}\big)
=
\Tr\ln(\io\mathbf 1+\Lambda_i^c)
\nonumber\\&\quad
+\Tr\ln\!\Big(\io\mathbf 1-\Lambda_i-W_i(\io\mathbf 1+\Lambda_i^c)^{-1}W_i^\dagger\Big)
\,.
\label{eq:Trln_0emb_step1}
\end{align}
Using $W_i=\R_i\D_i^T$ and the definition \eqref{eq:Deltai_req} gives
$W_i(\io\mathbf 1+\Lambda_i^c)^{-1}W_i^\dagger=\R_i\,\Delta_i(\io)\,\R_i^\dagger$.
Applying Eq.~\eqref{eq:det_lemma} to the second term in Eq.~\eqref{eq:Trln_0emb_step1} with
$A=\io\mathbf 1-\Lambda_i$, $U=\R_i$, and $C=\Delta_i(\io)$, and using again
Eq.~\eqref{eq:Sigmai_req} to write
\begin{align}
\R_i^\dagger(\io\mathbf 1-\Lambda_i)^{-1}\R_i
&=
\big[\io\mathbf 1-\eps_i-\Sigma_i(\io)\big]^{-1},
\label{eq:RLR_0emb}
\end{align}
one obtains
\begin{align}
&\Tr\ln\!\big(\io\mathbf 1-h_i^{0,{\rm emb}}\big)
=
\Tr\ln(\io\mathbf 1+\Lambda_i^c)
\nonumber\\&\qquad
+\Tr\ln(\io\mathbf 1-\Lambda_i)
\nonumber\\&\qquad
+\Tr\ln\!\big(\io\mathbf 1-\eps_i-\Sigma_i(\io)-\Delta_i(\io)\big)
\nonumber\\&\qquad
-\Tr\ln\!\big(\io\mathbf 1-\eps_i-\Sigma_i(\io)\big)
\,,
\label{eq:Trln_0emb_FT}
\end{align}
which rewrites $\Omega_{0,{\rm emb}}^i$ through Eq.~\eqref{eq:Omega_0emb_trlog}.

\subsubsection{Collecting terms}

Substituting Eqs.~\eqref{eq:Trln_qp_FT} and \eqref{eq:Trln_0emb_FT} into Eq.~\eqref{eq:Lbeta_def_req}
(using Eqs.~\eqref{Omega-qp} and \eqref{Omega-Hemb0}), and using
\begin{align}
&\Tr\ln(\io\mathbf 1-\Lambda)
=
\sum_{i=1}^{\N}\Tr\ln(\io\mathbf 1-\Lambda_i),
\\
&\Tr\ln\!\big(\io\mathbf 1-\eps-\Sigma(\io)\big)
\\&\qquad\qquad
=
\sum_{i=1}^{\N}\Tr\ln\!\big(\io\mathbf 1-\eps_i-\Sigma_i(\io)\big)
\,,
\nonumber
\end{align}
while the additive constants $\Tr(\Lambda_i^c)$ are exactly canceled by the corresponding terms in $\Omega_{\rm bath}^i$ [Eq.~\eqref{eq:Omega_bath_def}], one finds
\begin{align}
\mathcal{L}_\beta
&=
-T\sum_{n}e^{i\omega_n0^+}\Tr
\ln\!\big(\io\mathbf 1-h_0-\Sigma(\io)\big)
\nonumber\\&
+T\!\sum_{i=1}^{\N}\!\sum_{n}\!e^{i\omega_n0^+}\!
\Tr\ln\!\big(\io\mathbf 1 \!-\! \eps_i \!-\! \Sigma_i(\io) \!-\!\Delta_i(\io)\big)
\nonumber\\&
+\sum_{i=1}^{\N}\Omega_{\rm emb}^i
-\sum_{i=1}^{\N}\Omega_{\rm bath}^i
\,,
\label{eq:Lbeta_intermediate_FT}
\end{align}
where $\Omega_{\rm bath}^i$ is given by Eq.~\eqref{eq:Omega_bath_def}.
Finally, using the definition \eqref{eq:Omega_imp_def} of $\Omega_{\rm imp}^i$ yields Eq.~\eqref{eq:Lbeta_simplified_Omega}.


\subsection{Functional viewpoint on the ghost-GA--DMFT correspondence}
\label{sec:dmft_stationary_Binfty}

Let us consider the functional $\mathcal{L}_\beta$, see Eq.~\eqref{eq:Lbeta_simplified_Omega} in the infinite-bath limit, considering it as an unrestricted functional of the dynamical variables
$\Sigma_i(i\omega_n)$ and $\Delta_i(i\omega_n)$.
\begin{align}
&\mathcal{L}_\beta[\{\Sigma_i(i\omega_n)\},\{\Delta_i(i\omega_n)\}]
=
\nonumber\\
&\quad
-\,T\sum_{n}e^{i\omega_n0^+}\Tr
\ln\!\big(\io\mathbf 1-h_0-\Sigma(\io)\big)
\nonumber\\
&\quad
+T\sum_{i=1}^{\N}\sum_{n}e^{i\omega_n0^+}\Tr
\ln\!\big(\io\mathbf 1 \!-\!  \eps_i \!-\! \Delta_i(\io) \!-\!  \Sigma_i(\io)\big)
\nonumber\\
&\quad
+\sum_{i=1}^{\N}\Omega_{\rm imp}^i[\Delta_i(\io)]
\,.
\label{eq:Lbeta_simplified_Omega-unrestricted}
\end{align}

We note that Eq.~\eqref{eq:Lbeta_simplified_Omega-unrestricted} becomes equivalent to the following DMFT functional, previously introduced in Ref.~\cite{LDA+U+DMFT}:
\begin{align}
&\Gamma_\beta[\{G_{{\rm loc},i}(i\omega_n)\},\{\Sigma_i(i\omega_n)\},\{\Delta_i(i\omega_n)\}]
\nonumber\\
&\quad =\sum_{i=1}^{\N}\Omega_{\rm imp}^i[\Delta_i(i\omega_n)]
\nonumber\\
&\quad
-\,T\sum_{n}e^{i\omega_n0^+}\Tr
\ln\!\big(i\omega_n\mathbf 1-h_0-\Sigma(i\omega_n)\big)
\nonumber\\
&\quad
-\,T\sum_{i=1}^{\N}\sum_{n}e^{i\omega_n0^+}\Tr
\ln\!\big(G_{{\rm loc},i}(i\omega_n)\big)
\nonumber\\
&\quad
+\,T\sum_{i=1}^{\N}\sum_{n}e^{i\omega_n0^+}
\Tr
\Big[
\big(i\omega_n\mathbf 1-\eps_i-\Delta_i(i\omega_n)
\nonumber\\
&\qquad\quad
-\Sigma_i(i\omega_n)-G_{{\rm loc},i}(i\omega_n)^{-1}\big)\,
G_{{\rm loc},i}(i\omega_n)
\Big]
\,,
\label{eq:Kbeta_threevar}
\end{align}
where the local Green's function variable $G_{{\rm loc},i}(i\omega_n)$ is considered as an independent variable.
In fact, stationarity of $\Gamma_\beta$ with respect to $G_{{\rm loc},i}(i\omega_n)$ yields
\begin{equation}
G_{{\rm loc},i}(i\omega_n)^{-1}
=
i\omega_n\mathbf 1-\eps_i-\Delta_i(i\omega_n)-\Sigma_i(i\omega_n)
\,,
\label{eq:G_loc_stationary}
\end{equation}
and substituting Eq.~\eqref{eq:G_loc_stationary} into Eq.~\eqref{eq:Kbeta_threevar} eliminates $G_{\rm loc}$ and gives
Eq.~\eqref{eq:Lbeta_simplified_Omega-unrestricted} identically.

This observation provides us with an alternative route to the connection between ghost-GA and DMFT.
In fact, from this perspective, the ghost-GA can be recovered restricting Eq.~\eqref{eq:Lbeta_simplified_Omega-unrestricted} to the manifold generated by the static matrices $(\R_i,\Lambda_i,\D_i,\Lambda_i^c)$ through the parametrizations in Eqs.~\eqref{eq:Sigmai_req}--\eqref{eq:Deltai_req}, and observing that taking variations within this restricted manifold recovers Eq.~\eqref{eq:Lbeta_simplified_Omega}.
It is this specific parametrization of $\Sigma(i\omega_n)$ and $\Delta_i(i\omega_n)$ that makes it possible to enforce the restricted stationarity conditions through purely static density-matrix matching in terms of the auxiliary problems $\hat H_{\rm qp}$, $\hat H_{\rm emb}^i$, and $\hat H_{0,{\rm emb}}^i$, and to connect the present gauge-invariant formulation to the variational ghost-GA viewpoint.

Note that the restricted-manifold formulation used here is not the only possible.
In particular, Eq.~\eqref{eq:Sigmai_req} allows variations for which $\R_i^\dagger\R_i\neq\mathbf{1}_{\nu_i}$, i.e., it admits a contribution to $\Sigma_i(i\omega_n)$ proportional to $i\omega_n$ and, in turn, deviations from the spectral sum rule (see Sec.~\ref{sec:Asymptotic-Sigma}).
One could have alternatively restricted the variational space by imposing $\R_i^\dagger\R_i=\mathbf{1}_{\nu_i}$ from the outset, thereby excluding such a term.
However, such a restriction would not be compatible with our ghost-GA variational construction and its formulation in terms of static auxiliary-Hamiltonian stationarity conditions.

The derivation in Sec.~\ref{sec:equivalence_gga_dmft_Binfty} complements the present functional argument by deriving the $B\to\infty$ correspondence in a fully constructive way and by identifying explicitly the key limiting identities that characterize the large-$B$ limit, namely Eqs.~\eqref{eq:isometry_from-fit} and \eqref{eq:Delta_latt_equals_Delta}. These relations underlie the DMET interpretation and motivate the finite-$B$ restart/extrapolation protocol discussed in Sec.~\ref{sec:B_extrapolation}.

It is also worth noting that the finite-temperature functional derived here is tied to the complete-bath limit \(B\to\infty\) underlying the ghost-GA/ghost-DMET/DMFT correspondence, and explicitly involves the auxiliary quadratic embedding Hamiltonian \(\hat H^{i}_{0,\mathrm{emb}}\) and the corresponding contribution \(\Omega^{i}_{0,\mathrm{emb}}\). It therefore does not reduce, even at \(B=1\), to the finite-temperature RISB functional previously proposed in Ref.~\cite{dmet-risb-1}.


\section{Numerical validation and benchmarks}
\label{sec:numerics}

Throughout this section we consider the single-band Hubbard model on the Bethe lattice,
\begin{equation}
\hat H
=
-
t\sum_{\langle i,j\rangle}\sum_{\sigma=\uparrow,\downarrow}
\left(c_{i\sigma}^\dagger c_{j\sigma}+\mathrm{H.c.}\right)
+
U\sum_i \hat n_{i\uparrow}\hat n_{i\downarrow}
\,,
\end{equation}
where
\begin{equation}
\hat n_{i\sigma}=c_{i\sigma}^\dagger c_{i\sigma},
\qquad
\hat n_i=\hat n_{i\uparrow}+\hat n_{i\downarrow}
\,.
\end{equation}
The corresponding noninteracting density of states is semicircular,
\begin{equation}
\rho(\omega)=\frac{2}{\pi D^2}\sqrt{D^2-\omega^2}
\,,
\label{eq:semicircular}
\end{equation}
with $D$ the half-bandwidth. Throughout this section we set $D=1$.

In Sec.~\ref{sec:B_extrapolation}
we derive a few important identities implied by the constructive proof of the ghost-GA--DMFT correspondence, and verify them numerically.
In Sec.~\ref{Sec:Benchmark-temperature} we present finite-temperature benchmark calculations, showing that the finite-$T$ extension of the ghost-GA framework reproduces the DMFT thermodynamics with high accuracy already at $B=3$.


\subsection{Limiting identities implied by the ghost-GA--DMFT correspondence}
\label{sec:B_extrapolation}

In Sec.~\ref{sec:equivalence_gga_dmft_Binfty}, Eq~\eqref{eq:Delta_latt_equals_Delta}, we showed that, in the limit $B\to\infty$, the lattice Weiss field
$\Delta_i^{\mathrm{latt}}(z)$ defined by the cavity construction in Eq.~\eqref{eq:Delta_latt_def}
coincides with the DMFT Weiss field $\Delta_i(z)$.
The fact that
\begin{equation}
    \lim_{B\rightarrow\infty}
    \Delta_i^{\mathrm{latt}}(z)
    =\Delta_i(z)
\end{equation}
sheds light on recent benchmarks showing that small-$B$ ghost-GA solutions act as efficient preconditioners for DMFT~\cite{Makaresz-2026}.
The constructive proof of the ghost-GA--DMFT correspondence of Sec.~\ref{sec:equivalence_gga_dmft_Binfty} implies further identities that must become exact as $B\to\infty$:
\begin{itemize}
\item From Eq.~\eqref{eq:G_ctilde_b_inverse} it follows that, in the complete-bath limit, the $(\tilde c_i,b_i)$ and $(c_i,b_i)$ Green's function matrices coincide. Therefore the full one-body reduced density matrix of the impurity--bath subsystem must coincide. Besides Eqs.~\eqref{e6} and \eqref{e7}, this also gives
\begin{align}
\langle \Phi_i^0|\,\tilde c_{i\alpha}^\dagger \tilde c_{i\beta}\,|\Phi_i^0\rangle
&=
\langle \Phi_i|\,\cc_{i\alpha}\ca_{i\beta}\,|\Phi_i\rangle
\,,
\label{eq:ctilde_density_match}
\end{align}
where $\tilde c_{i\alpha}$ is defined in Eq.~\eqref{eq:setup_ctilde_def}. Equivalently,
\begin{align}
\sum_{a,b=1}^{B\nu_i}
[\R_i]_{a\alpha}\,
\langle \Phi_i^0|\,\fc_{ia}\fa_{ib}\,|\Phi_i^0\rangle\,
[\R_i^\dagger]_{\beta b}
&=
\langle \Phi_i|\,\cc_{i\alpha}\ca_{i\beta}\,|\Phi_i\rangle
\,.
\label{eq:physical_density_match}
\end{align}

\item Extending Eq.~\eqref{eq:Sigma_linear_term} of Sec.~\ref{sec:Asymptotic-Sigma} by one order, we write
\begin{align}
\Sigma_i(z)
&=
z\,\Sigma_i^{\mathrm{lin}}
+
\Sigma_i^{(0)}
+
\frac{\Sigma_i^{(1)}}{z}
+
\mathcal O(z^{-2})
\,,
\label{eq:Sigma_tail_compact}
\end{align}
with
\begin{align}
\Sigma_i^{\mathrm{lin}}
&=
\mathbf{1}_{\nu_i}
-
(\R_i^\dagger\R_i)^{-1},
\\
\Sigma_i^{(0)}
&=
(\R_i^\dagger\R_i)^{-1}
(\R_i^\dagger\Lambda_i\R_i)
(\R_i^\dagger\R_i)^{-1}
-
\eps_i\,,
\\
\Sigma_i^{(1)}
&=
(\R_i^\dagger\R_i)^{-1}
\Big[
\R_i^\dagger\Lambda_i^2\R_i
\\&\quad
-
(\R_i^\dagger\Lambda_i\R_i)
(\R_i^\dagger\R_i)^{-1}
(\R_i^\dagger\Lambda_i\R_i)
\Big]
(\R_i^\dagger\R_i)^{-1}
\,.\nonumber
\label{eq:Sigma_tail_coeffs}
\end{align}
Since the DMFT self-energy has no linear term, the $B\to\infty$ limit implies:
\begin{align}
\Sigma_i^{\mathrm{lin}} &\longrightarrow 0
\,,
\\
\Sigma_i^{(0)} &\longrightarrow \Sigma_{i,\mathrm{emb}}^{(0)}
\,,
\\
\Sigma_i^{(1)} &\longrightarrow \Sigma_{i,\mathrm{emb}}^{(1)}
\,.
\label{eq:Sigma_tail_matching}
\end{align}
The embedding coefficients are obtained directly from equal-time expectation values of $\hat{H}^{i}_{\mathrm{emb}}$ through the standard identities~\cite{Tail-1,Tail-2,Tail-3, Tail-DMFT-constraints}:
\begin{align}
\big[\Sigma_{i,\mathrm{emb}}^{(0)}\big]_{\alpha\beta}
&=
\bra{\Phi_i}\Big\{\,[c_{i\alpha},\hat{H}_{\mathrm{int}}^{i}],\,c_{i\beta}^\dagger\Big\}\ket{\Phi_i}
\,,
\label{eq:Sigma_0_emb}
\\
\big[\Sigma_{i,\mathrm{emb}}^{(1)}\big]_{\alpha\beta}
&=
\bra{\Phi_i}\Big\{\,[c_{i\alpha},\hat{H}_{\mathrm{int}}^{i}],\,[\hat{H}_{\mathrm{int}}^{i},c_{i\beta}^\dagger]\Big\}\ket{\Phi_i}
\nonumber\\
&\quad
-
\sum_{\gamma=1}^{\nu_i}
\big[\Sigma_{i,\mathrm{emb}}^{(0)}\big]_{\alpha\gamma}
\big[\Sigma_{i,\mathrm{emb}}^{(0)}\big]_{\gamma\beta}
\,.
\label{eq:Sigma_1_emb}
\end{align}

\end{itemize}


\begin{figure}[t]
\centering
\includegraphics[width=\columnwidth]{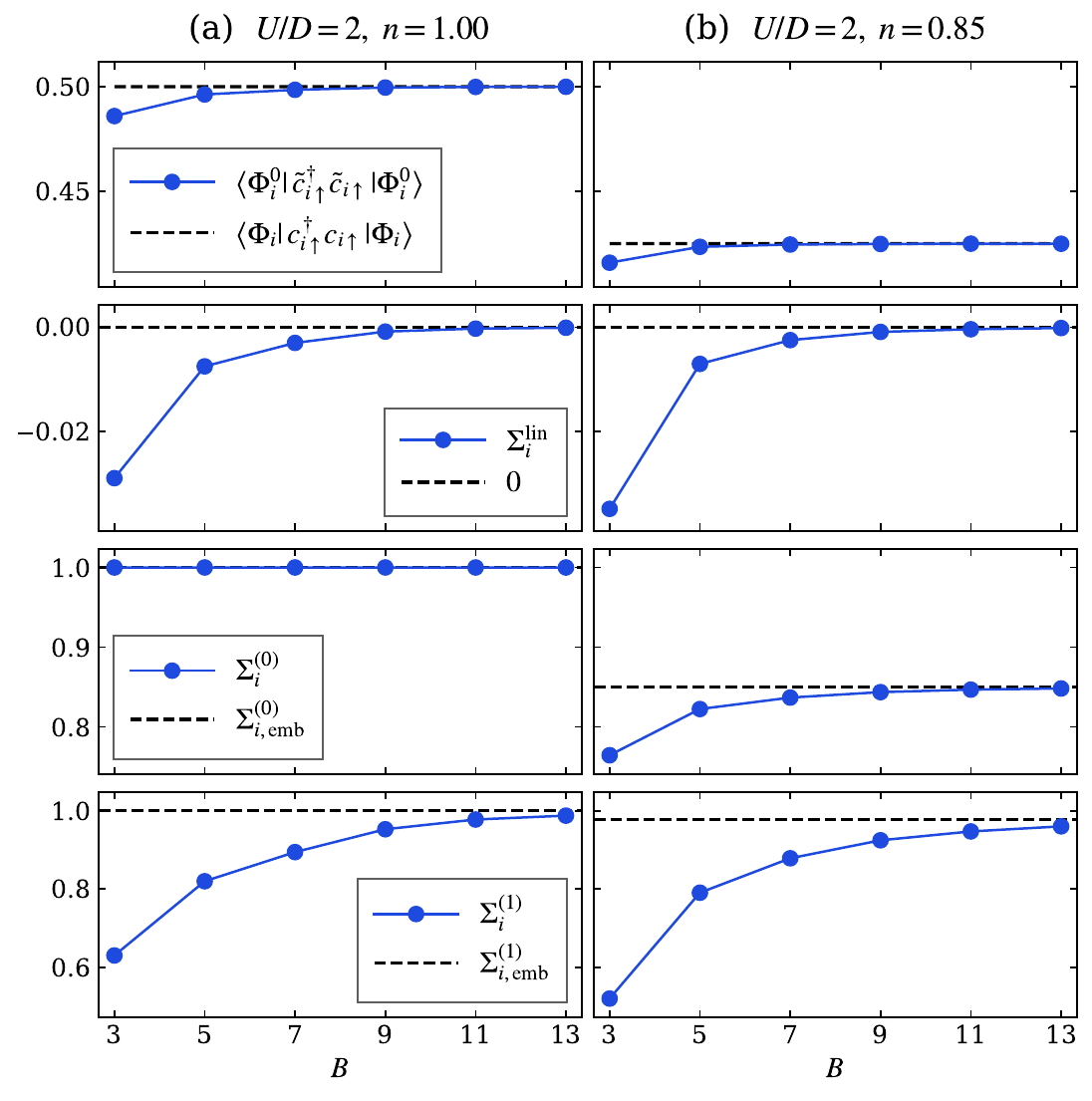}
\caption{
Zero-temperature large-$B$ behavior of the quantities entering Eqs.~\eqref{eq:physical_density_match} and \eqref{eq:Sigma_tail_matching} for the paramagnetic single-band Hubbard model on the Bethe lattice at $U/D=2$. The left and right columns correspond to $n=1.00$ and $n=0.85$, respectively.
}
\label{Figure6}
\end{figure}

Figure~\ref{Figure6} numerically verifies the limiting identities derived above for the paramagnetic single-band Hubbard model on the Bethe lattice at $T=0$ and $U/D=2$, for $n=1.00$ and $n=0.85$. For each bath size $B$, the interacting embedding Hamiltonian was solved with DMRG. The figure reports the quantities entering Eqs.~\eqref{eq:physical_density_match} and \eqref{eq:Sigma_tail_matching}; the black dashed lines denote the corresponding reference quantities indicated in the legend.

For the present single-band interaction
\begin{equation}
\hat{H}_{\mathrm{int}}^{i}=U\,\hat{n}_{i\uparrow}\hat{n}_{i\downarrow}
\end{equation}
one has
\begin{equation}
[c_{i\sigma},\hat{H}_{\mathrm{int}}^{i}]
=
U\,\hat{n}_{i,-{\sigma}}\,c_{i\sigma}
\,,
\end{equation}
so that, for a homogeneous paramagnetic solution with density
\begin{equation}
n=\langle \hat n_i\rangle
=
\langle \hat n_{i\uparrow}+\hat n_{i\downarrow}\rangle
\,,
\end{equation}
the commutator expressions in Eqs.~\eqref{eq:Sigma_0_emb} and \eqref{eq:Sigma_1_emb} reduce to
\begin{equation}
\Sigma_{i,\mathrm{emb}}^{(0)}
=
\big\langle \{[c_{i\sigma},\hat{H}_{\mathrm{int}}^{i}],c_{i\sigma}^{\dagger}\}\big\rangle
=
U\langle \hat{n}_{i,-{\sigma}}\rangle
=
\frac{U\,n}{2}
\,,
\end{equation}
and
\begin{align}
\Sigma_{i,\mathrm{emb}}^{(1)}
&=
\big\langle \{[c_{i\sigma},\hat{H}_{\mathrm{int}}^{i}],[\hat{H}_{\mathrm{int}}^{i},c_{i\sigma}^{\dagger}]\}\big\rangle
-
\left(\Sigma_{i,\mathrm{emb}}^{(0)}\right)^2
\nonumber\\
&=
U^2\langle \hat{n}_{i,-{\sigma}}\rangle
-
\left(\frac{U\,n}{2}\right)^2
=
U^2\,\frac{n}{2}\left(1-\frac{n}{2}\right)
\,.
\end{align}
Similarly, Eq.~\eqref{eq:physical_density_match} implies that, in the same limit,
\begin{equation}
\langle \Phi_i^0|\,\tilde c_{i\sigma}^{\dagger}\tilde c_{i\sigma}\,|\Phi_i^0\rangle
=
\langle \Phi_i|\,c_{i\sigma}^{\dagger}c_{i\sigma}\,|\Phi_i\rangle
=
\frac{n}{2}
\,.
\end{equation}

The figure shows the expected systematic convergence with increasing bath size. In particular, the occupation matching displayed in the upper row becomes increasingly accurate, the unphysical linear term $\Sigma_i^{\mathrm{lin}}$ is progressively suppressed and tends to zero, and the zeroth- and first-order coefficients approach the corresponding embedding values $\Sigma_{i,\mathrm{emb}}^{(0)}$ and $\Sigma_{i,\mathrm{emb}}^{(1)}$, in agreement with Eqs.~\eqref{eq:physical_density_match} and \eqref{eq:Sigma_tail_matching}.

Taken together, these results provide direct numerical verification of the large-$B$ identities derived above from Sec.~\ref{sec:equivalence_gga_dmft_Binfty}.
More importantly, they indicate that finite-$B$ ghost-GA solutions already encode nontrivial spectral information. Indeed, the self-energy tails and the cavity Weiss field are constrained by equal-time identities, and therefore by quantities that, within a variational framework, are expected to converge accurately already at relatively small $B$.
This suggests a concrete numerical strategy: starting from a converged small-$B$ solution, one can enlarge the self-energy representation while enforcing the known asymptotic constraints, and then update the hybridization sector through the cavity construction. In this way, small-$B$ ghost-GA solutions can be used as principled preconditioners for larger-$B$ calculations. Concrete numerical implementations of this strategy will be explored in future work.


\subsection{Benchmark calculations at finite temperature}
\label{Sec:Benchmark-temperature}

In this section we focus on the paramagnetic half-filled case, so that $\langle \hat n_i\rangle = 1$.
For the homogeneous single-site solution, the local variational matrices are site independent:
\begin{equation}
\R,\D\in\mathbb C^{2B\times 2},
\qquad
\Lambda,\Lambda^c\in\mathbb C^{2B\times 2B}
\,.
\end{equation}
The corresponding local embedding Hamiltonian is
\begin{align}
\hat H_{\rm emb}[\D,\Lambda^c]
&=
U\,\hat n_{\uparrow}\hat n_{\downarrow}
+
\sum_{a=1}^{2B}\sum_{\sigma=\uparrow,\downarrow}
\Big([\D]_{a\sigma}\,\cc_{\sigma}\bba_{a}+\mathrm{H.c.}\Big)
\nonumber\\
&\quad
+
\sum_{a,b=1}^{2B}
[\Lambda^c]_{ab}\,\bba_{b}\bbc_{a}
\,.
\end{align}

We denote the double occupancy by:
\begin{equation}
d=\langle \hat n_{\uparrow}\hat n_{\downarrow}\rangle
\,,
\end{equation}
which is computed as
\begin{equation}
d
=
\frac{
\Tr\!\left(
e^{-\beta \hat H_{\rm emb}[\D,\Lambda^c]}\,
\hat n_{\uparrow}\hat n_{\downarrow}
\right)}
{
\Tr\!\left(
e^{-\beta \hat H_{\rm emb}[\D,\Lambda^c]}
\right)
}
\,.
\label{eq:d_finiteT_benchmark}
\end{equation}

The internal energy per site is
\begin{align}
e
&=\frac{\mathcal{E}}{N}=
U\,d+e_{\rm kin},
\\
e_{\rm kin}
&=
\int d\eps\,\rho(\eps)\,\eps\,
\Tr\!\Big[
\R^\dagger\,
n_F\!\big(\eps\,\R\R^\dagger+\Lambda\big)\,
\R
\Big]
\,,
\label{eq:e_finiteT_benchmark}
\end{align}
where
\begin{equation}
n_F(X)=\big(e^{\beta X}+\mathbf 1\big)^{-1}
\end{equation}
is the Fermi function and $\rho(\eps)$ is the semicircular density of states defined in Eq.~\eqref{eq:semicircular}.

\begin{figure}[t]
\centering
\includegraphics[width=\columnwidth]{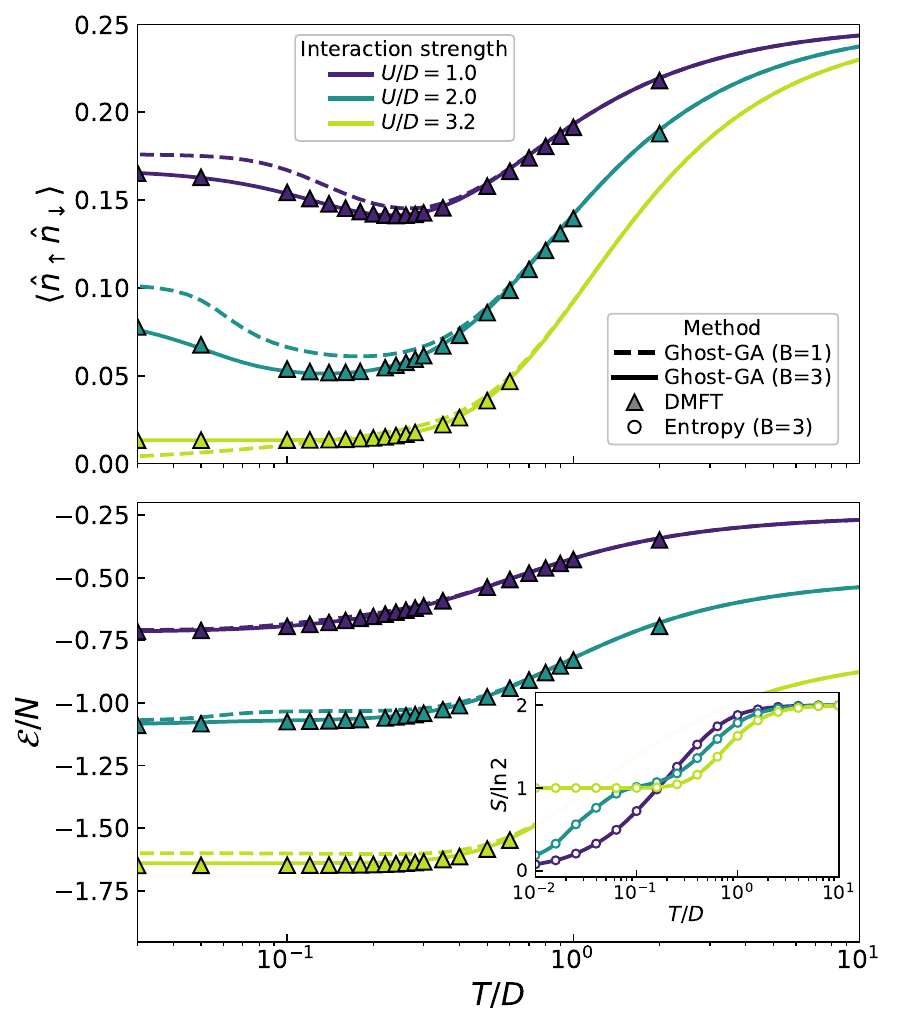}
\caption{
Finite-temperature ghost-GA results for the half-filled paramagnetic single-band Hubbard model on the Bethe lattice. Upper panel: double occupancy $d=\langle \hat{n}_{i\uparrow}\hat{n}_{i\downarrow}\rangle$ as a function of temperature for $U/D=1.0$, $2.0$, and $3.2$. Lower panel: $(\mathcal{E}-\mu N)/N$ for the same interaction strengths. Dashed and solid lines denote ghost-GA results for $B=1$ and $B=3$, respectively, while triangles denote DMFT data produced using NRGLjubljana~\cite{zitko_2021_4841076,Zitko2009}, an impurity solver employing the numerical renormalization group~\cite{Bulla2008}. The inset shows the entropy $S/\ln 2$ obtained for $B=3$.
}
\label{Figure7}
\end{figure}

Figure~\ref{Figure7} shows the temperature evolution of the double occupancy, energy, and entropy for three representative interaction strengths. 
We compare the $B=1$ and $B=3$ ghost-GA solutions with DMFT data obtained using a state-of-the-art numerical renormalization group~\cite{Bulla2008} impurity solver, NRGLjubljana~\cite{zitko_2021_4841076,Zitko2009}. The $B=3$ ghost-GA captures the DMFT behavior with high accuracy in all parameter regimes.

At fixed temperature, $d$ is progressively suppressed as $U/D$ increases, reflecting the reduction of charge fluctuations. For $U/D=1.0$ and $2.0$, the $B=3$ curves display a nonmonotonic temperature dependence: $d$ is reduced upon heating out of the low-temperature coherent regime, and then increases again as the system crosses over toward the high-temperature atomic limit. For $U/D=3.2$, by contrast, the double occupancy remains strongly suppressed over a broad temperature window, consistent with the onset of Mott-like local-moments.

The entropy shown in the inset is evaluated directly from the stationary value of the finite-temperature ghost-GA functional. This follows from the DMFT functional structure derived above: \(\mathcal L_\beta\) is the restriction of the DMFT functional to the finite-pole manifold generated by \((\R,\Lambda,\D,\Lambda^c)\), so its saddle-point value \(\mathcal L_\beta^\star\) provides the corresponding approximation to the grand potential. Accordingly, we evaluate it as:
\begin{equation}
S=\beta\left[{(E-\mu N)-\Omega_\beta}\right]
=\beta\left[(E-\mu N)-\mathcal L_\beta^\star\right]
\,.
\end{equation}
The value $\ln 4$ corresponds to all four local states of a single Hubbard site being thermally accessible. At stronger coupling, the $B=3$ results develop a broad regime with $S\simeq\ln 2$, which signals that charge fluctuations are frozen while the local spin-$\tfrac{1}{2}$ doublet remains active. 

These results indicate that, already at $B=3$, our finite-temperature ghost-GA extension captures the correct thermodynamic structure of both the double occupancy and the entropy in all parameter regimes, including the Mott phase.
Another remarkable feature of the present formalism is that, in contrast with previous finite-temperature extensions of the GA~\cite{finiteT-GA-Wang,finiteT-GA-fabrizio,Temperature-GA-NL}, its implementation remains very close to the $T=0$ ghost-GA scheme, the only formal modification being the replacement of ground-state expectation values by thermal averages. 
This provides a concrete route to low-cost finite-temperature calculations over broad parameter ranges, including the study of competing order parameters down to low temperatures.


\section{Conclusions}
\label{sec:conclusions}

We introduced a functional reformulation of the variational problem in ghost-GA~\cite{Ghost-GA,ALM_g-GA,gRISB}, in which the stationary point is determined by density-matrix matching conditions among three auxiliary Hamiltonians, recovering the ghost-DMET quantum-embedding formulation of the theory~\cite{gDMET}.
Using this formulation, we proved that, in the limit of infinitely many auxiliary bath modes, ghost-GA becomes strictly equivalent to DMFT~\cite{dmft_book,Anisimov_DMFT,Held-review-DMFT,DMFT,xidai_impl_LDA+DMFT,LDA+U+DMFT}.
This unified picture of seemingly distinct QE frameworks has immediate consequences.

A first remarkable consequence of this theorem is that it places on rigorous footing a central physical insight behind ghost-GA: although formulated as a ground-state variational theory, it does not discard the dynamical content of the correlated problem. Rather, the full local dynamical information recovered in DMFT is encoded in the same static variational parameters that optimize the ghost-GA ground state, so that the same variationally optimized local map that defines the ground state also yields an explicit wavefunction representation of the physical single-particle excitations, including both quasiparticle and Hubbard bands~\cite{Ghost-GA,ALM_g-GA,Gebhard-FL}. This, in turn, gives a rigorous foundation to the quasiparticle wavefunction perspective enabled by ghost-GA, which has already proved fruitful in the description of topologically nontrivial Hubbard bands with protected edge states and of neutral spinon excitations that re-emerge as heavy-fermion bands by proximity~\cite{Pasqua-2025,gGA-PhysRevB.111.125110}.
Our theorem also opens the possibility of calculating both ground-state observables and spectral properties with highly efficient ground-state impurity solvers, including MPS~\cite{itensor,block2,DMRG-REVIEW,DMRG-original-White-PRL,DMRG-original-White-PRB,DMRG_PhysRevB.104.115119,TH2,TH3}, neural quantum states (NQS)~\cite{carleo2017solving,sharir2022neural,chen2024empowering,ghost-NeuralNetworks,GOLDSHLAGER2024113351,levine2019quantum,yu2024solving,Zhou-2025}, methods based on coupled-cluster (CC) theory~\cite{CC1,CC2,CC3,CCSD-T-1,CCSD-T-2,CCSD-T-3,Sun2026-stochastic-CCSD}, variational impurity solvers based on superpositions of Gaussian states~\cite{BravyiGosset2017,Bauer-impurity-solver,Hogan2025EfficientQuantumImplementationDMFT}, quantum-assisted methods~\cite{Sriluckshmy2025,AVQITE,Error-mitigation-GPR,Chen2025,Kirby2026ObservationImprovedAccuracy,Rigo2025OperatorLanczosNQS}, and machine learning frameworks~\cite{surrogate-Marius,Linear-foundation-model-ghostGA-2025,Rende-foundation-NNQS,Zaklama2025AttentionBased-foundation-attention}.

A further immediate consequence of our theorem connecting ghost-GA and DMFT is a principled finite-temperature extension of ghost-GA. 
The algorithmic structure of this extension is unchanged, except that the ground-state expectation values of the auxiliary reference systems are replaced by the corresponding thermal averages. 
Our benchmark calculations for the single-band Hubbard model on the Bethe lattice indicate that this construction reproduces DMFT thermodynamics with high accuracy already at small bath size.
This framework opens a promising opportunity: because the finite-temperature ghost-GA self-consistency conditions can be enforced from static expectation values of finite-size impurity Hamiltonians, without explicitly computing spectra, one can seek and develop impurity solvers specifically tailored to this structure, with the prospect of substantially reducing computational cost while retaining high accuracy. Promising candidate approaches already available in this direction include tensor-network thermal-state methods~\cite{Verstraete2004MatrixPDO,Feiguin2005FiniteTDMRG,White2009METTS,stoudenmire2010METTS}, thermal pure quantum-state approaches~\cite{sugiura2012TPQ,Sugiura2013CanonicalTPQ}, finite-temperature extensions of the coupled-cluster formalism~\cite{sanyal1992thermal,white2018timeTCC,hummel2018finiteTCC,harsha2019thermofieldCC}, and quantum algorithms~\cite{wu2019variational,motta2020determining,getelina2023adaptiveQMETTS,white2020,Harsha2022}, while potentially avoiding major bottlenecks of conventional impurity solvers, such as the fermionic sign problem in continuous-time quantum Monte Carlo~\cite{ctqmc-sign-problem_PhysRevB.101.045108,ctqmc-sign-problem_PhysRevLett.126.216401}.
In particular, this could enable explicit ab-initio calculations of the superconducting transition temperature \(T_c\) in realistic cluster embedding studies of high-\(T_c\) materials~\cite{Haule-Cuprates-PhysRevX.15.021071}. 
Likewise, this could enable correlated finite-temperature phase-diagram calculations over broad pressure-temperature ranges in planetary-interior materials~\cite{Ho-FeO-NatCommun-2024,Blesio-FeO-outer-core-2025,Rao-Zhu-2025}.
It could also enable quantitative predictions of magnetic anisotropy energies in rare-earth–transition-metal intermetallic compounds, as well as the associated temperature-driven spin-reorientation transitions. Addressing this problem across the rare-earth series requires going beyond conventional crystal-field approaches and treating crystal-field effects, hybridization, spin–orbit coupling, strong electron correlations, and finite-temperature fluctuations on equal footing~\cite{Patrick2018CalculatingMA,lee2025importance,zhang2025electronic,eps-Iron-PRL}. Our method could provide such a concrete route to predictive, materials-specific calculations of anisotropy and magnetic phase behavior in technologically important rare-earth magnets.
More broadly, this work provides a concrete starting point for future controlled extensions beyond DMFT within a variational quantum-embedding framework.

\section*{Author contributions}

N.L. led the project, developed the theoretical framework, and carried out the analytic derivations.
S.G., T.-H.L., Y.-X.Y., O.G. and A.R. implemented all algorithms numerically and performed numerical validations and benchmarks. 
All authors discussed the results and contributed to editing the manuscript.

\section*{Acknowledgments}
The authors acknowledge helpful discussions with M.~Fabrizio and A.~Georges. 
N.L. gratefully acknowledges funding from the National Science Foundation under Award No. DMR-2532771 and from the Simons Foundation (Grant No. 00024037). The Flatiron Institute is a division of the Simons Foundation.
Part of this work by Y.Y. was supported by the US Department of Energy (DOE), Office of Science, Basic Energy Sciences, Materials Science and Engineering Division, including the grant of computer time at the National Energy Research Scientific Computing Center (NERSC) in Berkeley, California. This part of research was performed at the Ames National Laboratory, which is operated for the US DOE by Iowa State University under Contract No. DE-AC02-07CH11358. T.-H.L. gratefully acknowledges funding from the National Science and Technology Council (NSTC) of Taiwan under Grant No. NSTC 112-2112-M-194-007-MY3 and the National Center for Theoretical Sciences (NCTS) in Taiwan.

\appendix

\section{Block-matrix identities}
\label{sec:woodbury_projected_inverses}

Let $A\in\mathbb{C}^{n\times n}$ and $C\in\mathbb{C}^{k\times k}$ be invertible, and let
$U\in\mathbb{C}^{n\times k}$.
Define the block matrix
\begin{align}
M
&=
\begin{pmatrix}
A & U\\
U^\dagger & C
\end{pmatrix}
\,.
\label{eq:block_matrix_M}
\end{align}
Assume that the matrices $A-U\,C^{-1}U^\dagger$ (equivalently $C-U^\dagger A^{-1}U$) are invertible.

\subsection{Block inverse of $M$}

Under the assumptions above, one has
\begin{widetext}
    
\begin{align}
M^{-1}
&=
\begin{pmatrix}
\big(A-U\,C^{-1}U^\dagger\big)^{-1}
&
-\big(A-U\,C^{-1}U^\dagger\big)^{-1}U\,C^{-1}
\\
-\,C^{-1}U^\dagger\big(A-U\,C^{-1}U^\dagger\big)^{-1}
&
C^{-1}+C^{-1}U^\dagger\big(A-U\,C^{-1}U^\dagger\big)^{-1}U\,C^{-1}
\end{pmatrix}
\,.
\label{eq:block_inverse_M}
\end{align}

\end{widetext}

\subsection{Projected inverse identity used in the main text}

Let $A\in\mathbb{C}^{n\times n}$ and $U\in\mathbb{C}^{n\times k}$, and let $C\in\mathbb{C}^{k\times k}$ be arbitrary.
Assume that $A$ and $A-UCU^\dagger$ are invertible, and define
\begin{align}
X &= U^\dagger A^{-1}U
\,,
\\
Y &= U^\dagger (A-UCU^\dagger)^{-1}U
\,.
\end{align}
Assuming that $X$ is invertible, one has
\begin{align}
\Big[U^\dagger(A\!-\!UCU^\dagger)^{-1}U\Big]^{-1}
&=
\big(U^\dagger A^{-1}U\big)^{-1}-C
\,.
\label{eq:woodbury_projected_inverse}
\end{align}

\subsubsection*{Proof.}
Let $Q=(A-UCU^\dagger)^{-1}$.
From $(A-UCU^\dagger)Q=\mathbf{1}$ it follows that
\begin{align}
Q\,U
&=
A^{-1}U+A^{-1}U\,C\,(U^\dagger Q U)
\,,
\\
Y
&=
X+XCY
\,.
\end{align}
Therefore $(\mathbf{1}-XC)Y=X$, i.e.\ $(X^{-1}-C)Y=\mathbf{1}$, which implies
$Y^{-1}=X^{-1}-C$ and yields Eq.~\eqref{eq:woodbury_projected_inverse}.
\hfill$\square$

\section{Schmidt decomposition theorem for Slater determinants in bipartite fermionic systems}
\label{app:schmidt_theorem}

Let $n_A,n_B\in\mathbb{N}$ and $n=n_A+n_B$. Consider fermionic modes
$\{a_i\}_{i=1}^{n_A}$ (subsystem $A$) and $\{a_{i+n_A}\}_{i=1}^{n_B}$ (subsystem $B$), and a quadratic Hamiltonian
\begin{align}
\hat{H}
&=\sum_{i,j=1}^{n} h_{ij}\,a_i^\dagger a_j
\,,
\qquad
h=h^\dagger\in\mathbb{C}^{n\times n}
\,,
\label{eq:app_H}
\end{align}
with block form
\begin{align}
h=
\begin{pmatrix}
h^{AA} & h^{AB}\\
h^{BA} & h^{BB}
\end{pmatrix}
\,,
\qquad
h^{BA}=(h^{AB})^\dagger
\,,
\label{eq:app_h_block}
\end{align}
i.e.:
\begin{align}
\hat{H}
&=
\sum_{i,j=1}^{n_A}[h^{AA}]_{ij}\,a_i^\dagger a_j
+\sum_{i,j=1}^{n_B}[h^{BB}]_{ij}\,a_{i+n_A}^\dagger a_{j+n_A}
\nonumber\\
&\quad+
\sum_{i=1}^{n_A}\sum_{j=1}^{n_B}
\Big([h^{AB}]_{ij}\,a_i^\dagger a_{j+n_A}+\text{H.c.}\Big)
\,.
\label{eq:app_H_blocks}
\end{align}

Let $\ket{\Psi_0}$ be the ground state of $\hat{H}$ with $N$ fermions, and assume
\begin{align}
n_A\le N\le n_B
\,.
\label{eq:app_N_assumption}
\end{align}

Define the $n_A\times n_A$ matrix
\begin{align}
[\rho_A^{0}]_{ij}= \langle\Psi_0|a_i^\dagger a_j|\Psi_0\rangle,
\qquad (i,j=1,\dots,n_A)
\,.
\label{eq:app_rhoA0}
\end{align}
Assume that $\rho_A^{0}(\mathbf{1}_{n_A}-\rho_A^{0})$ is invertible.
Define the $n_B\times n_A$ matrix $\B$ with entries
\begin{align}
\B_{jk}
&=
\sum_{l=1}^{n_A}
\left[\frac{1}{\sqrt{\rho_A^{0}(\mathbf{1}_{n_A}-\rho_A^{0})}}\right]_{kl}
\langle\Psi_0|a_l^\dagger a_{j+n_A}|\Psi_0\rangle
\,,
\label{eq:app_B}
\end{align}
where $j=1,\dots,n_B;\,k=1,\dots,n_A$.

Using $\B$, define $n_A$ modes in $B$ by
\begin{align}
b_k^\dagger=\sum_{j=1}^{n_B}\B_{jk}\,a_{j+n_A}^\dagger
\,,
\qquad (k=1,\dots,n_A)
\,.
\label{eq:app_b}
\end{align}
The modes $\{b_k\}_{k=1}^{n_A}$ are independent canonical fermionic modes.

Then there exist additional independent modes $\{c_{B,l}^\dagger\}_{l=n_A+1,\dots,N}$ within $B$ such that, defining
\begin{align}
\ket{\psi_C}=c_{B,n_A+1}^\dagger\cdots c_{B,N}^\dagger\ket{0}
\,,
\label{eq:app_psiC}
\end{align}
one can write
\begin{align}
\ket{\Psi_0}=\ket{\Psi^{\mathrm{emb}}_0}\otimes\ket{\psi_C}
\,,
\label{eq:app_tensor_product}
\end{align}
where $\ket{\Psi^{\mathrm{emb}}_0}$ is the ground state (in the sector with $n_A$ fermions) of the quadratic Hamiltonian acting on the $2n_A$ active modes $\{a_i\}_{i=1}^{n_A}$ and $\{b_i\}_{i=1}^{n_A}$:
\begin{align}
\hat{H}^{\mathrm{emb}}_A
&=
\sum_{i,j=1}^{n_A}[h^{AA}]_{ij}\,a_i^\dagger a_j
+\sum_{q,q'=1}^{n_A}\big[\B^\dagger h^{BB}\B\big]_{qq'}\,b_q^\dagger b_{q'}
\nonumber\\
&\quad+
\sum_{i,q=1}^{n_A}\Big(\big[h^{AB}\B\big]_{iq}\,a_i^\dagger b_q+\text{H.c.}\Big)
\,.
\label{eq:app_HembA}
\end{align}
Moreover, the one-body correlators of the active modes satisfy
\begin{align}
\langle\Psi_0|a_i^\dagger a_j|\Psi_0\rangle&=
[\rho_A^{0}]_{ij}
\,,
\label{eq:app_active_correlator1}
\\
\langle\Psi_0|a_i^\dagger b_j|\Psi_0\rangle&=
\big[\rho_A^{0}(\mathbf{1}_{n_A}-\rho_A^{0})\big]^{\frac{1}{2}}_{ij}
\,,
\label{eq:app_active_correlator2}
\\
\langle\Psi_0|b_i^\dagger b_j|\Psi_0\rangle&=
\big[\mathbf{1}_{n_A}-\rho_A^{0}\big]_{ij}
\,.
\label{eq:app_active_correlator3}
\end{align}

\subsection{Expectation-value identity for the $A$--$B$ coupling}
\label{app:O_identity}

Assume that the one-body matrix $h$ depends on a complex matrix parameter $R$ only through
\begin{equation}
h^{AB}(R)=R\,\bar h^{AB},
\qquad
h^{BA}(R)=(h^{AB}(R))^\dagger,
\label{eq:app_hAB_factor_3}
\end{equation}
where $\bar h^{AB}$ is independent of $R$, while $h^{AA}$ and $h^{BB}$ are also independent of $R$.
Let $\ket{\Psi_0}$ be the $N$-fermion ground state of $\hat H$ and construct the bath modes $\{b_k\}_{k=1}^{n_A}$ from $\ket{\Psi_0}$ using Eqs.~\eqref{eq:app_B} and \eqref{eq:app_b}, so that Eq.~\eqref{eq:app_tensor_product} holds with the embedded state $\ket{\Psi^{\mathrm{emb}}_0}$.

For any $i=1,\dots,n_A$ and any index $\mu$ labeling the rows of $\bar h^{AB}$, we define the following operators:
\begin{align}
\hat O_{i\mu}= \sum_{j=1}^{n_B}[\bar h^{AB}]_{\mu j}\,a_i^\dagger a_{j+n_A},
\label{eq:app_O_def_2}
\\
\hat O^{\mathrm{emb}}_{i\mu}=
\sum_{k=1}^{n_A}\left(\sum_{j=1}^{n_B}[\bar h^{AB}]_{\mu j}\,\B_{jk}\right)a_i^\dagger b_k
\,.
\label{eq:app_Oemb_def_2}
\end{align}

Below we prove that
\begin{equation}
\langle\Psi_0|\hat O_{i\mu}|\Psi_0\rangle
=
\langle\Psi_0^{\mathrm{emb}}|\hat O^{\mathrm{emb}}_{i\mu}|\Psi_0^{\mathrm{emb}}\rangle
\,.
\label{eq:app_O_expect_identity_2}
\end{equation}

\subsubsection*{Proof.}

\begin{itemize}
\item First we establish the identity
\begin{equation}
\langle\Psi_0|a_i^\dagger a_{j+n_A}|\Psi_0\rangle
=
\sum_{k=1}^{n_A}\langle\Psi_0|a_i^\dagger b_k|\Psi_0\rangle\,\B_{jk}
\,.
\label{eq:app_AB_as_abB_2}
\end{equation}
Indeed, Eq.~\eqref{eq:app_B} implies
\begin{equation}
\langle\Psi_0|a_i^\dagger a_{j+n_A}|\Psi_0\rangle
=
\sum_{k=1}^{n_A}\big[\rho_A^{0}(\mathbf{1}_{n_A}-\rho_A^{0})\big]^{\frac{1}{2}}_{ik}\,\B_{jk},
\label{eq:app_AB_correlator_decomposition_3}
\end{equation}
and Eq.~\eqref{eq:app_active_correlator2} gives
\begin{equation}
\big[\rho_A^{0}(\mathbf{1}_{n_A}-\rho_A^{0})\big]^{\frac{1}{2}}_{ik}
=
\langle\Psi_0|a_i^\dagger b_k|\Psi_0\rangle
\,,
\label{eq:app_ab_equals_sqrt_3}
\end{equation}
so that Eq.~\eqref{eq:app_AB_as_abB_2} follows.

\item Now, using Eq.~\eqref{eq:app_O_def_2} and substituting Eq.~\eqref{eq:app_AB_as_abB_2},
\begin{align}
\langle\Psi_0|\hat O_{i\mu}|\Psi_0\rangle
&=
\sum_{j=1}^{n_B}[\bar h^{AB}]_{\mu j}\,\langle\Psi_0|a_i^\dagger a_{j+n_A}|\Psi_0\rangle
\nonumber\\
&=
\sum_{k=1}^{n_A}\left(\sum_{j=1}^{n_B}[\bar h^{AB}]_{\mu j}\,\B_{jk}\right)
\langle\Psi_0|a_i^\dagger b_k|\Psi_0\rangle
\,.
\label{eq:app_O_expect_intermediate_3}
\end{align}
Finally, since $\ket{\Psi_0}=\ket{\Psi_0^{\mathrm{emb}}}\otimes\ket{\psi_C}$ in Eq.~\eqref{eq:app_tensor_product} and $a_i^\dagger b_k$ acts only on the active $(A,b)$ modes, we have
\begin{equation}
\langle\Psi_0|a_i^\dagger b_k|\Psi_0\rangle
=
\langle\Psi_0^{\mathrm{emb}}|a_i^\dagger b_k|\Psi_0^{\mathrm{emb}}\rangle
\,.
\label{eq:app_ab_factorization_3}
\end{equation}
Substituting Eq.~\eqref{eq:app_ab_factorization_3} into Eq.~\eqref{eq:app_O_expect_intermediate_3} and comparing with the definition \eqref{eq:app_Oemb_def_2} proves Eq.~\eqref{eq:app_O_expect_identity_2}.

\end{itemize}

We note that, treating $R$ and $R^*$ as independent variables and using only the explicit dependence \eqref{eq:app_hAB_factor_3}, Eq.~\eqref{eq:app_O_def_2} coincides with the formal operator derivative 
\begin{equation}
\hat O_{i\mu}=\partial \hat H/\partial R_{i\mu}
\,.
\end{equation}
Similarly, viewing Eq.~\eqref{eq:app_HembA} as an operator-valued expression in which $\B$ enters as a coefficient matrix and the only explicit dependence on $R$ is through the $(A,b)$ hybridization term $h^{AB}(R)\B=R(\bar h^{AB}\B)$, Eq.~\eqref{eq:app_Oemb_def_2} coincides with 
\begin{equation}
\hat O^{\mathrm{emb}}_{i\mu}=\partial \hat H_A^{\mathrm{emb}}/\partial R_{i\mu}
\,.
\end{equation}

With the notation above, Eq.~\eqref{eq:app_O_expect_identity_2} can be formally rewritten as:
\begin{equation}
\left\langle\Psi_0\left|\frac{\partial \hat H}{\partial R_{i\mu}}\right|\Psi_0\right\rangle
=
\left\langle\Psi_0^{\mathrm{emb}}\left|\frac{\partial \hat H_A^{\mathrm{emb}}}{\partial R_{i\mu}}\right|\Psi_0^{\mathrm{emb}}\right\rangle
\,.
\label{eq:app_derivative_identity}
\end{equation}

\subsection{Expectation-value identity for the $A$--$A$ block}
\label{app:O_identity_AA}

Below we prove that, for any $i,j=1,\dots,n_A$,
\begin{equation}
\langle\Psi_0|a_i^\dagger a_j|\Psi_0\rangle
=
\langle\Psi^{\mathrm{emb}}_0|a_i^\dagger a_j|\Psi^{\mathrm{emb}}_0\rangle
\,.
\label{eq:app_AA_expect_identity}
\end{equation}

\subsubsection*{Proof.}

Since $\ket{\Psi_0}=\ket{\Psi^{\mathrm{emb}}_0}\otimes\ket{\psi_C}$ in Eq.~\eqref{eq:app_tensor_product} and $a_i^\dagger a_j$ acts only on the active modes $\{a_i\}_{i=1}^{n_A}$, Eq.~\eqref{eq:app_AA_expect_identity} follows immediately.
\hfill$\square$

We note that, treating $h^{AA}$ and $(h^{AA})^*$ as independent variables in Eqs.~\eqref{eq:app_H_blocks} and \eqref{eq:app_HembA}, the operator $a_i^\dagger a_j$ coincides with the formal derivatives
\begin{align}
a_i^\dagger a_j&=\partial \hat H/\partial h^{AA}_{ij}
\,,
\\
a_i^\dagger a_j&=\partial \hat H^{\mathrm{emb}}_A/\partial h^{AA}_{ij}
\,.
\end{align}

With the notation above, Eq.~\eqref{eq:app_AA_expect_identity} can be formally rewritten as
\begin{equation}
\left\langle\Psi_0\left|\frac{\partial \hat H}{\partial h^{AA}_{ij}}\right|\Psi_0\right\rangle
=
\left\langle\Psi^{\mathrm{emb}}_0\left|\frac{\partial \hat H^{\mathrm{emb}}_A}{\partial h^{AA}_{ij}}\right|\Psi^{\mathrm{emb}}_0\right\rangle
\,.
\label{eq:app_derivative_identity_AA}
\end{equation}

\subsection{Algebraic simplification of the bath block $\B^\dagger h^{BB}\B$}

Because $|\Psi_0\rangle$ is a Slater determinant ground state of the quadratic Hamiltonian
$\hat{H}=\sum_{i,j=1}^{n}h_{ij}\,a_i^\dagger a_j$, its one-body correlations can be written in terms of the
(zero-temperature) Fermi projector $n_F(h)$ associated to $h$:
\begin{align}
\langle\Psi_0|a_i^\dagger a_j|\Psi_0\rangle
&=
[n_F(h)]_{ji}.
\label{eq:corr_as_fermi_projector}
\end{align}
In the $A/B$ block notation,
\begin{align}
h
&=
\begin{pmatrix}
h^{AA} & h^{AB}\\
h^{BA} & h^{BB}
\end{pmatrix}.
\label{eq:block_h}
\end{align}
\begin{align}
n_F(h)
&=
\begin{pmatrix}
n_F^{AA} & n_F^{AB}\\
n_F^{BA} & n_F^{BB}
\end{pmatrix}.
\label{eq:block_nF}
\end{align}
Equation~\eqref{eq:corr_as_fermi_projector} implies $\rho_A^{0}=(n_F^{AA})^{T}$.
We also define the matrix:
\begin{align}
S_{lk}
&=
\Big[\rho_A^{0}(\mathbf{1}_{n_A}-\rho_A^{0})\Big]^{-1/2}_{kl}
\nonumber\\
&=
\Big[n_F^{AA}\big(\mathbf{1}_{n_A}-n_F^{AA}\big)\Big]^{-1/2}_{lk}
\,.
\label{eq:def_S}
\end{align}

Having established this notation, we have:
\begin{align}
\B_{jk}
&=
\sum_{l=1}^{n_A}
S_{lk}\,\langle\Psi_0|a_l^\dagger a_{j+n_A}|\Psi_0\rangle
\nonumber\\
&=
\sum_{l=1}^{n_A}
S_{lk}\,[n_F(h)]_{j+n_A,l}.
\label{eq:B_from_corr_and_S}
\end{align}
Hence, in matrix form,
\begin{align}
\B
&=
n_F^{BA}S
\,.
\label{eq:B_equals_fBA_S}
\end{align}
and the bath-bath block entering $H_{\rm emb}^A$ and the embedded hybridization matrix are:
\begin{align}
\B^\dagger h^{BB}\B
&=
S^\dagger\,n_F^{AB}\,h^{BB}\,n_F^{BA}\,S,
\label{eq:Bdag_hBB_B_start}
\\
h^{AB}\B
&=
h^{AB}\,n_F^{BA}\,S.
\label{eq:hyb_matrix}
\end{align}

Let $\Pi_A$ be the projector onto $A$ (so $\Pi_B=\mathbf{1}-\Pi_A$ projects onto $B$). Then
\begin{align}
\B^\dagger h^{BB}\B
&=
S^\dagger\,
\Pi_A n_F(h)\,\Pi_B\,
h\,
\Pi_B n_F(h)\,\Pi_A\,
S.
\label{eq:Bdag_hBB_B_projector_form}
\end{align}

Below we derive a useful alternative form for Eq.~\eqref{eq:Bdag_hBB_B_projector_form}.

\subsubsection*{Step 1}

We use the algebraic identity
\begin{align}
\Pi_B\,h\,\Pi_B
&=
\Pi_B\,h
+
h\,\Pi_B
+
\Pi_A\,h\,\Pi_A
-
h
\,,
\label{eq:projector_identity}
\end{align}
which follows from $\Pi_B=\mathbf{1}-\Pi_A$ by direct expansion.
Using Eq.~\eqref{eq:projector_identity} in Eq.~\eqref{eq:Bdag_hBB_B_projector_form} gives
\begin{align}
&\Pi_A n_F(h)\,\Pi_B h\Pi_B\,n_F(h)\,\Pi_A
=
\Pi_A n_F(h)\,\Pi_B h\,n_F(h)\,\Pi_A
\nonumber\\
&\qquad\qquad\qquad
+
\Pi_A n_F(h)\,h\Pi_B\,n_F(h)\,\Pi_A
\nonumber\\
&\qquad\qquad\qquad
+
\Pi_A n_F(h)\,\Pi_A h\Pi_A\,n_F(h)\,\Pi_A
\nonumber\\
&\qquad\qquad\qquad
-
\Pi_A n_F(h)\,h\,n_F(h)\,\Pi_A.
\label{eq:after_projector_identity}
\end{align}

\subsubsection*{Step 2}

We have
\begin{align}
n_F(h)^2
&=
n_F(h),
\nonumber\\
[n_F(h),h]
&=
0.
\label{eq:nF_proj_comm}
\end{align}
Therefore,
\begin{align}
\Pi_A n_F(h)\,h\,n_F(h)\,\Pi_A
&=
\Pi_A n_F(h)\,h\,\Pi_A.
\label{eq:nFhnF_simpl}
\end{align}
Moreover, the first two terms on the right-hand side of Eq.~\eqref{eq:after_projector_identity} are Hermitian conjugates.

\subsubsection*{Step 3}

We insert $\mathbf{1}=\Pi_A+\Pi_B$ once:
\begin{align}
\Pi_A n_F(h)\,\Pi_B h\,n_F(h)\,\Pi_A
&=
\Pi_A n_F(h)\,\Pi_B h\,\Pi_A n_F(h)\,\Pi_A
\nonumber\\
&
+
\Pi_A n_F(h)\,\Pi_B h\Pi_B\,n_F(h)\,\Pi_A.
\label{eq:term1_split}
\\
\Pi_A n_F(h)\,h\Pi_B\,n_F(h)\,\Pi_A
&=
\Pi_A n_F(h)\,\Pi_A h\Pi_B\,n_F(h)\,\Pi_A
\nonumber\\
&
+
\Pi_A n_F(h)\,\Pi_B h\Pi_B\,n_F(h)\,\Pi_A.
\label{eq:term2_split}
\end{align}
Denoting
\begin{align}
X
&=
\Pi_A n_F(h)\,\Pi_B h\Pi_B\,n_F(h)\,\Pi_A
\nonumber\\
&=
n_F^{AB}h^{BB}n_F^{BA},
\label{eq:def_X-B}
\end{align}
and using Eqs.~\eqref{eq:term1_split}--\eqref{eq:term2_split} in Eq.~\eqref{eq:after_projector_identity}, together with
Eq.~\eqref{eq:nFhnF_simpl}, we obtain an identity for $X$:
\begin{align}
X
&=
\Pi_A n_F(h)\,h\,\Pi_A
-
\Pi_A n_F(h)\,\Pi_A h\Pi_A\,n_F(h)\,\Pi_A
\nonumber\\
&\quad
-
\Pi_A n_F(h)\,\Pi_B h\,\Pi_A n_F(h)\,\Pi_A
\nonumber\\
&\quad
-
\Pi_A n_F(h)\,\Pi_A h\Pi_B\,n_F(h)\,\Pi_A.
\label{eq:X_from_identity}
\end{align}
Writing Eq.~\eqref{eq:X_from_identity} in $A/B$ block form yields
\begin{align}
n_F^{AB}h^{BB}n_F^{BA}
&=
\Pi_A n_F(h)\,h\,\Pi_A
-
n_F^{AA}h^{AA}n_F^{AA}
\nonumber\\
&\quad
-
n_F^{AB}h^{BA}n_F^{AA}
-
n_F^{AA}h^{AB}n_F^{BA}.
\label{eq:X_block_intermediate}
\end{align}

\subsubsection*{Step 4}

Since $[n_F(h),h]=0$ and both operators are Hermitian,
\begin{align}
\Pi_A n_F(h)\,h\,\Pi_A
&=
\frac{1}{2}\,\Pi_A\Big(n_F(h)\,h+h\,n_F(h)\Big)\Pi_A
\nonumber\\
&=
\frac{1}{2}\,\Pi_A n_F(h)\,(\Pi_A+\Pi_B)\,h\,\Pi_A
\nonumber\\
&\quad
+
\frac{1}{2}\,\Pi_A h\,(\Pi_A+\Pi_B)\,n_F(h)\,\Pi_A
\nonumber\\
&=
\frac{1}{2}\Big(
n_F^{AA}h^{AA}
+
n_F^{AB}h^{BA}
\nonumber\\
&\quad\qquad\qquad
+
h^{AA}n_F^{AA}
+
h^{AB}n_F^{BA}
\Big).
\label{eq:PiAnhPiA_symmetric}
\end{align}
Substituting Eq.~\eqref{eq:PiAnhPiA_symmetric} into Eq.~\eqref{eq:X_block_intermediate} and regrouping gives
\begin{align}
n_F^{AB}h^{BB}n_F^{BA}
&=
\frac{1}{2}\Big[
n_F^{AA}h^{AA}\big(\mathbf{1}_{n_A}-n_F^{AA}\big)
\nonumber\\
&\quad\qquad\qquad
+
\big(\mathbf{1}_{n_A}-n_F^{AA}\big)h^{AA}n_F^{AA}
\Big]
\nonumber\\
&\quad
+
\Big(\tfrac{\mathbf{1}_{n_A}}{2}-n_F^{AA}\Big)\,h^{AB}n_F^{BA}
\nonumber\\
&\quad
+
n_F^{AB}h^{BA}\Big(\tfrac{\mathbf{1}_{n_A}}{2}-n_F^{AA}\Big).
\label{eq:key_symmetric_identity}
\end{align}

\subsubsection*{Step 5}

From Eqs.~\eqref{eq:Bdag_hBB_B_start} and \eqref{eq:key_symmetric_identity},
\begin{align}
\B^\dagger h^{BB}\B
&=
S^\dagger\,
\Big(n_F^{AB}h^{BB}n_F^{BA}\Big)\,
S.
\label{eq:Bdag_hBB_B_via_X}
\end{align}
Since $S$ is a matrix function of $n_F^{AA}$, it commutes with $n_F^{AA}$, and we can rewrite
$h^{AB}n_F^{BA}S=h^{AB}\B$ using Eq.~\eqref{eq:hyb_matrix}.
Therefore,
\begin{widetext}
\begin{align}
\B^\dagger h^{BB}\B
&=
\frac{1}{2}\left[
\sqrt{\frac{n_F^{AA}}{\mathbf{1}_{n_A}-n_F^{AA}}}\;h^{AA}\;
\sqrt{\frac{\mathbf{1}_{n_A}-n_F^{AA}}{n_F^{AA}}}
+\text{H.c.}
\right]
+
\left[
\frac{\tfrac{\mathbf{1}_{n_A}}{2}-n_F^{AA}}
{\sqrt{n_F^{AA}\big(\mathbf{1}_{n_A}-n_F^{AA}\big)}}\;
(h^{AB}\B)
+\text{H.c.}
\right]
\nonumber\\
&=
\frac{1}{2}\left[
\sqrt{\frac{(\rho_A^{0})^{T}}{\mathbf{1}_{n_A}-(\rho_A^{0})^{T}}}\;h^{AA}\;
\sqrt{\frac{\mathbf{1}_{n_A}-(\rho_A^{0})^{T}}{(\rho_A^{0})^{T}}}
+\text{H.c.}
\right]
+
\left[
\frac{\tfrac{\mathbf{1}_{n_A}}{2}-(\rho_A^{0})^{T}}
{\sqrt{(\rho_A^{0})^{T}\big(\mathbf{1}_{n_A}-(\rho_A^{0})^{T}\big)}}\;
(h^{AB}\B)
+\text{H.c.}
\right]
\,.
\label{eq:Bdag_hBB_B_final}
\end{align}
\end{widetext}

Note that Eq.~\eqref{eq:Bdag_hBB_B_final} expresses $\B^\dagger h^{BB}\B$ solely in terms of $\rho_A^{0}$ and the hybridization matrix $h^{AB}\B$ [Eq.~\eqref{eq:hyb_matrix}]. Consequently, once $h^{AB}\B$ has been constructed, $\B^\dagger h^{BB}\B$ can be evaluated by simple $n_A\times n_A$ matrix operations, without explicitly forming $n_F^{AB}$ and $n_F^{BA}$ or carrying out additional sums over the $B$ degrees of freedom.


%

\end{document}